\documentclass{emulateapj}
\usepackage{natbib}
\newcommand{\onecolumn}{}

\newcommand{\Draft}{false}

\usepackage{amsmath}
\usepackage{amssymb}
\usepackage{xspace}
\usepackage{color}
\usepackage{ifthen}
\usepackage{xspace}

\definecolor{gray}{rgb}{0.5,0.5,0.5}

\newcommand{\XRBaiii}{zM}
\newcommand{\XRBaiv}{zm}
\newcommand{\XRBav}{ZM}
\newcommand{\XRBavi}{Zm}
\newcommand{\XRBa}[1]{%
\ifthenelse{#1 = 3}{\XRBaiii}{}%
\ifthenelse{#1 = 4}{\XRBaiv}{}%
\ifthenelse{#1 = 5}{\XRBav}{}%
\ifthenelse{#1 = 6}{\XRBavi}{}%
\xspace}

\newcommand{\Model}[1]{Model~\XRBa{#1}}
\newcommand{\Models}[1]{Models~\XRBa{#1}}

\newcommand{\powersep}{{\ensuremath{\times}}}

\newcommand{\g}{{\ensuremath{\mathrm{g}}}\xspace}
\newcommand{\K}{{\ensuremath{\mathrm{K}}}\xspace}
\newcommand{\cm}{{\ensuremath{\mathrm{cm}}}\xspace}
\newcommand{\yr}{{\ensuremath{\mathrm{yr}}}\xspace}
\newcommand{\km}{{\ensuremath{\mathrm{km}}}\xspace}
\newcommand{\Msun}{{\ensuremath{\mathrm{M}_{\odot}}}\xspace}
\newcommand{\Sec}{{\ensuremath{\mathrm{s}}}\xspace}
\newcommand{\hour}{{\ensuremath{\mathrm{h}}}\xspace}
\newcommand{\erg}{{\ensuremath{\mathrm{erg}}}\xspace}
\newcommand{\ergs}{{\ensuremath{\mathrm{erg}\,\mathrm{s}^{-1}}}\xspace}

\newcommand{\gcc}{{\ensuremath{\mathrm{g}\,\mathrm{cm}^{-3}}}\xspace}

\newcommand{\gscs}{{\ensuremath{\mathrm{g}\,\mathrm{cm}^{-2}\,\mathrm{s}^{-1}}}\xspace}

\newcommand{\keV}{{\ensuremath{\mathrm{keV}}}\xspace}
\newcommand{\MeV}{{\ensuremath{\mathrm{MeV}}}\xspace}
\newcommand{\ms}{{\ensuremath{\mathrm{ms}}}\xspace}

\newcommand{\gunit}{{\ensuremath{\mathrm{cm}^{-2}\,\mathrm{s}^{-1}}}\xspace}
\newcommand{\mue}{{\ensuremath{\mu_{\mathrm{e}}}}\xspace}
\newcommand{\Lb}{{\ensuremath{L_{\mathrm{b}}}}\xspace}
\newcommand{\Qb}{{\ensuremath{Q_{\mathrm{b}}}}\xspace}
\newcommand{\Qnuc}{{\ensuremath{Q_{\mathrm{nuc}}}}\xspace}
\newcommand{\LCNO}{{\ensuremath{L_{\mathrm{CNO}}}}\xspace}
\newcommand{\epsCNO}{{\ensuremath{\varepsilon_{\mathrm{CNO}}}}\xspace}
\newcommand{\epsta}{{\ensuremath{\varepsilon_{3\alpha}}}\xspace}
\newcommand{\epscool}{{\ensuremath{\varepsilon_{\mathrm{cool}}}}\xspace}
\newcommand{\MH}{{\ensuremath{M_{\El{H}}}}\xspace}
\newcommand{\trec}{{\ensuremath{\tau_{\mathrm{recur}}}}\xspace}
\newcommand{\Xav}{{\ensuremath{\left<X\right>}}\xspace}
\newcommand{\Yav}{{\ensuremath{\left<Y\right>}}\xspace}

\newcommand{\Msunyr}{{\ensuremath{\Msun\,\yr^{-1}}}\xspace}

\newcommand{\lSect}[1]{{\label{sec:#1}}}
\newcommand{\lFig}[1]{{\label{fig:#1}}}

\newcommand{\lTab}[1]{{\label{tab:#1}}}

\newcommand{\Tabff}[1]{{\ref{tab:#1}}}
\newcommand{\Tab}[1]{{Table~\Tabff{#1}}}
\newcommand{\Tabs}[1]{{Tables~\Tabff{#1}}}
\newcommand{\pan}[1]{{\textit{#1}}}
\newcommand{\PAN}[1]{{{#1}}}
\newcommand{\Pan}[1]{{Panel~\PAN{#1}}}
\newcommand{\Panff}[1]{{\PAN{#1}}}

\newcommand{\FIGFF}[2]{{\ref{fig:#2}\PAN{#1}}}
\newcommand{\Figff}[1]{{\FIGFF{}{#1}}}
\newcommand{\FIG}[2]{{Fig.~\FIGFF{#1}{#2}}}
\newcommand{\Fig}[1]{{\FIG{}{#1}}}

\newcommand{\Figure}[1]{{Figure~\FIGFF{}{#1}}}

\newcommand{\Sectff}[1]{{\ref{sec:#1}}}
\newcommand{\Sect}[1]{{\S~\Sectff{#1}}}

\newcommand{\isofont}[1]{{\mathrm{#1}}}
\newcommand{\isomass}[1]{{\ensuremath{\isofont{^{#1}}}}}
\newcommand{\isocharge}[1]{{\ensuremath{\isofont{_{#1}}}}}
\newcommand{\isotope}[3]{{\ensuremath{\isocharge{#1}\isomass{#2}\isofont{#3}}}}

\newcommand{\I}[2]{{\isotope{}{#1}{#2}}}
\newcommand{\El}[1]{{\I{}{#1}}}

\newcommand{\Ep}[1]{{\ensuremath{10^{#1}}}}
\newcommand{\Epp}[2]{{\ensuremath{10^{#1#2}}}}
\newcommand{\Eppp}[3]{{\ensuremath{10^{#1#2#3}}}}
\newcommand{\E}[1]{{\ensuremath{\powersep\Ep{#1}}}}
\newcommand{\EE}[2]{{\ensuremath{\powersep\Ep{#1#2}}}}

\newcommand{\Ye}{{\ensuremath{Y_e}}\xspace}
\newcommand{\Mdot}{{\ensuremath{\dot{M}}}\xspace}
\newcommand{\MdotEdd}{{\ensuremath{\dot{M}_{\mathrm{Edd}}}}\xspace}

\newcommand{\nue}{{\ensuremath{\nu_{\El{e}}}}\xspace}
\newcommand{\anue}{{\ensuremath{\bar{\nu}_{\El{e}}}}\xspace}
\newcommand{\betap}{{\ensuremath{e^{+}}}\xspace}

\newcommand{\Rag}{{($\alpha$,$\gamma$)}\xspace}
\newcommand{\Rga}{{($\gamma$,$\alpha$)}\xspace}
\newcommand{\Rpg}{{(\El{p},$\gamma$)}\xspace}

\newcommand{\Rap}{{($\alpha$,\El{p})}\xspace}
\newcommand{\Rng}{{(\El{n},$\gamma$)}\xspace}
\newcommand{\Rgn}{{($\gamma$,\El{n})}\xspace}

\newcommand{\Ran}{{($\alpha$,\El{n})}\xspace}

\newcommand{\Rpn}{{(\El{p},\El{n})}\xspace} 
\newcommand{\Rppg}{{(2\El{p},$\gamma$)}\xspace}
\newcommand{\Rbp}{{(\ensuremath{e^+}\nue)}\xspace}
\newcommand{\Rbm}{{(\ensuremath{e^-}\anue)}\xspace}

\newcommand{\Rpb}{{(\El{p},\betap\nue)}\xspace}

\newcommand{\REC}{{($e^-,\nu_{\rm e}$)}\xspace}

\newcommand{\Rad}{{($\alpha$)}\xspace}

\slugcomment{Draft for ApJ, \today} 

\shorttitle{Models for X-Ray Bursts}
\shortauthors{Woosley et al.}

\begin{document}

%% LaTeX will automatically break titles if they run longer than
%% one line. However, you may use \\ to force a line break if
%% you desire.

\title{Models for Type I X-Ray Bursts with Improved Nuclear Physics}

%% Use \author, \affil, and the \and command to format
%% author and affiliation information.
%% Note that \email has replaced the old \authoremail command
%% from AASTeX v4.0. You can use \email to mark an email address
%% anywhere in the paper, not just in the front matter.
%% As in the title, you can use \\ to force line breaks.

\author{S.~E.\ Woosley\altaffilmark{1}, A.~Heger\altaffilmark{2}, A.~Cumming\altaffilmark{1}, R.~D.~Hoffman\altaffilmark{3}, J.~Pruet\altaffilmark{3}, T.~Rauscher\altaffilmark{4}, H.~Schatz\altaffilmark{5}, B.~A.~Brown\altaffilmark{5}, \ \& M.~Wiescher\altaffilmark{6}}

\altaffiltext{1}{Department of Astronomy and Astrophysics,
University of California, Santa Cruz, CA 95064; woosley@ucolick.org, cumming@ucolick.org}
\altaffiltext{2}{Department of Astronomy and Astrophysics, University
of Chicago, 5640 S.\ Ellis Ave, Chicago, IL 60637, and Theoretical
Astrophysics Group, MS B227, Los Alamos National Laboratory, Los
Alamos, NM 87545; 1@2sn.org}
\altaffiltext{3}{N-Division, Lawrence Livermore National Laboratory, Livermore, CA 94550; rdhoffman@llnl.gov, pruet1@popcorn.llnl.gov}
\altaffiltext{4}{Department of Physics and Astronomy, University of Basel, Switzerland; Thomas.Rauscher@unibas.ch}
\altaffiltext{5}{Department of Physics and Astronomy and National Superconducting Cyclotron Laboratory, Michigan State University, East Lansing, MI 48824; schatz@nscl.msu.edu, brown@nscl.msu.edu}
\altaffiltext{6}{Physics Department, University of Notre Dame, Notre Dame, Indiana, 46556; Wiescher.1@nd.edu} 

\begin{abstract} 
Multi-zone models of Type~I X-ray bursts are presented that use an
adaptive nuclear reaction network of unprecedented size, up to 1300
isotopes, for energy generation and include the most recent
measurements and estimates of critical nuclear physics.  Convection
and radiation transport are included in calculations that carefully
follow the changing composition in the accreted layer, both during the
bursts themselves and in their ashes.  Sequences of bursts, up to 15
in one case, are followed for two choices of accretion rate and
metallicity, up to the point where quasi-steady state is achieved.
For $\Mdot=1.75\E{-9}\,\Msunyr$ (and $\Mdot=3.5\E{-10}\,\Msunyr$, for
low metallicity), combined hydrogen-helium flashes occur.  These
bursts have light curves with slow rise times (seconds) and long
tails. The rise times, shapes, and tails of these light curves are
sensitive to the efficiency of nuclear burning at various waiting
points along the \textsl{rp}-process path and these sensitivities are
explored. Each displays ``compositional inertia'' in that its
properties are sensitive to the fact that accretion occurs onto the
ashes of previous bursts which contain left-over hydrogen, helium and
CNO nuclei.  This acts to reduce the sensitivity of burst properties
to metallicity.  Only the first anomalous burst in one model produces
nuclei as heavy as $A=100$. For the present choice of nuclear physics
and accretion rates, other bursts and models make chiefly nuclei with
$A\approx64$.  The amount of carbon remaining after hydrogen-helium
bursts is typically $\lesssim 1$\% by mass, and decreases further as
the ashes are periodically heated by subsequent bursts.  For
$\Mdot=3.5\E{-10}\,\Msunyr$ and solar metallicity, bursts are ignited
in a hydrogen-free helium layer.  At the base of this layer, up to
90\% of the helium has already burned to carbon prior to the unstable
ignition of the helium shell. These helium-ignited bursts have a)
briefer, brighter light curves with shorter tails; b) very rapid rise
times ($<0.1\,$s); and c) ashes lighter than the iron group.
\end{abstract}

\keywords{neutron stars, X-ray bursts, nucleosynthesis}

\section{Introduction}
\lSect{intro}

Since the 1970's, when the nuclear instability of accretion onto
neutron stars was noted \citep{HH75}, and the connection with short,
transient X-ray flashes (Type I X-ray bursts) pointed out
\citep{WT76}, there have been numerous studies of thermonuclear
flashes on neutron stars. For reviews, see
\citealt{LE93,LE95,BI98,sb03}.  Studies in the mid-1980's elucidated
the relevant nuclear physics, known as the \textsl{rp}-process
\citep{WW81}, and showed that the most critical parameter determining
burst properties was the accretion rate, with several different
regimes of bursting behavior expected \citep{fuj81,fus87}. In
particular, combined helium and hydrogen burning powers flashes with
the lowest critical masses for accretion rates between
$4$--$11\E{-10}$ and $2\E{-8}$ $\Msunyr$ (depending upon the
metallicity of the accreted matter), whereas pure helium flashes occur
beneath a stable hydrogen shell for rates between $2\E{-10}$ and
$4$--$11\E{-10}\,\Msunyr$. Weakly flashing hydrogen shells occur for
rates less than $2\E{-10},\Msunyr$.  The generic properties predicted
for these flashes --- intervals, durations, energies, brightness,
etc. --- were in good agreement with observations, though the trends
in burst properties with changing accretion rate were often at odds
with the simple theory (e.g., \citealt{vplewin,bil00}).

Previous theoretical studies can generally be criticized, however,
either for oversimplification of the nuclear physics, especially the
use of small approximation networks for the energy generation, e.g.,
\citealt{Taa80,AJ82,WW84}, or for inadequate attention to the stellar
model, especially convection, in single-zoned studies that use large
reaction networks \citep{sch01a,sch01b,bro02a}.  In this paper, we
combine both: large networks and current nuclear data with leading edge
(albeit one-dimensional) stellar models that include convection and
multi-zone radiation transport.

Any such attempt to simulate Type I X-ray bursts realistically meets
with four challenges.  First is the physics of the accretion process
and the geometry of the runaway. Is the accretion uniformly
distributed over the surface of the neutron star prior to runaway and
does the burst commence almost simultaneously over that surface?
Observations of nearly-coherent oscillations during Type I bursts in
the last six years have brought these questions into focus (see
\citealt{sb03} for a review).  The oscillations, which are interpreted
as rotational modulation of brightness asymmetries, indicate rapid
rotation, and perhaps non-uniform ignition. If so, rotation may be key
to balancing the strong transverse pressure gradients that arise near
a local hot spot \citep{spit02,zing03}.  We do not address these
issues in this paper; our one-dimensional calculations address only the
local physics per unit area. While comparison to observables like the
light curve could be affected by the geometry, the nuclear physics of
the runaway is not.

Second, the nuclear physics is complex with no single or even several
reactions governing the energy generation rate. Recent years have
brought significant advances in our understanding of the major flows
in the \textsl{rp}- and \textsl{$\alpha$p}-processes and the
properties of the nuclei that govern them (e.g.,
\citealt{sch01a,sch01b,bro02a}).  These advances should be included in
any modern study.

A particularly exciting development has been the discovery of very
energetic, long duration Type I bursts, known as ``superbursts'' (see
\citealt{kuu02} for a review of superburst properties).  These flashes
have 1000 times the energy and duration of normal Type~I bursts, and
have been interpreted as unstable ignition of a thick carbon layer
\citep{CB01,SB02}, originally proposed by \citet{WT76} as a
gamma-ray burst model. Calculating the amount of carbon produced
during unstable hydrogen/helium burning, and how much carbon survives
to great depth, requires an accurate multi-zone calculation with a
large nuclear network.

The third challenge is following not just one, but many bursts.  As
was pointed out by \citet{Taa80}, the thermal and compositional
``inertia'' of the neutron star envelope can have important
implications for the properties of subsequent bursts. It is only by
computing a succession of X-ray bursts that the heating
associated with the previous thermonuclear flashes and compositional
change in the accreted layer can be taken into account
\citep{AJ82,WW84,taa93}.  Here we not only carry out fine-zoned
studies of individual bursts, but follow each case until a steady
state is achieved, including in one study, 15 consecutive bursts.

The fourth challenge is making contact with the rich archive of
photometry and spectra for observed X-ray bursts.  The models presented
here are limited to single-temperature black bodies calculated using
flux-limited radiative diffusion, although more detailed studies of
the spectrum and color temperature could use our results as input to a
more superior treatment of radiation transport.  An immediate
application of our results is direct comparisons with observed light
curves to study, for example, the physics underlying burst rise and
decay times. We do not attempt such a comparison in this initial
paper, but leave this for future work.

An outline of the paper is as follows.  In \Sect{kepler} we describe
modifications to the one-dimensional implicit hydrodynamics code --
\textsc{Kepler} -- used for this study. This code has been used before
to study X-ray bursts (e.g., \citealt{WW84,fus92,taa93,TWL96}) and we
need only clarify recent modifications to the nuclear reaction library
and the implementation of energy generation from an extended
network. A novel approach to evolving the abundance vector in each
zone, called an ``adaptive reaction network'', is employed in which
the number of isotopes employed in each cycle of each stellar model
may vary according to the active flows and significant abundances
present. At maximum, we employed over 1300 isotopes in the reaction
network for over 1000 zones at a time. The energy generation obtained
from this network is used in calculating the stellar model.

In subsequent sections we describe the results for four models that
crudely sample the accretion rates and compositions relevant for X-ray
bursts.  Two accretion rates, $3.5$ and $17.5\E{-10}\,\Msunyr$ and two
metallicities, $Z=0.001$ and $0.02$ are examined.  As expected, in
three of these combined hydrogen helium runaways were observed. In the
fourth, $Z= 0.02$ and $\Mdot= 3.5\E{-10}\,\Msunyr$, hydrogen burns away
before helium ignites in a thick shell that has already burned, at its
base, mostly to carbon.

In the results and conclusions sections we describe some of the
novel aspects of our results, the sensitivity to nuclear rates
employed, the light curves and intervals for the bursts, and the
composition of the ashes after many bursts have occurred, addressing
both the expected composition of the crust, and whether enough carbon
is produced to power a superburst.

\section{The Kepler Code and the Nuclear Data Employed}
\lSect{kepler}

\textsc{Kepler} is a one-dimensional implicit hydrodynamics code
adapted to the study of stellar evolution and explosive astrophysical
phenomena \citep{Wea78}.  It has been used for twenty years to
simulate X-ray bursts on neutron stars. The equation of state allows
for a general mixture of radiation, ions, and electrons of arbitrary
degeneracy and relativity. Electron-positron pair formation is also
accurately included. Convective mixing is modeled using mixing length
theory in a time-dependent manner. That is, the composition is
diffusively mixed with a diffusion coefficient calculated from the
convective velocity.  The convective criterion is Ledoux but with a
substantial semi-convective diffusion coefficient, about $10\,\%$ of
the radiative diffusion coefficient, in regions that are stable by the
Ledoux criterion and unstable by Schwarzschild.  A relatively small
amount of convective overshoot is included by flagging convective
boundary zones as semi-convective. The opacity, neutrino losses, and
other aspects of the code have been discussed in \citet{WHW02}.  In
particular, the recent OPAL and Los Alamos opacity tables are used
wherever the helium mass fraction exceeds \Ep{-5} and the temperature
is less than $\Ep9\,\K$.

Accretion is handled as in \citet{WW84} and \citet{TWL96}. In the
Lagrangian code, the outer boundary pressure is increased to simulate
the weight of the accreted matter at the given gravitational
potential. This continues until the pressure equivalent to the mass of
a new zone has accumulated. At that point a new zone is added to the
grid with conditions like those in the previous outer zone. The
surface boundary pressure is zeroed and the process begun anew.  All
runs accrete zones of $2\EE19\,\g$ and all accreted zones are
retained until the end of the run, i.e., there was no re-zoning of the
accreted layers.  Some models accumulated more than 1000 zones this
way.  Typically, the zones at the base of the accreted layer become
significantly thinner than $1\,\cm$!

During the accretion process, nuclear energy generation and
composition change are calculated at every time step. Convection is
always ``on'' in the sense that, if the Ledoux criterion for
instability is satisfied, the code responds. In this way we found
that, even though the strongest convection accompanied the onset of a
burst, there were interesting episodes of both convection and
thermohaline mixing in between the bursts.

\subsection{Adaptive network}
\lSect{adaptive}

The major improvement in the present study was a more
detailed treatment of the nuclear physics and energy
generation. Unlike previous multi-zone calculations that carried only
a very sparse ``approximation network'', we computed energy generation
and composition changes with a much larger network previously used only
to study nucleosynthesis.

Reaction rates were extracted from a library of nuclear data
(\Sect{rates}) carried in the calculation for isotopes ranging from
hydrogen to polonium ($Z=1$ to $84$).  When the abundance of an
isotope exceeded a threshold mass fraction of \Ep{-17}, all neighboring
isotopes that could be accessed by \Rng, \Rpg, \Rag, \Ran, \Rap, or \Rpn
reactions and their inverse, as well as \Rppg were added.  For \Rng and
\Rgn links the limit is somewhat smaller, \Ep{-18}. Additionally, all
possible decays, \REC, \Rbp, \Rbm, and \Rad, were included for all
isotopes in the network independent of their abundance.  Conversely,
when the mass fraction of an isotope dropped below \Ep{-23}, and its
presence is not necessary to satisfy any of the above conditions for
adding it, it was removed from the network. These criteria were
applied in all zones during each stage of the calculation and only one
network, the sum of all these local conditions, was used at any point
in time throughout the star. This is necessary since zones may mix at
unpredictable times and locations during the evolution.

\subsection{Binding energies}
\lSect{binding}

Nuclear flow through the \textsl{rp}-process waiting point nuclei is
exponentially sensitive to the proton separation energies ($S_p$) of a
few nuclei near the proton-drip line.  Experimental values for the
binding energies were taken from \citeauthor{aud95} (\citeyear{aud95};
AW95) where available (\Fig{spfig}). For many of the relevant nuclei
however, no experimental information is available and one must rely on
theoretical mass predictions.  For nuclei with $Z>N$ in the mass range
$A=41$--$75$, we used the compilation of \cite{bro02a}, in which the
experimentally determined masses of $N>Z$ nuclei, together with a
Skyrme Hartree-Fock calculation of Coulomb displacement energies, are
used to estimate masses for the $Z>N$ mirrors. These mass
predictions have an uncertainty of about 100 keV for the mass
difference to the mirror nucleus. This results in an average
uncertainty of the proton separation energies of about $140\,\keV$ if
the mirror masses are sufficiently well known.  As \Tab{sptbl} shows,
for many of the relevant \textsl{rp}-process nuclei this is not the
case and uncertainties are substantially larger, though still within a
few hundred \keV.

For another large set of nuclei, we used the unpublished calculations
of \citep{bro02b} (\Fig{spfig}).  These are obtained using the same
displacement energy method as in \citet{bro02a}, but the theoretical
error in the displacement energy is larger than 100 keV (though still
within a few 100 keV) because of the need to apply a spherical basis
to a mass region where some nuclei are strongly deformed. Furthermore,
for many of these heavier nuclei, the masses of the $N > Z$ mirrors
are not known experimentally. We then used the theoretical
extrapolations of AW95 as a basis for the displacement energy
method. The errors in the AW95 extrapolations for $A > 76$ introduced
an additional error of order 500 keV for nuclei near the Z = N
line. When no information is available either from AW95 or Brown, we
employed the theoretical estimates of \citet{mol95}.

\Tab{sptbl} shows the proton separation energies of several key nuclei
important in determining flow past the waiting point nuclei with long
half-lives against positron decay -- \I{64}{Ge}, \I{68}{Se}, \I{72}{Kr}, and
\I{76}{Sr}.  As discussed by \cite{orm97} and \cite{bro02a},
uncertainties in the masses of \I{64}{Ge} and \I{68}{Se} dominate
uncertainties in the proton separation energies of \I{65}{As} and
\I{69}{Br}, key nuclei in the \textsl{rp}-process.

\subsection{Reaction rates}
\lSect{rates}

Nuclear reaction rates were taken from experiment, shell-model
calculations, and statistical model (Hauser-Feshbach) calculations. An
experimentally determined rate was adopted whenever
possible\footnote{See \texttt{http://www-pat.llnl.gov/Research/RRSN/}
for sources of experimentally determined rates.}.  Largely, though,
experimental information was unavailable for the proton-rich nuclei
important in X-ray bursts.  For nuclei in the mass range $A=44$--$63$,
experimentally-undetermined \Rpg rates were taken from the shell model
calculations of \cite{fis01}.  Those authors also provide a critical
discussion of uncertainties inherent in the different methods of
calculating rates near the proton drip line.  All other reaction rates
were calculated using the Hauser-Feshbach code \textsc{Non-Smoker} as
described by \cite{rau00}.

\subsection{Weak rates}
\lSect{weakr}

Our calculations include electron capture
\begin{equation}
e^-  + \I{A}Z\rightarrow \I{A}{(Z-1)} + \nue,
\end{equation}
nuclear $\betap$ decay (positron emission)
\begin{equation}
\I{A}Z\rightarrow \I{A}{(Z-1)} + e^+ + \nue,
\end{equation}
and the neutrino energy loss rates associated with the above
processes.  For low temperature and density ($T_9<0.5$, $\rho
\Ye<10^5\,\gcc$), we adopt the ground state rates.  This is
appropriate for the proton rich nuclei of interest during
\textsl{rp}-process burning. These nuclei have decays characterized by
large Q-values, for which positron emission dominates over electron
capture. We take ground state weak decay rates from experiment where
available and from the compilation of \cite{mol97}
otherwise. \cite{mol97} calculated weak decay assuming only the
presence of Gamow-Teller (GT) transitions. As is discussed below, Fermi
transitions dominate the lifetimes of many near-proton-drip line
nuclei. For such nuclei the lifetime is typically a factor of $4$--$6$
shorter than that estimated from a consideration of GT transitions
alone. This issue and others relating to the weak physics of $N\sim Z$
nuclei will be addressed in a future study.

For nuclei with $A\leq60$ and for higher temperature and density
($T_9>0.5$, $\rho \Ye>10^5\,\gcc$), we include the influence on the
weak rates of thermal population of excited states as well as
continuum electron capture.  These rates are taken from the
compilation of \citeauthor{lan00} (\citeyear{lan00}, LMP), and from
the compilation of \citeauthor{ffn} (\citeyear{ffn}, FFN) for nuclei
not studied by \citeauthor{lan00}.  With the exception of a slightly
different interpolation of neutrino energy loss rates, interpolation
of the weak rates is done following the prescription of \cite{ful85}.

Thermal effects on weak rates for near-proton-drip line nuclei in the
mass range $A\sim 60$--$100$ follow simple systematics.  Consideration
of these systematics can be used to determine if our use of ground state
lifetimes for these nuclei is reasonable.  \Fig{square} illustrates
important nuclear and weak flows near the waiting point nucleus
\I{72}{Kr}.  These flows are typical of those near the other waiting
point nuclei in this mass region.

As can be seen from \Fig{square}, important weak decay parents fall
into four categories.  Nuclei with $Z=N+1$ undergo \betap-decay to
their isospin mirrors. For these nuclei, the ground state of the
(\betap) daughter is the isobaric analog state (IAS) of the parent
ground state, the first excited state of the daughter is the IAS of
the first excited state in the parent, and so on for all of the
levels. Because the nuclear Hamiltonian commutes with the isospin
raising and lowering operators, the Q-values and rates for all of
these (parent level)$\rightarrow$(IAS in daughter) Fermi transitions
are approximately the same.  GT transitions are not expected to
dominate the decay rates because the GT strength is typically spread
out over a wide range in daughter excitation energy. In addition,
electron capture cannot compete with these large Q-value $\beta^+$
decays at the small temperatures and electron Fermi energies achieved
in X-ray bursts. Consequently, the decay rates of $Z=N+1$ nuclei are
essentially temperature and density independent (for the range of
conditions found in X-ray bursts). For nuclei with $Z=N+2=$ even, the
situation is similar. Each parent state decays via a Fermi transition
to an IAS in the daughter and the decay rates for these nuclei are
well approximated by the ground state decay rate.

Nuclei with $N=Z=$ odd also decay principally via Fermi transitions.
However, while for these nuclei it is true that every low-lying
daughter state has a low-lying IAS in the parent, the converse is not
true.  Parent states in these $N=Z=$ odd nuclei that do not decay via
a Fermi transition typically decay about an order of magnitude slower
than do states with a low lying IAS in the even-even
daughter. Consequently, \betap-rates for these nuclei are
approximately proportional to $G_{\mathrm{daughter}}
/G_{\mathrm{parent}}$ \citep{pf02}.  Here $G$ is the partition
function and the ratio is a rapidly decreasing function of temperature
because of the high level density of odd-odd nuclei (relative to
even-even nuclei).

The last, and most important, class of $Z\sim N$ nuclei are those with
$Z=N=$ even.  Of these, only two have partition functions at $T_9=1.5$
(\I{76}{Sr} with $G\approx 1.7$ and \I{80}{Zr} with $G\approx 1.6$)
large enough for the decay of excited states to determine the lifetime
(see also \cite{sch98} for a discussion of $2^+$ lifetimes in X-ray
bursts and estimates of $2^+$ excitation energies for $Z>82$ nuclei).
For these two nuclei the thermal decay rate may be larger by a factor
of $\sim 2-3$ than the ground state decay rate, though a more
reasonable estimate is probably in the $30\,\%$ range.  As with the
other important decays, electron capture is negligible for $N=Z=$ even
nuclei during \textsl{rp}-process burning.

Detailed calculations of thermal effects on the weak decays of $A>60$
nuclei should eventually be incorporated into X-ray burst studies.
However, the above considerations suggest that the adoption of
laboratory ground state rates is not too bad. One shortcoming of our
procedure concerns $N=Z=$ odd nuclei. We typically underestimate the
lifetime of these nuclei in the X-ray burst environment by about a
factor of 5. For $A<80$ this error is not expected to be important
because so little of the nuclear flow ($\sim 1\,\%$ or less) proceeds
through the \betap-decay of these nuclei.  With increasing atomic
mass number the proton drip line tends closer to $N=Z$. Errors in the
lifetimes of $N=Z=$ odd nuclei may have a larger influence on nuclear
flow for $A>80$.  However, a qualitative change in the flow is
unlikely because the typical lifetimes of important $A>80$ nuclei are
so short ($\sim1\,\Sec$) compared to lifetimes of long-lived waiting
point nuclei ($\approx 64\,\Sec$ for \I{64}{Ge}).

\subsection{Neutrino Losses from Radioactive Decays}
\lSect{vogel}

Neutrino losses during weak decays are an important part of the total
energy budget, since neutrinos typically carry away a sizeable
fraction (half or more) of the energy available in a decay.  When a
weak decay rate is estimated from the compilation of LMP or FFN, we
also adopt the LMP or FFN value for the neutrino energy loss rate
associated with that decay.  For a few light nuclei (\I7{Be}, \I{13}N,
\I{14,15}O, \I{17,18}F), neutrino energy loss rates are calculated
using experimentally determined ground state weak strength
distributions.  Other neutrino energy loss rates were estimated using
a code provided to us by Petr Vogel.  This code uses an empirical form
of the GT$^+$ strength distribution to estimate the average energy of
neutrinos emitted in a decay.  The neutrino energy loss rate is then
taken to be the product of the weak decay rate and the average
neutrino energy.  Though empirical strength distributions cannot
reliably reproduce ground state decay rates, they can do a fair job of
estimating average neutrino energies because phase space so heavily
favors those transitions with the most energetic outgoing neutrinos.
For decays characterized by large Q-values, Vogel's estimates of
average neutrino energies typically agree with more detailed estimates
(FFN, LMP) to within $10\,\%$ or $15\,\%$.

\section{Initial Models and Setup}
\lSect{models}

Four different combinations of accretion rate and metallicity were
examined (\Tab{summary}).  The first model, \XRBa3, employs conditions
very similar to \citet{sch01a}: an accretion rate of $0.1\,\MdotEdd =
8.8\E3\,\gscs = 1.75\EE-9\,\Msunyr$ for a $10\,\km$ \ radius neutron
star.  For comparison, we also calculated a similar model, \XRBa4,
with an accretion rate one-fifth as large. The composition in both
cases was taken to be $75.9\,\%$ hydrogen, $24\,\%$ helium, and
$0.1\,\%$ \I{14}N by mass, corresponding approximately to the $Z
=\Epp-3$ used by \citeauthor{sch01a}. Hereafter this is referred to as
the $5\,\%$ solar composition, since the total mass fractions of CNO
in the sun are about $20$ times larger.  For comparison, we also
calculated similar models \XRBa5 and \XRBa6 with solar metallicity
($70.48\,\%$ \I1H, $27.52\,\%$ \I4{He}, $2\,\%$ \I{14}N) and high and
low accretion rates respectively. These are similar to the
\citet{ag89} solar abundances, but the rearrangement of the CNO
isotopes into nitrogen that naturally occurs in the CNO cycle was
skipped since this will occur rapidly at the high temperatures of the
accretion process and the energy added is inconsequential compared
with that of accretion.

To save time and improve energy conservation, only the outer
$2\EE25\,\g$ of neutron star crust is carried in the calculation.
This is orders of magnitude larger than the mass of all X-ray bursts
combined, so the layer essentially acts as a large repository of
thermal inertia. Its composition is taken to be iron and no nuclear
reactions are followed in this layer. We take the luminosity at the
base of the substrate to be $3.2\EE33\,\ergs$ for \Models4 and \XRBa6
and $1.6\EE34\,\ergs$ for \Models3 and \XRBa5, i.e., the accretion
rate times $0.15\,\MeV$/nucleon as in \citeauthor{sch01a}.  Before
accretion is switched on, we allow the substrate to relax to thermal
steady state with the power input at its base balanced by the
luminosity flowing into the accreted zones.

A $1.4\,\Msun$ neutron star with radius $R=10\,\km$ is adopted giving
a surface gravity $g=GM/R^2=1.9\E{14}\,\gunit$.  We do not include the
effects of general relativity in our simulations, we address the
relativistic corrections that must be applied to our results in
\Sect{gr}.

\section{Results}
\lSect{results}

\subsection{Model \XRBaiii}
\lSect{a3}

The model considered in greatest detail and followed though the
largest number of flashes was \Model3, whose parameters
duplicated those of previous studies by \cite{sch01a,sch01b}. The
accretion rate and metallicity (\Tab{summary}) imply that hydrogen
will survive to the depth where helium ignites, so that this should be
a representative case of a combined hydrogen-helium runaway
(\Sect{intro}). Previous studies have also suggested that the nuclear
flows should lead to the creation of quite heavy nuclei, with the
\textsl{rp}-process terminating in the \El{Sb}\El{Sn}\El{Te} cycle,
making this an interesting test case for the large adaptive network.

\subsubsection{The first burst in Model \XRBaiii}

Following accretion for $41,640\,\Sec$, during which a layer of
$4.66\EE21\,\g$ accumulated, the temperature at the base of the
accreted material began to rise rapidly compared with the accretion
time scale. The hydrogen and helium mass fractions at the base of the
accreted layer at that sample time (\Fig{comp1}) were 0.693
(down from an initial 0.759 due to stable hydrogen burning) and
0.283.  The temperature there was $2.67\E8\,\K$, and the density,
$8.87\E5\,\gcc$.  Most of the energy generation at this point,
$3.5\EE35\,\ergs$, was still coming from the $\beta$-limited CNO
cycle, though helium burning had increased the abundance of catalytic
nuclei to mass fractions $1.19\,\%$ of \I{14}O and $1.29\,\%$ \I{15}O,
respectively.

Fifty-three seconds later, at $41,693\,\Sec$, the temperature at the
base of the hydrogen had increased to $3.27\E8\,\K$ and the maximum
was starting to move outwards. The energy generation had risen by a
factor of 5.  Six seconds later, at $41,699\,\Sec$, energy began to be
transported by convection (\Fig{Aconv1}) when the maximum temperature
was $3.97\E8\,\K$.  This convection began $5\EE20\,\g$, or about half a
meter above the base of the hydrogen layer.  At this point, the mass
fractions at the base of the accreted layer were \I1H: $0.683$,
\I4{He}: $0.256$, \I{14}O: $0.037$, and \I{15}O: $0.016$.  Over the
next 7 seconds the convective region grew to encompass the entire
accreted layer, with the exception of the outer 2 or 3 zones and a
mass of $\sim 5\E{19}\,\g$.  As the convection neared the surface the
observable transient commenced (\Fig{lite1}).  At this time
$41,705\,\Sec$ (\Fig{comp2}), the composition still consisted almost
entirely of hydrogen and helium though appreciable amounts of heavy
elements were beginning to be synthesized. The nuclear power being
generated was $4.5\E{38}\,\ergs$, but only a small fraction of this
had appeared at the surface ($L = 1.5\E{35}\,\ergs$).  Convection
ceased during the next second.

From that point on the luminosity rose slowly, by diffusion, to
nearly the Eddington value, $2\E{38}\,\ergs$.  Qualitatively, it seems
that the energy is transported by convection until an adiabatic
temperature gradient is established to the surface. Accomplishing this
requires raising the temperature which leads to expansion.  The
necessary $P\mathrm{d}V$ work against the enormous gravitational
potential of the neutron star uses up most of the nuclear energy
release in the early stages of the run away. Once this has been
accomplished, convection shuts off and is unimportant during the
remainder of the burst.  The light curve, for this model, is powered by
diffusion in near steady state with the nuclear power. As the burning
region cools gravitational potential energy is converted back into
heat, but this is a small fraction of nuclear energy until the burst
is essentially over.
 
At $41,715\,\Sec$, for example, about $5$--$10\,\Sec$ into the burst,
the luminosity was $1.5\E{38}\,\ergs$ and all convection had
ceased. Nuclear energy generation, $1.6\E{38}\,\ergs$, was in near
steady state with the luminosity.  An appreciable gradient in hydrogen
was beginning to develop.  At the bottom of the layer, the most
abundant mass fractions were \I1H ($0.279$), \I4{He} ($0.077$),
\I{60}{Zn} ($0.460$), and \I{64}{Ge} ($0.116$), but half way out (in
mass) the most abundant mass fractions were \I1H ($0.593$), \I4{He}
($0.175$), \I{56}{Ni} ($0.030$), and \I{60}{Zn} ($0.156$).  The
temperature and density at the bottom of the layer were $1.51\E9\,\K$
and $4.37\E5\,\gcc$, near the maximum developed during the burst.

Energy generation from the \textsl{rp}-process and transport by
radiative diffusion continued to power a brilliant display with a long
tail lasting several hundred seconds (\Fig{lite1}) (n.b., the
luminosity from the burst must be added to the accretion luminosity,
$2\E{37}\,\ergs$ here, in all plots of the light curve).  The
composition $290\,\Sec$ after the burst began (at $41,992\,\Sec$) is
shown in \Fig{comp3}. At this point energy is continuing to be
generated by nuclear reactions happening in unburned hydrogen well
above the base of the accreted layer --- where hydrogen has burned
away. Energy generation from the decay of radioactivities produced at
the bottom of the layer is negligible.  The temperature and density of
the layer at this time are shown in \Fig{dntn1}.

\subsubsection{The second burst in Model \XRBaiii}

Following the first burst, accretion continued at
$1.75\E{-9}\,\Msunyr$ until a second critical mass had accumulated
(\Tab{a3p}). However, this time accretion occurred onto the ashes of
the previous burst, which contained unburned hydrogen and helium,
rather than onto an inert substrate. This distinction greatly altered
the conditions for, and nature of, the second and subsequent
bursts. 

\Fig{compp2} shows the major abundances at $54,592\,\Sec$ after
accretion began, about $30\,\Sec$ before the second runaway. Unburned
hydrogen and helium are abundant in the outer ashes of burst 1 - as
they must be in any burst that has not remained fully convective
throughout its burning.  Not only are hydrogen and helium abundant,
but so are \I{14}O and \I{15}O from helium burning during the
inter-burst period.  This production of CNO nuclei was previously
discussed by \cite{HF84}.  The mass fractions at maximum of $^{14}$O
and \I{15}O are $3.5\,\%$ and $4.3\,\%$ respectively.  Most of these
abundances are large because the second combined hydrogen-helium
runaway is already beginning, but even $1900\,\Sec$ earlier, at
$52,668\,\Sec$, the mass fractions were already $0.56\,\%$ and
$0.99\,\%$, about what the CNO processing would give for accreted
matter with solar metallicity - even though this was a ``low''
metallicity study.

Because of this, the critical mass for all bursts except the first one
is smaller, giving a shorter interval between bursts. This implies
bursts with shorter durations and less energy. Less extreme values of
density and temperature are also achieved and the products of the
\textsl{rp}-process are not so heavy. Since the second runaway
commences in the ashes of the first and quickly becomes convective,
some of the ashes of the first burst, as well as \I{14}O and \I{15}O
catalysts are mixed out into the second accreted layer. Since decays
have gone on during the inter-pulse period, this would be an
opportunity for the second burst to make even heavier nuclei than the
first one. However, the lower temperature and density are more
critical, and the final ashes of the second (and subsequent) pulses
are actually considerably lighter (\Fig{adistp13}).

Because the second burst is more typical for the assumed accretion
rate and metallicity, it is of some interest to examine its light
curve. \Fig{lite2r} shows that the rise to maximum is again relatively
slow (compare with \Fig{Aconv1}), occurring on a radiative diffusion
time for the accreted layer, about $10\,\Sec$.  This is the same
behavior seen in burst~1 and happens because the convection so
apparent in \Fig{lite2r} dies out before the surface luminosity rises
above the background value given by accretion. During this convective
stage, the star is far from steady state. The luminosity at the base
of the convective layer is orders of magnitude greater than at the
top, enabling the convection zone to extend throughout the layer
(e.g., \citealt{joss77}).  After this adiabatic condition has been
established, the envelope is able to carry the necessary flux by
radiative diffusion and the rise time for the luminosity is slow. As
we shall see this contrasts with the situation when a hydrogen
depleted helium shell runs away (\Sect{mod6}) and the luminosity rises
almost instantly.

We shall discuss details of the nuclear physics affecting the rise
time and tail of the light curve in \Sect{weak}. The full light curve
of the second burst using standard settings is given in \Fig{lite2}.

\subsubsection{Bursts numbers 3 through 14 in Model \XRBaiii}

Bursts $3$ through $14$ closely resembled burst $2$.  The composition
at the onset of the third burst is given in \Fig{compp3}. The burst
once again ignites in the outer ashes of the previous one.  An
interesting development is the gradual depletion of \I{12}C which had
significant abundance in the ashes of the first burst.  This is
processed to heavier elements by $\alpha$-captures when the heat wave
from subsequent bursts penetrates into the ashes. \Fig{comp3} and
\Figff{adistp13} show that these subsequent bursts also produce an
abundance distribution centered on lighter iron group elements,
notably $^{64}$Ge (which decays to $^{64}$Zn), though a tail of
significant production still extends to $A \sim 100$.

\Fig{conv3} shows the convective histories for the first three
flashes, once again emphasizing the difference between the first
violent, large mass burst and subsequent, mutually similar weaker
bursts.

The cumulative
bursting history is displayed in \Fig{Arep2} and shows that steady
state is achieved after only one burst. The composition after 14
bursts is shown in \Fig{lastcompa3} and the density-temperature
structure then is given in \Fig{lastdntn}.  At that time, the only
abundances lighter than $A = 60$ having mass fraction greater than
$0.001$ are \I{12}C ($0.0016$), \I{28}{Si} ($0.0068$), \I{32}S
($0.0037$), and \I{60}{Ni} ($0.0052$).  This very low carbon abundance
is insufficient to undergo unstable ignition in deeper layers and
power a superburst \citep{CB01}.

\subsubsection{Sensitivity of results to nuclear physics at the waiting points}
\lSect{weak}

No single reaction rate governs energy generation by the
\textsl{rp}-process.  Early on, energy is produced by helium burning
and the break out from the beta-limited CNO cycle. After the initial
destruction of \I{14}O and \I{15}O, the burning follows the reaction
sequence 3$\alpha\rightarrow$\I{12}C\Rpg\I{13}N\Rpg\I{14}O\-\Rap\I{17}F
\Rpg\I{18}{Ne}\Rap\I{21}{Na}\Rpg\I{22}{Mg}\Rap\I{25}{Al}$\ldots$.
Later, depending upon how far up in mass the flow has gone, hydrogen
burns at a rate sensitive to the capture cross sections,
photodisintegration rates (hence particle separation energies), and
weak-decay rates for progressively heavier nuclei. The
\textsl{rp}-process is very similar to the $r$-process in both its
dependence upon the properties of waiting point nuclei all along the
process path, and in that it involves nuclei whose properties are
poorly determined.

We undertook a study of the sensitivity of the burst light curve to
nuclear uncertainties. This is by no means the first such study. For
other recent efforts see \citealt{rem99,thi01,rau00b}.  Rather than vary
a large number of individual rates however, we chose, in this initial
survey, to vary groups of rates. One group is the collection of all
electron-capture and positron emission rates for nuclei heavier than
\I{56}{Ni}, i.e., the lifetimes of \I{57}{Cu} and all heavier unstable
nuclei. This set affects the flow from the iron group to heavier
nuclei, ultimately near $A\sim100$ for the first burst.  We varied the
rates up and down by one order of magnitude compared to the standard
values, most of which were due to measurements or estimates of the
ground state lifetime.  It is likely that our standard values
overestimate these lifetimes, so multiplying the rates by ten is
probably more reasonable than dividing them by ten, though we did both
(\Fig{weaktest1}).

It is to be emphasized that the key weak decay rates, e.g., for
\I{60}{Zn}, \I{64}{Ge}, \I{68}{Se}, etc., are not themselves uncertain
to an order of magnitude, but are probably known, even at these high
temperatures and densities, to a factor of two (\Sect{weakr}).  When
we vary these decay rates, we are really exploring how efficiently the
flow goes through critical waiting points by a variety of nuclear
reactions and equilibrium links, e.g., not just
\I{64}{Ge}\Rbp\I{64}{Ga}, but also, to some extent,
\I{64}{Ge}\Rpg\I{65}{As}\Rbp\I{65}{Ge},
\I{64}{Ge}\Rpg\I{65}{As}\Rpg\I{66}{Se} \Rbp\I{66}{As}, etc.  The rate
of these weak flows is sensitive to the quasi-equilibrium abundances
of nuclei like \I{65}{As} and \I{66}{Se} as well as their half-lives
against positron emission.

Flows may also stagnate before reaching \I{56}{Ni}, so we also carried
out a set of runs where all rates for unstable nuclei heavier than
$A=27$ were similarly varied.  Finally, to target more specifically
key reaction rates we varied individually the rates for groups of
nuclei in the vicinity of key waiting points around $A = 60$, $64$,
and $68$.  The results are shown in \Fig{zn60}.  In all cases, we
follow three bursts in model \XRBa3.  The first burst is more directly
comparable with the earlier work of \citet{sch01a}, while the third
may be regarded as more typical.

Our first observation is that the nuclear rates can affect the
\emph{rise} time of the burst as well as the tail of the light curve
(\Fig{rist3}). In the hydrogen flashes studied here, convection dies
out before the light curve has risen to a small fraction of its
maximum. One might expect the rise time to then be a consequence of
radiative diffusion. We find, however, that varying the decay rates
between $A=27$ and $56$ also has a direct effect on the rise.  Some
key nuclear physics affecting the flow from \I{26}{Si} to $A\gtrsim56$
during the rise of the burst are the decay rates of \I{33}{Ar}, and
\I{37,38}{Ca} and proton capture on \I{30}S and \I{34}{Ar}, both
inhibited by photodisintegration of\I{31}{Cl} and \I{35}K
respectively.  The flow to \I{56}{Ni} passes through \I{34}{Ar},
\I{35-38}K, \I{36-39}{Ca},\I{40-42}{Sc}, \I{41-43}{Ti}, \I{44,45}V,
\I{45,46}{Cr}, \I{47-49}{Mn}, \I{48-50}{Fe}, \I{51-55}{Co}, and
\I{52-56}{Ni}.

It is in the peaks and tails of bursts though that the nuclear physics
has its most dramatic consequences.  \Fig{weaktest1} and \Tab{a3p10}
show the results for the first burst from \Model3. The tail of
the burst involves continued hydrogen burning at a rate sensitive to
the positron-decay lifetimes and proton separation energies of nuclei
along the \textsl{rp}-process path. After the rise, the values for
rates between \El{Si} and \El{Ni} do not appear to be critical, unless
they are very small, but different choices for the rates above Ni can
have dramatic consequences on the shape, duration, and peak brightness
of the light curve.  Most important are the flows around $A = 60$,
$64$, and $68$.  The leakage through \I{60}{Zn} at a relevant
temperature and density ($1.5\E9\,\K$; $3\E5\,\gcc$) proceeds by
\I{60}{Zn}\Rpg\I{61}{Ga}\Rpg\I{62}{Ge}\Rpg\I{62}{Ga}\Rpg\I{63}{Ge}\Rbp
\I{63}{Ga}\Rpg\I{64}{Ge}, with the \Rpg reactions at \I{60}{Zn},
\I{61}{Ga} and \I{62}{Ga} strongly hindered by photodisintegration.
Most critical then are the half-lives of \I{62}{Ge} and \I{63}{Ge} and
proton separation energies of \I{61}{Ga}, \I{62}{Ge}, and \I{63}{Ge}.

In the vicinity of \I{64}{Ge} the critical flows at $T = 1.0\E9\,\K$
and $\rho = 2.5\E5\,\gcc$ have a different character.  The dominant
reactions are
\I{64}{Ge}\Rbp\I{64}{Ga}\Rpg\I{65}{Ge}\Rpg\I{66}{As}\Rpg\I{67}{Se}\Rpb
\I{67}{As}\Rpg\I{68}{Se} with critical decays at \I{64}{Ge} and
\I{67}{Se}.  Unlike at \I{60}{Zn}, the critical proton captures in
this case are not equilibrated, except for \I{66}{As}\Rpg\I{67}{Se},
and depend on cross section more than separation energies. The nucleus
\I{65}{As}, long thought to play a critical role in admitting
\textsl{rp}-process flow to heavier nuclei \citep{wal84} is not
critical here so long as its proton separation energy is as negative
as indicated in \Tab{sptbl}.  The nucleus \I{66}{Se} is in
quasiequilibrium with \I{65}{As} at $1.5\E9\,\K$ but drops out by
$1.0\E9\,\K$ so the small abundance of \I{65}{As} is a hindrance in
its production.

Near \I{68}{Se}, the flows are similar to \I{64}{Ge}:
\I{68}{Se}\Rpb\I{68}{As}\Rpg\I{69}{Se}\Rpg\I{70}{Br}\Rpg\I{71}{Kr}\Rbp
\I{71}{Br}\Rpg\I{72}{Kr} with critical decays at \I{68}{Se} and
\I{71}{Kr}.

For calculations using faster weak rates, the burst is
brighter and decays more quickly, as one may expect. The converse is
true for slower weak rates. \Tab{a3p10} shows that the effects of
compositional inertia persist regardless of the choice of rates. It is
diminished just a little for faster weak rates because more hydrogen
burns away in the outer layers. Conversely it is more dramatic in the
unlikely case that the rates are much slower.

The effective lifetime of the waiting point nuclei \I{64}{Ge},
\I{68}{Se}, \I{72}{Kr} also depends on the rates of two-proton-capture
reactions discussed by \cite{sch98}. The reaction formalism applied
here treats these processes as two sequential proton capture
reactions. The reaction rates depend very sensitively on the
associated masses as already pointed out by \cite{bro02a}. While the
Audi-Wapstra masses used here result only in a weak flow through the
two-proton capture link, the experimental uncertainties allow the
possibility of a much higher reaction flow if \I{65}{As}, \I{69}{Br},
and \I{73}{Rb} are less unbound. The recent discovery of a 0$^+$ shape
isomer in \I{72}{Kr} \citep{bou03} opens the possibility of a
substantial additional reaction flow through such isomeric states in
the $N = Z$ waiting point nuclei. More experimental and theoretical work
is needed to investigate that possibility.

The termination of nuclear flows in \XRBa3 are very similar to
Figure~1 of \citet{sch01a} including the closed loop caused by
\I{107}{Te}\Rga\I{103}{Sn}. 

\subsection{Model \XRBavi}
\lSect{mod6}

As an example of a qualitatively different sort of burst, we next
consider \Model6.  This model had an accretion rate five times lower,
$3.5\E{-10}\,\Msunyr$, than \Model3, and a metallicity ten times
higher (nominally ``solar'').  This combination of longer time and a
higher abundance of CNO catalyst leads to hydrogen depletion at the
base of the accreted layer long before the first burst ignites.
Indeed a substantial fraction of the helium also burns.  At the time
of the first burst, $273,780.9\,\Sec$ after the onset of accretion,
the helium and carbon mass fractions at the base of the accreted layer
are $30\,\%$ and $67\,\%$ respectively (\Fig{compa6a}).

The runaway commences in the helium shell, near its middle for the
first burst (\Fig{lite6a1}), and somewhat higher up for the later
bursts.  Vigorous convection moves outwards colliding after about
$0.4\,\Sec$ with the hydrogen shell ($-0.32\,\Sec$ in
\FIG{C}{lite6a1}).  The mass of the entire accreted layer at this
point is $6.14\E{21}\,\g$, of which the hydrogen layer ($X>0.01$) is
$7.6\E{20}\,\g$.  At this time the highest temperature, $5.0\E8\,\K$,
is at the base of the helium convection zone, where the density is
$1.2\E6\,\gcc$.  This collision leads to heating if the hydrogen layer
and an explosion by \I{12}C\Rpg as (hot) carbon is mixed out into
hydrogen.  Time steps as small as nanoseconds are required to follow
this interaction.  The maximum in energy generation shifts from deep
in the helium shell to the base of the hydrogen shell in which a
second convective region now grows. \Fig{lite6a1} shows this
interaction.  As time passes, the base of the helium zone grows hotter
at the same time as the hydrogen convective shell digs deeper into the
helium and carbon.

The progression of the hydrogen convective shell as it follows the
helium shell inwards involves some interesting physics.  As the
temperature, energy generation (\FIG{F}{lite6a1}), and luminosity of
the helium layer increases, it becomes buoyant with respect to the
overlaying hydrogen layer and mixing sets in.  When the two shells
connect, briefly mixing \I{12}C and \I1H, rapid energy release occurs
locally by \I{12}C\Rpg and a subsequent \textsl{rp}-process.  This
raises the entropy over that of the layers below and temporarily shuts
off the convective mixing between them.  But because part of the
initial mixing dredges down hydrogen, some of this large energy
generation occurs below the original interface, just resolved by one
or two zones in the present model.  With time, burning raises the
entropy of these zones so that they eventually link up with the
hydrogen convective shell.  The rising helium energy generation also
keeps the helium convection zone in close proximity the the hydrogen
shell.  In this way, the interface moves downward, piece by piece.
While the specifics of this merger of the two convective shells may be
model-dependent, it seems inevitable that such a merger happens on a
very short time scale once the helium convection zone encounters the
hydrogen layer.  The interface becomes, at least episodically,
Rayleigh-Taylor unstable, resulting in burning, mixing ``mushrooms''.
In contrast, the single interface propagating downward observed in the
present calculation is due to the limitations of spherical symmetry
imposed by the one-dimensional \textsc{Kepler} code.  A
multi-dimensional study would be both interesting and, given the short
time scale, feasible.

As time passes, the base of the helium zone grows hotter at the same
time as the hydrogen convective shell digs deeper into the helium and
carbon.  The base of the hydrogen convective layer also grows hotter
and denser as the shell becomes more massive.  At the same time the
concentration of heavy elements in the hydrogen rises.  After $2500$
cycles, $70\,\ms$ later, the surface luminosity climbs to
$\Ep{37}\,\ergs$.  At this point the density and temperature at the
base of the helium convection region, $2.5\E{21}\,\g$ above the bottom
of the accreted layer, are $3.4\E5\,\gcc$ and $1.7\E9\,\K$.  The
temperature and density at the base of the hydrogen convective layer,
$3.7\E{21}\,\g$ above the the iron substrate, are $2.5\E5\,\gcc$ and
$1.5\E9\,\K$.  Both shells are generating comparable energy.

One millisecond ($600$ cycles) later, the helium convective shell has
been entirely eaten away by the inwardly growing hydrogen convective
shell.  When the the luminosity is $\Ep{38}\ergs$, the hydrogen shell
has moved in to $2.5\E{21}\,\g$, where its temperature and density, at
the base, are $1.8\E9\,\K$ and $2.0\E5\,\gcc$.  Owing to
time-dependent convection, there are gradients in the abundances in
this convective shell, but at the base where energy generation is a
maximum, the mass fractions of \I1H, \I4He, \I{26}{Si}, and \I{30}S
are $0.036$, $0.38$, $0.12$, and $0.42$ respectively.  The inner
$2.5\E{21}\,\g$ does not participate in a major way in the burst.  At
this point it is still mostly helium and carbon.  Helium and carbon
burn, in radiative equilibrium, by an inward moving flame to
\I{28}{Si} (\Fig{lite6a1}).  Since the fraction of heavier elements
(\I{28}{Si}) produced by the flame decreases as it proceeds inward,
the thermohaline instability sets in immediately behind the flame.
Though this instability is too slow to affect the flame itself, it
mixes these layers and the even heavier ashes of the
\textsl{rp}-process above, resulting in a chemically homogeneous layer
of ash from each burst.  The flame consumes essentially all of the
helium, but $\sim 10\,\%$ of carbon remains that continues burning on
a longer time scale, down to a few percent, as the ashes cool
(\FIG{B}{lite6a1} and \Panff{C}).

The heat wave from the convective helium runaway also initiates some
burning in the non-convective helium and carbon beneath before the
arrival of the helium burning fame.  Remaining \I{14,15}{N} (from
decayed \I{14,15}{O}) is first converted by \Rag reactions to
\I{18,19}F which are converted by subsequent $\betap$-decay and \Rag
and \Rap reactions to \I{22}{Ne}.  This, in turn, is burned into
\I{25,26}{Mg} by another \Rag or \Ran reaction.  This latter reaction,
in particular, is responsible for the small ``arc'' of increased
energy generation visible in \FIG{D}{lite6a1} just before the helium
flame moves in.  This arc starts at $-0.01\,\Sec$ and is caught up by
the helium flame at $\sim0.015\,\Sec$ and $m=1.75\EE21\,\g$.

Seven tenths of a second after the hydrogen and helium convective
shells first collided (\Fig{lite6a1}, the temperature at the base of
the entire accreted layer is $1.75\E9\,\K$ and the density is
$7.6\E5\,\gcc$.  Convection is subsiding and most of the hydrogen is
gone.  The luminosity is $2.2\E{38}\,\ergs$.  Integrated over the
entire accreted layer the dominant abundances are \I4{He} ($0.12$),
\I{12}C ($0.055$), \I{16}O ($0.055$), \I{28}{Si} (0.14),\I{34}{Ar}
($0.077$), \I{38}{Ca} ($0.15$), \I{39}{Ca} ($0.14$), and \I{40}{Ca}
($0.05$).  The remainder of the burst will be powered by helium
burning and, in the tail of the burst, the Kelvin-Helmholtz
contraction of the cooling ashes.

After the main burst was over, for example $78\,\Sec$ after its onset,
the composition consisted chiefly of helium burning ashes with
appreciable helium itself remaining near the outside.  Principal mass
fractions integrated over the accreted layer were \I4{He} ($0.038$),
\I{12}C ($0.037$), \I{16}O ($0.032$), $^{40}$Ca ($0.27$), \I{28}{Si}
($0.18$), and traces of other abundances extending up to the iron
group (\Fig{compa6f}).

Subsequent bursts had slightly increasing critical masses, larger
values of $\alpha$, but observable properties closely resembling those
of burst 1.  The cause here is once again compositional inertia, but
of a different sort.  In the other models, the runaway ignited, to
varying degrees, in the ashes of the previous burst where the
composition played a major role in either catalyzing or directly
powering the initial nuclear energy generation.  Here, where ignition
always occurs in the freshly accreted layer, the effect of the
accreted ashes upon the \textsl{opacity} of the substrate is more
important.  Because of the lower electron scattering opacity
($\propto\left<Z^2/A\right>$) and higher heat capacity (ionic
contribution $\propto1/\left<A\right>$) of the ashes of the helium
burning layer(s) compared with the (assumed) iron substrate at the
beginning of accretion, the layers below the helium runaway are cooled
more efficiently in subsequent bursts.  Additional helium must then
burn before the runaway temperature is reached.  A larger critical
mass accumulates to compensate for the lower helium mass fraction.  At
the bottom of the freshly accreted layer just prior to the onset of
bursts $1$--$7$ the helium mass fraction was $0.298$, $0.266$,
$0.226$, $0.187$, $0.150$, $0.137$, and $0.105$.  The remainder was
mostly \I{12}C, some ($<1\,\%$ each) \I{14}C, \I{14,15}N, \I{16,18}O,
\I{18,19}F, \I{22}{Ne} and species less abundant than $1\EE-4$.

This raises the interesting possibility that continued burst activity
in \Model6 might eventually lead to a stable helium burning shell
and the accumulation of a carbon layer that could power a superburst.
Unfortunately, this was an ``expensive'' model to follow, requiring
about $20,000$ cycles per burst and several hundred zones.  After
burst 4 we followed the evolution using a ``dezoning'' algorithm that
joined neighboring zones when even their combined thickness was
smaller than $1\,\cm$ (mostly in the ashes) as they get increasingly
compressed by the weight of new zones and ashes layers added atop.
The continued evolution of this model is planned.

The light curve for the second burst \Model6 is shown in \Fig{lite6as}
and \Fig{lite6a}.  Its characteristics, including the rapid rise time,
were typical for all seven bursts followed.  The diffusion time of our
outer zone (roughly $150\,\cm$ at $\Ep4\,\gcc$) is approximately
$1\,\ms$ and this sets a lower bound on our resolution of the rise of
the light curve.  It seems likely that the rapid rise times seen here
will characterize all bursts in which a pure helium flash occurs
beneath a hydrogen layer of appreciable mass.

Note that in burst 6 the convection reached and stayed at the surface,
making a more luminous burst (\Tab{a6p}).

\subsection{Models \XRBaiv \ and \XRBav}

These were similar to \Model3, each being combined
hydrogen-helium runaways.  The effects of compositional inertia, though
still present were diminished relative to \Model3 because of the
higher initial metallicity (\Model5; \Tab{a5p}) and longer
accretion interval (\Model4; \Tab{a4p}).  The higher metallicity
makes the critical mass of the freshly accreted layer less sensitive
to the (nearly solar) metallicity created in the outer ashes of the
previous burst by primary reactions.  A longer burst interval allows
the more complete combustion of the ashes of the previous burst during
the inter-pulse period.

In fact, the global properties of \Models5 and \XRBa3 are similar in
terms of recurrence interval, burst energy, and duration emphasizing
the greater dependence of outcome on accretion rate than composition.
After the first flash, burst properties are very similar
(\Fig{Brep1}). There are differences though.  \Model5 has burst
intervals about $10\,\%$--$20\,\%$ less than \XRBa3, but total burst
energies ($E_{0.01}$ in the tables) $25\,\%$ smaller.  The ratio
$E_{0.01}/\mathrm{Delay}$ is also significantly less.  Both these
effects reflect the greater amount of burning in between pulses for
the model with the higher metallicity.
 
The recurrence times and $\alpha$ values are more regular for model ZM 
than for model zM. This is consistent with the fact that the heating in
model ZM is determined by the accreted metallicity, whereas in model zM,
the heating is determined by metallicity produced by the previous burst.
Taam et al.~(1993) also concluded that lower metallicity led to less 
regular burst properties.

The burst energies for \Model4 are higher by about a factor of
three than \XRBa3 reflecting the larger critical mass required for the
lower accretion rate.  The fraction of material that burns between
bursts is also greater so that the combination $E_{0.01}/(\Mdot \times
\mathrm{Delay})$ is smaller.

Bursts can also be ignited by compositional inertia due to helium
ignition in the ashes layer.  In \Model4 the energy released by
the triple-$\alpha$ reaction in the hydrogen-free helium layer of the
ashes generated enough energy to initiate convective mixing between
the ashes layer and the overlaying hydrogen layer at runaway.

\subsection{Corrections for General Relativity}
\lSect{gr}

The calculations in this paper were carried out assuming Newtonian
gravity. \citet{taa93} discuss the corrections that can be expected
for general relativity (see also \citealt{AJ82}; \citealt{LE93};
\Sect{a3} of \citealt{cum02}).  For the $1.4\,\Msun$, $10\,\km$
neutron star we assume here, the gravitational redshift correction is
$1+z=1.31$.  The recurrence time scales and burst durations in all the
tables and figures should be increased by $(1+z)$, the energies
decreased by $1+z$, and the luminosities decreased by $(1+z)^2 =
1.72$.  In addition, note that the mass co-ordinate used in this paper
refers to the baryonic mass (number of baryons multiplied by proton
mass) rather than rest mass.  Also the accretion rates we give are in
baryonic masses in the frame of the simulated layer and have to be
decreased by $1+z$ in the observer frame.

\subsection{Mixing by Thermohaline Convection}
\lSect{thal}

Bursts typically leave behind a radial composition gradient because
burning can reach different compositions in different zones.  For
example, a more extensive \textsl{rp}-process may occur in outer
layers than further in. This sometimes happens because of the
different density- and temperature-dependencies of the $3 \alpha$
reaction, \Rap reactions along the \textsl{rp}-process path, and the
\textsl{rp}-process itself.  If ``heavier'' material - according to
its composition - is located above ``lighter'' material, but the
compositional inversion is less destabilizing than the temperature
stratification, thermohaline convection occurs ( ``salt finger
instability'').  Otherwise, if the thermal stratification is too week,
convection according to the Ledoux criterion sets in.  In an ideal gas
the compositional buoyancy is determined by the mean molecular weight
gradient (``$\mu$''), but in degenerate regions, the mean molecular
weight per electron (``\mue'') is more important.  As a result, when
the matter becomes more degenerate, thermohaline convection can switch
on or off.  Additionally, weak decays change $\mu$ and $\mue$ and
their gradients over time.  We model this according to \citet{KRT80},
but using a generalization for degenerate equation of state.

Thermohaline mixing mostly occurs within the ashes of each burst,
slowly homogenizing its chemical composition, frequently only after
several subsequent bursts have occurred.  Eventually, after many
bursts, the mixing of neighboring layers of ash is also observed.

For example, thermohaline convection is indicated in \FIG{A}{lite6a1}
for \Model6.  Mixing is first facilitated by the behavior of the flame
in the helium layer which makes less silicon as it moves in.  This
leads to thermohaline convection immediately behind the flame.  On the
other hand, in the hydrogen-rich layer on top, the \textsl{rp}-process
produces heavier ashes than pure helium burning, leading to mixing
later on.  This later mixing leads to the spikes of energy generation
after $10\,\Sec$ and between $2\EE21\,\g$ and $3\EE21\,\g$ in
\FIG{B}{lastcompa3} and are due to mixing events at the upper edge of
this thermohaline convective region.

In all other figures showing the convective structure we omitted
thermohaline convection in the plot in order to retain the visibility
of burning regions, but it was included everywhere.

\section{Comparison with Analytic Calculations}
\lSect{analytic}

We now compare our results with semi-analytic ignition models,
following \citet{CB00}.  These models apply a one-zone ignition
criterion to simple models of the accumulating layer.  This approach
allows a survey of parameter space to be made while still giving a
good estimate of the ignition depth, and has recently been applied to
observations of the regular Type I bursters 4U~1820-30 \citep{cum03}
and GS~1826-24 \citep{gal03}.  A comparison with the time-dependent
simulations presented in this paper is of value for two
reasons. First, it provides a cross-check. Second, it highlights cases
in which additional physics not included in the semi-analytic models
(such as thermal and compositional inertia) is needed.

The models are constructed as follows (see \citealt{CB00} for a
detailed description).  We integrate the steady-state entropy and heat
flux equations down through the layer, following the composition
change as hydrogen burns to helium via the hot CNO cycle.  Both hot
CNO burning and a heat flux from below heat the layer.  Helium burning
reactions during accumulation are not included.  We adjust the
thickness of the layer until the ``one-zone'' ignition criterion
$\partial\epsta/\partial T=\partial\epscool/\partial T$
\citep{fuj81,fus87,BI98} is met at the base.  Here, \epsta is the
triple alpha energy production rate (augmented by a factor of $1.9$ to
account for proton captures on carbon), and \epscool is a local
approximation of the radiative cooling.

The thickness of the layer at ignition sets the burst recurrence time.
The burst energy is estimated by assuming complete burning of the
layer, and a nuclear energy release $\Qnuc=1.6+4.0\,\Xav\,\MeV$ per
nucleon, where \Xav is the mass-weighted mean $X$ in the layer at
ignition.  This expression for \Qnuc includes $35\,\%$ energy loss in
neutrinos during hydrogen burning \citep{fuj87}, a little more than we
find in our simulations (see \Sect{vogel}).

\Tab{theory} compares burst 1 and 2 from each run with the analytic
predictions.  The agreement is very good in general.  By far the
largest discrepancy is for \Model3. While the 1st burst agrees
well with the analytic predictions, the recurrence times and burst
energies for subsequent bursts are much less than expected. This
demonstrates very clearly the role of compositional inertia in
decreasing the recurrence time.  As discussed in \Sect{results},
triple alpha reactions in the low density layers during the cooling
tail of the burst give rise to a layer with solar abundance of CNO (or
larger).  Residual hydrogen burning in this layer during accumulation
gives an additional heat source, reducing the mass needed for
ignition.

Compositional inertia is most important for \Model3 because of the low
metallicity.  The luminosity from deeper layers, \Lb, in terms of the
energy per accreted nucleon, \Qb, is
$\Lb\approx\Ep{34}\,\ergs\;\times\;(\Qb/0.1\,\MeV)\;(\Mdot/0.1\,\MdotEdd)$.
The hot CNO luminosity is roughly $\LCNO\approx\epsCNO\Delta \MH$,
where $\epsCNO=5.8\E{13}\,\ergs\;\times\;(Z/0.01)$ is the hot CNO
energy production rate \citep{fow66,WW81} ($Z$ is the CNO mass
fraction), and $\Delta \MH$ is the mass of the layer containing
hydrogen.  This gives $\LCNO\approx
\Ep{35}\,\ergs\;\times\;(Z/0.02)\;(\Delta \MH/\Ep{21}\,\g)$.  For the
solar metallicity models, the hot CNO flux dominates the heating of
the layer as it accumulates ($\LCNO\gg\Lb$), whereas for low
metallicity, the heat flux from below plays an important role in
heating the layer ($\LCNO\sim\Lb$).  This makes the low metallicity
models most sensitive to any extra heating, e.g., compositional
inertia effects.  \Figure{Brep1} shows very clearly the difference
between the luminosity prior to the 1st burst ($\sim\Ep{34}\,\ergs$
from deeper layers) and between subsequent bursts ($\sim\Ep{35}\,\ergs$
from hot CNO burning of residual hydrogen beneath the freshly accreted
matter).

Also apparent in Table 10 is that for Model Zm, the \Qnuc
values are less than expected from burning pure helium to iron group
(which gives $\approx 1.6$ MeV per nucleon). This is due to the fact
that much of the helium burns to carbon prior to ignition. Our
analytic models do not include helium burning reactions, but still
give a good estimate of the recurrence time in this case. However,
they are unable to address the question of how much carbon is burned
before the runaway occurs, and whether a stable helium burning layer
might form. Recently, \cite{nar03} calculated ignition conditions,
including helium burning before ignition by approximating the
composition profile as that due to steady helium burning. They also
found that some helium burned to carbon before instability occurred,
and referred to these bursts as "delayed bursts".

\section{Conclusions}

By coupling nuclear reaction networks of unprecedented size directly
to the calculation of zoned stellar models of X-ray bursts we have
calculated the most realistic models to date of Type I X-ray bursts on
neutron stars.  Two values of accretion rate and metallicity were
explored and, in all cases, at least $4$ and up to $15$ repetitions
were followed, assuring that the properties of the bursts had reached
steady state.  The effects of varying the nuclear physics were also
examined, in particular key lifetimes at waiting points along the path
of the \textsl{rp}-process.

For the conditions studied, we find, in agreement with \citet{sch01a},
that when the full reaction network is included, no hydrogen enters
the ocean and crust of the neutron star. Deep hydrogen burning
(e.g.~as in \citealt{TWL96}), that has also been discussed as a
possible energy source for superbursts \citep{kuu02} does not happen
and may have been an artifact of the limited nuclear reaction networks
used in previous studies.

The first burst in each sequence had different properties to
subsequent bursts. This is because accretion initially proceeds on an
inert substrate. However, hydrogen, helium, and CNO nuclei are not
depleted in the outer layers of each burst and burning continues
between bursts.  This leads to both ``thermal inertia'' and
``compositional inertia'' \citep{Taa80}.  Subsequent runaways
(\Tab{a3p}) ignite in the ashes of previous ones, require less
critical mass and are therefore less energetic.  The critical mass
becomes nearly a constant independent of the composition of the
accreted matter (at least for matter with sub-solar initial
composition --- see \Tab{a3p} and \Tabs{a6p} through
\Tabff{a4p}). This complicates attempts to infer the metallicity of
the accreted material from burst properties (e.g., \citealt{gal03}).

These effects have important implications for the composition of the
ashes, which eventually become incorporated into the neutron star
crust.  In \Model3, which has the same conditions as the one-zone
model of \citet{sch01a}, only the first, energetic, burst ends in a
SnSbTe cycle.  Subsequent bursts ignite sooner, reach lower
temperatures, and do not produce isotopes much heavier than mass $64$
(\Tab{comp}).  The matter that accretes into the neutron crust has
mean mass in the 60's rather than $100$.  However, it may well be that
similarly violent bursts to \citet{sch01a} will still be found in
steady state for different accretion rates.

As has been pointed out frequently, X-ray bursts are marvelous
laboratories for the study of nuclear astrophysics, especially of
nuclei near the proton-drip line up to mass $110$.  Our studies
confirm (e.g., \Sect{weak}) the sensitivity of the light curves of
bursts powered by hydrogen burning to nuclear flows above the iron
group \citep{sch01a,koi99}. We additionally find that the \emph{rise} times
of such bursts are also sensitive to nuclear decays \emph{below} the
iron group (\Fig{weaktest1}).  Separating these effects out from those
due to thermal diffusion and, possibly, the spreading of the burning
on the neutron star will be difficult and will rely on accurate data
from the nuclear laboratory.  However, the light curves for \Model3,
for example, compare favorably with observations of bursts from
GS~1826-24 \citep{gal03}, which \citet{bil00} proposed were powered by
the \textsl{rp}-process.  Particularly noteworthy are the long ($\sim
10\,\Sec$) observed rise times, which compare well with \Fig{rist3},
for example.  There is much to be learned from a detailed comparison
of observations and theory for this source.

In \Model6 (\Tab{a6p}), the runaway was initiated by helium burning
beneath a stable hydrogen shell. In contrast to the hydrogen-helium
flashes, in which convection had ceased by the time the surface
luminosity began to increase, convection in these bursts continued for
$\approx 5$ seconds after ignition.  Also, when the convection zone
driven by unstable helium burning first broached the overlying
hydrogen layer, a virtual explosion occurred initiated by \I{12}C\Rpg
reactions.  The resulting rise time was very short, approximately
$10\,\ms$.  The burst duration was shorter than the others, as
expected for bursts where most of the fuel is helium, and there was a
period of super-Eddington luminosity (which would drive radius
expansion, although this is not followed in our models).  All four
bursts in \Model6 had these same characteristics --- very rapid rise
time, brief duration, and super-Eddington luminosities.  An
interesting question is whether these characteristics are shared by
all helium initiated bursts capped by a layer of accreted hydrogen.

Our results have implications for carbon-powered superbursts.  For all
of the hydrogen-helium flash models (\XRBa3, \XRBa4, \XRBa5), we find
very little carbon remains after each flash ($\lesssim 1$\% by mass)
in agreement with one-zone calculations \citep{sbc03b}.  However, in
addition, we find that carbon is depleted further by alpha-captures as
the ashes are heated by subsequent bursts.  This is because some
helium remains as well as carbon, and alpha-capture is efficient at
converting carbon to heavier nuclei such as \I{28}{Si}.  This implies
that unstable hydrogen and helium burning at these compositions and
accretion rates does not lead to accumulation of sufficient carbon to
power an unstable runaway \citep{WT76} leading to a superburst
\citep{CB01}.  In addition, the less extensive \textsl{rp}-process
found in \Model3 compared with \citet{sch01a} means that
photo-disintegration of heavy elements \citep{sbc03} will play a less
important role in superburst energetics. Carbon production at these
accretion rates may rely on some fraction of stable burning of the
accreted fuel \citep{sbc03b}.

In the helium flash model, \Model6, we find that the accumulating
helium very nearly burns stably, with a substantial amount of helium
burning to carbon prior to the runaway.  This carbon is burned up
after helium ignition, leaving only $\sim 3\,\%$ by mass in the deep
ashes after several bursts.  However, the carbon mass fraction at the
base before the flash was found to increase as subsequent bursts were
followed, reaching $90$\% after the seventh burst.  As we continue to
evolve this model, it may be that helium burning will ultimately
stabilize, allowing production of enough carbon to power a superburst.
However, we note that the accretion rate in \Model6 is much less than
inferred for the superburst sources ($\approx 0.1$--$0.3$ Eddington;
\citealt{kuu02}).

\acknowledgements

We thank Ron Taam for helpful discussions concerning the GR effects
for interpreting the results obtained in the Newtonian frame of our
calculation with respect to observational quantities and the issue of
super-burst fuel.  We also appreciate the assistance and calculations
of neutrino loss rates by Petr Vogel (\Sect{vogel}). 

At UCSC, this research has been supported by the NSF (AST 02-06111),
NASA (NAG5-12036), and the DOE Program for Scientific Discovery
through Advanced Computing (SciDAC; DE-FC02-01ER41176).  AH is
supported, in part, by the Department of Energy under grant B341495 to
the Center for Astrophysical Thermonuclear Flashes at the University
of Chicago, and a Fermi Fellowship at the University of Chicago. AC is
supported by NASA Hubble Fellowship grant HF-01138 awarded by the
Space Telescope Science Institute, which is operated for NASA by the
Association of Universities for Research in Astronomy, Inc.~under
contract NAS~5-26555. RH and JP performed this work under the auspices
of the U.S. Department of Energy at the University of California
Lawrence Livermore National Laboratory under contract
No. W-7405-Eng-48.  AB is supported by the NSF, PHY-0244453. TR
acknowledges support by the Swiss NSF (grant 2000-061031.02, PROFIL
professorship 2024-067428.02). HS is an Alfred P. Sloan Fellow and
supported by the NSF under grants PHY 01-10253 and PHY 00-72636.  MW
is supported at Notre Dame by NSF PHY-0072711 (Joint Institute for
Nuclear Astrophysics) and PHY-0140324.

\clearpage

\onecolumn

\begin{deluxetable}{ccc}
\tablecaption{Proton separation energies of isotones near
the long-lived waiting point nuclei
$^{64}{\rm Ge}$, $^{68}{\rm Se}$, $^{72}{\rm Kr}$, 
and $^{76}{\rm Sr}$.
\label{tab:sptbl}}
\tablewidth{0pt} 
\tablehead{ 
\colhead{Nucleus} & 
\colhead{$S_p$(MeV)\tablenotemark{a}} & 
\colhead{Uncertainty(keV)}} 
\startdata 
$^{65}{\rm As}$ & -0.43 & 290 \\ 
$^{66}{\rm Se}$ & 2.43 & 180 \\ 
$^{69}{\rm Br}$ & -0.73 & 320 \\ 
$^{70}{\rm Kr}$ & 2.14 & 190 \\ 
$^{73}{\rm Rb}$ & -0.55 & 320 \\ 
$^{74}{\rm Sr}$ & 1.69 & 210 \\ 
$^{77}{\rm Y}$ & -0.23 & unknown \\ 
$^{78}{\rm Zr}$ &1.28 & unknown 
\tablenotetext{a}{Taken from the compilation of \cite{bro02a}, except
for the proton separation energies of \I{77}Y and \I{78}{Zr},
which were taken from the unpublished
calculations of \cite{bro02b}.}  
\enddata
\end{deluxetable}

%\clearpage

\begin{deluxetable}{cccc}
\tablecaption{Summary of Model Properties
\label{tab:summary}}
\tablewidth{0pt}
\tablehead{
\colhead{Model} & 
\colhead{$Z$} & 
\colhead{Acc Rate} & 
\colhead{\# bursts} \\  
\colhead{ } & 
\colhead{($\mathrm{Z}_\odot)$} & 
\colhead{($\Eppp-10\,\Msunyr$)} & 
\colhead{ }}   
\startdata
\XRBa4 &  0.05 & 3.5  &   4 \\
\XRBa3 &  0.05 & 17.5 &  15 \\
\XRBa6 &  1    & 3.5  &   7 \\
\XRBa5 &  1    & 17.5 &  12 
\enddata
\end{deluxetable}

%\clearpage

\begin{deluxetable}{cccccccc}
\tablecaption{Composition\tablenotemark{a} After Burst
\label{tab:comp}}
\tablewidth{0pt}
\tablehead{
\colhead{Model} & 
\colhead{Burst num.} &
\colhead{$^A$Z} & 
\colhead{X} &
\colhead{$^A$Z} & 
\colhead{X} &
\colhead{$^A$Z} & 
\colhead{X}}
\startdata
\XRBaiii & 1  & $^{106}$Sn & 0.18 & $^{104}$In  & 0.09 & $^{106}$In & 0.06  \\
\XRBaiii & 3  & $^{64}$Zn  & 0.16 & $^{68}$Se   & 0.09 & $^{32}$S   & 0.06  \\
\XRBavi  & 1  & $^{28}$Si & 0.27 & $^{40}$Ca & 0.23 & $^{24}$Mg & 0.11 \\
\XRBavi  & 3  & $^{28}$Si & 0.27 & $^{40}$Ca & 0.18 & $^{24}$Mg & 0.15  
\tablenotetext{a}{Three most abundant isotopes at the bottom of the
most recently accreted critical mass shortly after the burst indicated
is over.}
\enddata
\end{deluxetable}
%  4.199173 x 10**4 s cycle  2700 for pulse 1 xrba3 zones 51, 60; 286 z in prob
%  1.095705 x 10**5 s cycle  9100 for pulse 3 xrba3 zones 500,510 659 z in prob
%  2.738601 x 10**5 s cycle 17100 for pulse 1 xrba6 zones 51, 60; 357 z in prob
%  8.734797 x 10**5 s cycle 57900 for pulse 3 xrba6 zones 680, 700 1018 z ``

\clearpage

\begin{deluxetable}{ccccccccc}
\tablecaption{Flash Properties for Model~\XRBaiii.
\label{tab:a3p}}
\tablewidth{0pt} 
\tablehead{ 
\colhead{Pulse} &
\colhead{Delay\tablenotemark{a}} & 
\colhead{$L_{\mathrm{peak}}$} &
\colhead{$\tau_{0.01}$\tablenotemark{b}} & 
\colhead{$E_{0.01}$} &
\colhead{$\tau_{1/2}$} & 
\colhead{$E_{1/2}$} &
\colhead{$\tau_\mathrm{rise}$\tablenotemark{c}} &
\colhead{$\alpha$\tablenotemark{d}} \\ 
\colhead{\#} &
\colhead{(h)} & 
\colhead{(\Ep{38}\,\ergs)} & 
\colhead{(\Sec)} &
\colhead{(\Ep{39}\,\erg)} & 
\colhead{(\Sec)} &
\colhead{(\Ep{39}\,\erg)}&
\colhead{(\Sec)} &
\colhead{}}
\startdata 
  1 &      11.59 &       1.78 &        527 &      20.92 &         43 &       6.10 &       2.64 &     62 \\
  2 &       3.58 &       0.89 &        382 &       6.50 &         31 &       2.02 &       2.24 &     61 \\
  3 &       3.14 &       1.14 &        326 &       6.37 &         25 &       2.08 &       2.40 &     55 \\
  4 &       2.84 &       0.97 &        308 &       5.36 &         25 &       1.77 &       2.49 &     59 \\
  5 &       2.96 &       1.18 &        315 &       6.23 &         20 &       1.80 &       2.37 &     53 \\
  6 &       3.21 &       0.98 &        342 &       6.00 &         26 &       1.85 &       2.26 &     59 \\
  7 &       3.08 &       1.02 &        339 &       6.01 &         26 &       1.91 &       2.50 &     57 \\
  8 &       2.90 &       1.09 &        307 &       6.03 &         23 &       1.86 &       2.38 &     53 \\
  9 &       3.46 &       0.97 &        359 &       6.41 &         29 &       2.02 &       2.26 &     60 \\
 10 &       3.10 &       1.13 &        330 &       6.41 &         24 &       1.98 &       2.38 &     54 \\
 11 &       3.17 &       1.03 &        331 &       6.11 &         24 &       1.82 &       2.26 &     58 \\
 12 &       3.17 &       1.07 &        325 &       6.24 &         25 &       1.95 &       2.28 &     56 \\
 13 &       3.13 &       1.02 &        339 &       6.03 &         26 &       1.90 &       2.37 &     58 \\
 14 &       3.00 &       1.13 &        308 &       6.22 &         24 &       1.96 &       2.26 &     53 \\
 15 &       3.52 &       1.01 &        366 &       6.73 &         28 &       2.02 &       2.16 &     58 
\enddata
\tablenotetext{a}{``Delay'' indicates the time from the last burst or
the time since start of accretion for the first burst.
$L_{\mathrm{peak}}$ gives the maximum luminosity.} 
\tablenotetext{b}{Duration
and total energy for the time during which the burst exceeds 1\,\% of
its peak luminosity are denoted as $\tau_{\mathrm{0.01}}$ and
$E_{\mathrm{0.01}}$. The time and energy for emission above 50\%
maximum are $\tau_{\mathrm{1/2}}$ and
$E_{\mathrm{1/2}}$ }
\tablenotetext{c}{Time for the burst to rise from $10\,\%$ to $50\,\%$
of its peak luminosity.}
\tablenotetext{d}{Persistent luminosity compared to integrated burst energy
(here we integrate over the time the burst has $1\,\%$ of it peak luminosity).}
\end{deluxetable}

\begin{deluxetable}{ccccccccc}
\tablecaption{Model~\XRBaiii \ with weak rates ($A > 56$) ten times faster.
\lTab{a3p10}}
\tablewidth{0pt} 
\tablehead{ 
\colhead{Pulse} &
\colhead{Delay} & 
\colhead{$L_{\mathrm{peak}}$} &
\colhead{$\tau_{0.01}$} & 
\colhead{$E_{0.01}$} &
\colhead{$\tau_{1/2}$} & 
\colhead{$E_{1/2}$} &
\colhead{$\tau_\mathrm{rise}$} &
\colhead{$\alpha$} \\ 
\colhead{\#} &
\colhead{(h)} & 
\colhead{(\Ep{38}\,\ergs)} & 
\colhead{(\Sec)} &
\colhead{(\Ep{39}\,\erg)} & 
\colhead{(\Sec)} &
\colhead{(\Ep{39}\,\erg)}&
\colhead{(\Sec)} &
\colhead{}}
\startdata 
  1 &      11.60 &       2.30 &        332 &      20.81 &         82 &      15.40 &       3.42 &     62 \\
  2 &       3.43 &       0.95 &        261 &       6.25 &         53 &       3.66 &       2.02 &     61 \\
  3 &       4.07 &       1.24 &        251 &       7.92 &         55 &       5.14 &       2.09 &     57 \\
  4 &       4.00 &       1.19 &        251 &       7.65 &         55 &       4.85 &       2.09 &     58 
\enddata
\end{deluxetable}

\begin{deluxetable}{ccccccccc}
\tablecaption{Model~\XRBaiii \ with weak rates ($A > 56$) ten times slower.
\lTab{a3pp1}}
\tablewidth{0pt} 
\tablehead{ 
\colhead{Pulse} &
\colhead{Delay} & 
\colhead{$L_{\mathrm{peak}}$} &
\colhead{$\tau_{0.01}$} & 
\colhead{$E_{0.01}$} &
\colhead{$\tau_{1/2}$} & 
\colhead{$E_{1/2}$} &
\colhead{$\tau_\mathrm{rise}$} &
\colhead{$\alpha$} \\ 
\colhead{\#} &
\colhead{(h)} & 
\colhead{(\Ep{38}\,\ergs)} & 
\colhead{(\Sec)} &
\colhead{(\Ep{39}\,\erg)} & 
\colhead{(\Sec)} &
\colhead{(\Ep{39}\,\erg)}&
\colhead{(\Sec)} &
\colhead{}}
\startdata 
  1 &      11.59 &       1.75 &        923 &      15.84 &         44 &       5.94 &       2.60 &     81 \\
  2 &       2.75 &       1.10 &        579 &      10.11 &         31 &       2.38 &       1.44 &     30 \\
  3 &       2.70 &       1.32 &        357 &       6.14 &         20 &       2.02 &       0.82 &     49 \\
  4 &       6.82 &       2.30 &        424 &      10.31 &         16 &       2.88 &       2.22 &     74 
\enddata
\end{deluxetable}

\begin{deluxetable}{ccccccccc}
\tablecaption{Flash Properties for Model~\XRBavi.
\label{tab:a6p}}
\tablewidth{0pt} 
\tablehead{ 
\colhead{Pulse} &
\colhead{Delay} & 
\colhead{$L_{\mathrm{peak}}$} &
\colhead{$\tau_{0.01}$} & 
\colhead{$E_{0.01}$} &
\colhead{$\tau_{1/2}$} & 
\colhead{$E_{1/2}$} &
\colhead{$\tau_\mathrm{rise}$\tablenotemark{a}} &
\colhead{$\alpha$} \\ 
\colhead{\#} &
\colhead{(h)} & 
\colhead{(\Ep{38}\,\ergs)} & 
\colhead{(\Sec)} &
\colhead{(\Ep{39}\,\erg)} & 
\colhead{(\Sec)} &
\colhead{(\Ep{39}\,\erg)}&
\colhead{(\ms)} &
\colhead{}}
\startdata 
  1 &      76.05 &       3.33 &        103 &       7.37 &         28 &       6.73 &      32.07\tablenotemark{b} &    231 \\
  2 &      80.43 &       4.73 &         83 &       7.64 &         13 &       6.06 &       1.38 &    235 \\
  3 &      86.14 &       5.06 &         87 &       7.95 &         12 &       6.19 &       0.91 &    242 \\
  4 &      96.19 &       5.10 &         93 &       8.07 &         12 &       6.14 &       1.04 &    267 \\
  5 &     107.00 &       4.55 &        103 &       8.54 &         15 &       6.86 &       1.82 &    280 \\
  6 &     111.20 &       6.33 &         27 &       7.98 &         11 &       6.52 &       0.61 &    312 \\
  7 &     130.84 &       5.12 &        107 &       9.20 &         14 &       6.87 &       0.51 &    318 
\enddata
\tablenotetext{a}{Note that the rise time in this table is in $\ms$
not $\Sec$ as in the other tables.}
\tablenotetext{b}{This first burst shows a triple peak.  If only the
time required to rise to $50\,\%$ of the first peak is considered, the
corresponding rise time is less than $0.7\,\ms$.  Indeed, if the rise
time to $30\,\%$ of peak is considered, all bursts have rise times
between $0.3\,\ms$ and $0.4\,\ms$, except for the first burst that rises
in $0.7\,\ms$, as  mentioned above. }
\end{deluxetable}

\clearpage

\begin{deluxetable}{ccccccccc}
\tablecaption{Flash Properties for Model~\XRBav.
\label{tab:a5p}}
\tablewidth{0pt} 
\tablehead{ 
\colhead{Pulse} &
\colhead{Delay} & 
\colhead{$L_{\mathrm{peak}}$} &
\colhead{$\tau_{0.01}$} & 
\colhead{$E_{0.01}$} &
\colhead{$\tau_{1/2}$} & 
\colhead{$E_{1/2}$} &
\colhead{$\tau_\mathrm{rise}$} &
\colhead{$\alpha$} \\ 
\colhead{\#} &
\colhead{(h)} & 
\colhead{(\Ep{38}\,\ergs)} & 
\colhead{(\Sec)} &
\colhead{(\Ep{39}\,\erg)} & 
\colhead{(\Sec)} &
\colhead{(\Ep{39}\,\erg)}&
\colhead{(\Sec)} &
\colhead{}}
\startdata 
  1 &       3.47 &       1.91 &        124 &       4.93 &         15 &       2.15 &       0.66 &     78 \\
  2 &       2.66 &       1.51 &        141 &       4.61 &         16 &       1.88 &       0.55 &     64 \\
  3 &       2.72 &       1.58 &        143 &       4.65 &         17 &       2.01 &       0.56 &     65 \\
  4 &       2.69 &       1.51 &        138 &       4.59 &         18 &       2.01 &       0.59 &     65 \\
  5 &       2.65 &       1.63 &        133 &       4.68 &         15 &       1.94 &       0.54 &     63 \\
  6 &       2.74 &       1.54 &        135 &       4.62 &         17 &       2.02 &       0.58 &     66 \\
  7 &       2.65 &       1.52 &        143 &       4.60 &         16 &       1.87 &       0.59 &     64 \\
  8 &       2.70 &       1.50 &        146 &       4.66 &         17 &       1.96 &       0.57 &     64 \\
  9 &       2.65 &       1.57 &        144 &       4.62 &         17 &       2.02 &       0.57 &     64 \\
 10 &       2.68 &       1.55 &        141 &       4.58 &         16 &       1.92 &       0.55 &     65 \\
 11 &       2.68 &       1.65 &        134 &       4.72 &         17 &       2.09 &       0.51 &     63 \\
 12 &       2.73 &       1.66 &        133 &       4.70 &         16 &       2.04 &       0.51 &     65 
\enddata
\end{deluxetable}

%\clearpage

\begin{deluxetable}{ccccccccc}
\tablecaption{Flash Properties for Model~\XRBaiv.
\label{tab:a4p}}
\tablewidth{0pt}
\tablehead{
\colhead{Pulse} &
\colhead{Delay} & 
\colhead{$L_{\mathrm{peak}}$} &
\colhead{$\tau_{0.01}$} & 
\colhead{$E_{0.01}$} &
\colhead{$\tau_{1/2}$} & 
\colhead{$E_{1/2}$} &
\colhead{$\tau_\mathrm{rise}$} &
\colhead{$\alpha$} \\ 
\colhead{\#} &
\colhead{(h)} & 
\colhead{(\Ep{38}\,\ergs)} & 
\colhead{(\Sec)} &
\colhead{(\Ep{39}\,\erg)} & 
\colhead{(\Sec)} &
\colhead{(\Ep{39}\,\erg)}&
\colhead{(\Sec)} &
\colhead{}}
\startdata
  1 &      60.16 &       2.97 &        313 &      20.92 &         37 &       8.23 &       1.09 &     64 \\
  2 &      57.27 &       2.36 &        333 &      20.43 &         55 &      11.14 &       0.87 &     62 \\
  3 &      54.91 &       2.61 &        301 &      19.48 &         48 &      10.20 &       0.93 &     63 \\
  4 &      53.03 &       3.13 &        275 &      18.88 &         36 &       8.44 &       1.08 &     62 
\enddata
\end{deluxetable}
 
%\clearpage
                                                                                                                
\begin{deluxetable}{lccccc}
\tablewidth{0pt} \tablecaption{Comparison of burst properties with
analytic models
\label{tab:theory}}
\tablehead{ \colhead{} & \colhead{\trec} &
\colhead{Energy} & \colhead{\Qnuc\tablenotemark{a}} &
\colhead{\Xav\tablenotemark{b}} &
\colhead{\Yav\tablenotemark{b}}\\
&\colhead{(\hour)} & \colhead{($\Ep{39}\,\erg$)}}
\startdata
\multicolumn{5}{c}{\Model4}\\ \noalign{\smallskip} \hline \noalign{\smallskip} 
burst 1  & 60.2 & 20.9 & 4.51 & &             \\ 
burst 3  & 54.9 & 19.5 & 4.60 & &             \\ 
analytic & 78.0 & 24.5 & 4.09 & 0.623 & 0.376 \\ 
\noalign{\smallskip} \hline \noalign{\smallskip} 
\multicolumn{5}{c}{\Model6}\\
\noalign{\smallskip} \hline \noalign{\smallskip} 
burst 1  & 76.1 & 7.37 & 1.26 & &             \\ 
burst 3  & 86.1 & 7.64 & 1.20 & &             \\ 
analytic & 70.3 & 9.73 & 1.80 & 0.051 & 0.929 \\
\noalign{\smallskip} \hline \noalign{\smallskip} 
\multicolumn{5}{c}{\Model3}\\
\noalign{\smallskip} \hline \noalign{\smallskip}  
burst 1  & 11.6 & 20.9 & 4.68 & &             \\
burst 3  & 3.14 & 6.37 & 5.25 & &             \\
analytic & 14.0 & 24.3 & 4.54 & 0.735 & 0.264 \\ 
\noalign{\smallskip} \hline \noalign{\smallskip} 
\multicolumn{5}{c}{\Model5}\\
\noalign{\smallskip} \hline \noalign{\smallskip} 
burst 1  & 3.47 & 4.93 & 3.68 &       &       \\ 
burst 3  & 2.72 & 4.65 & 4.43 &       &       \\ 
analytic & 4.64 & 6.72 & 3.77 & 0.544 & 0.436  
\enddata

\tablenotetext{a}{The nuclear energy release in \MeV per accreted
nucleon.  For the analytic models, we set
$\Qnuc=1.6+4\,\Xav\,\MeV$.  For the time-dependent models,
$\Qnuc=E_{0.01}/(\trec\Mdot)=2.59\,\MeV/$nucleon$\;\times\;
(E_{0.01}/\Ep{39}\,\erg)\; (\trec/1\,\hour)^{-1} (\Mdot/1.75\times
10^{-9}\,\Msunyr)^{-1}$.}  
\tablenotetext{b}{The mass-weighted hydrogen or helium fraction
immediately before the burst, e.g., $\Xav=\frac1M\int X(M)
\mathrm{d}M$.}
\end{deluxetable}

\clearpage 
%1 
\begin{figure} 
\centering 
\includegraphics[bb=48 66 732 542,draft=\Draft,angle=0,width=\columnwidth]{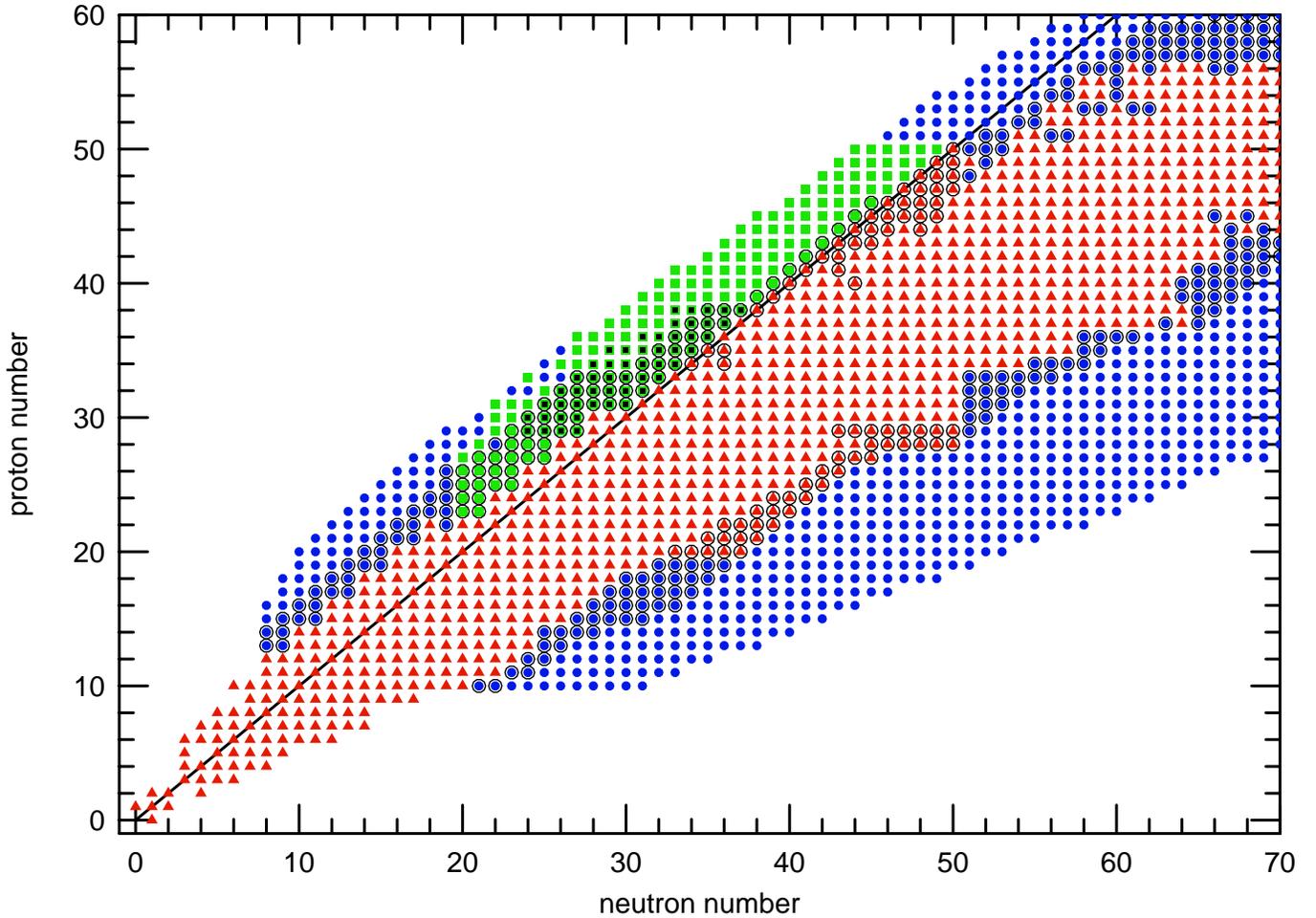}
\caption{Color-coded nuclear mass excesses taken from different data
sources in the region of interest for the \textsl{rp}-process.  A
\textsl{dark line} shows the location of nuclei with $Z=N$.  The
actual size of the network used in these studies varied with time and
model (\Sect{adaptive}).  \textsl{Red triangles} indicate nuclei with
experimentally determined mass excesses \citep{aud95}.  Nuclei with
\textsl{black circles} surrounding a symbol indicate that an
extrapolated or interpolated mass is available from
\citeauthor{aud95}.  For those circles enclosing a \textsl{red
triangle}, these estimated values were used.  \textsl{Green squares},
circled or not, show where data from \citet{bro02a} and \citet{bro02b}
was used.  These add a calculated displacement energy to the
Audi-Wapstra masses for $N\ge Z$ to obtain the masses of mirror nuclei
with $Z>N$.  For a subset of these (\textsl{black squares inside green
squares}) from \citet{bro02a}, the errors in the calculated
displacement energies are $\sim 100\,\keV$.  The \textsl{solid green
squares} which cover a wider region are from \citet{bro02b} and have a
larger error of several hundred keV in the calculated displacement
energy due to the wider extrapolation and the possible effects of
deformation in the $A=80$ mass region.  In the region of $N=Z$ from
$A=76-100$, the error in the mass excess is dominated by the error in
the \citeauthor{aud95} extrapolations, which are on the order of
$0.5\,\MeV$.  Finally, \textsl{blue circles} indicate mass excess data
taken from \cite{mol95}.  These were used wherever shown,
(circled or uncircled).  \lFig{spfig}}
\end{figure}

\clearpage
%2
\begin{figure} \centering
\includegraphics[draft=\Draft,angle=0,width=\columnwidth]{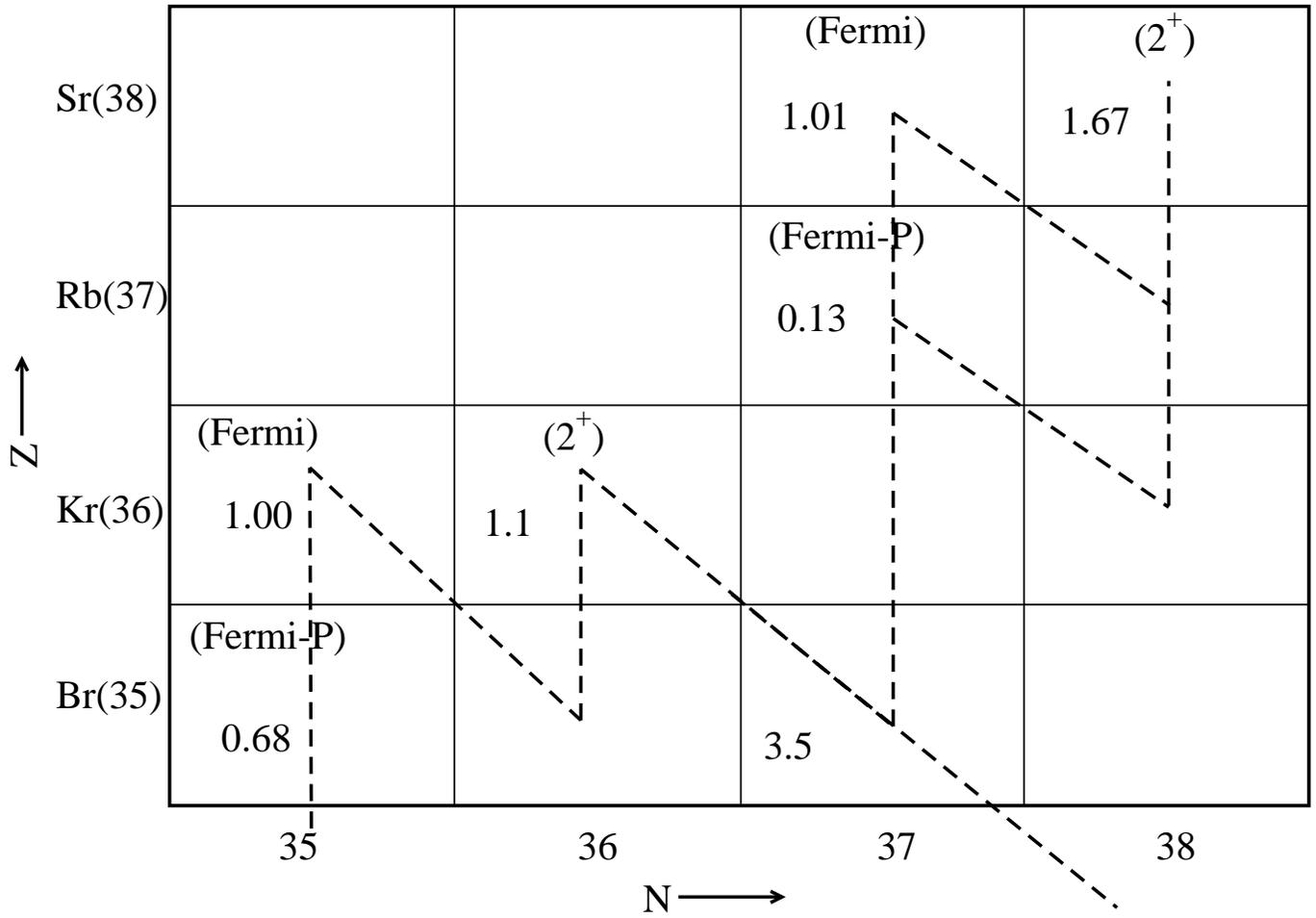}
\caption{Systematics of weak decays for nuclei along the
\textsl{rp}-process path near \I{72}{Kr}.  Nuclei labeled ``(Fermi)''
in this figure are those whose decay rates are dominated by Fermi
transitions and expected to be insensitive to the temperature.  Nuclei
labeled ``(Fermi-P)'' are those whose decay rates decrease quickly
with temperature. Estimates of the low-temperature decay rates for
these nuclei depend largely on estimates of the nuclear partition
functions. Nuclei labeled ``($2^+$)'' are those even-even nuclei whose
low-temperature decay can be sensitive to the decay rate of the first
excited $2^+$ state. The number in each box is the estimated ratio of
the total (\betap+ electron capture) decay rate at $T=1.5\E9\,\K$,
$\rho=\Ep5\,\gcc$ to the ground state \betap decay rate.  The dashed
black line illustrates important weak and nuclear flows during the
\textsl{rp}-process.\lFig{square}}
\end{figure}

\clearpage
%3
\begin{figure} \centering
\includegraphics[draft=\Draft,angle=0,width=\columnwidth]{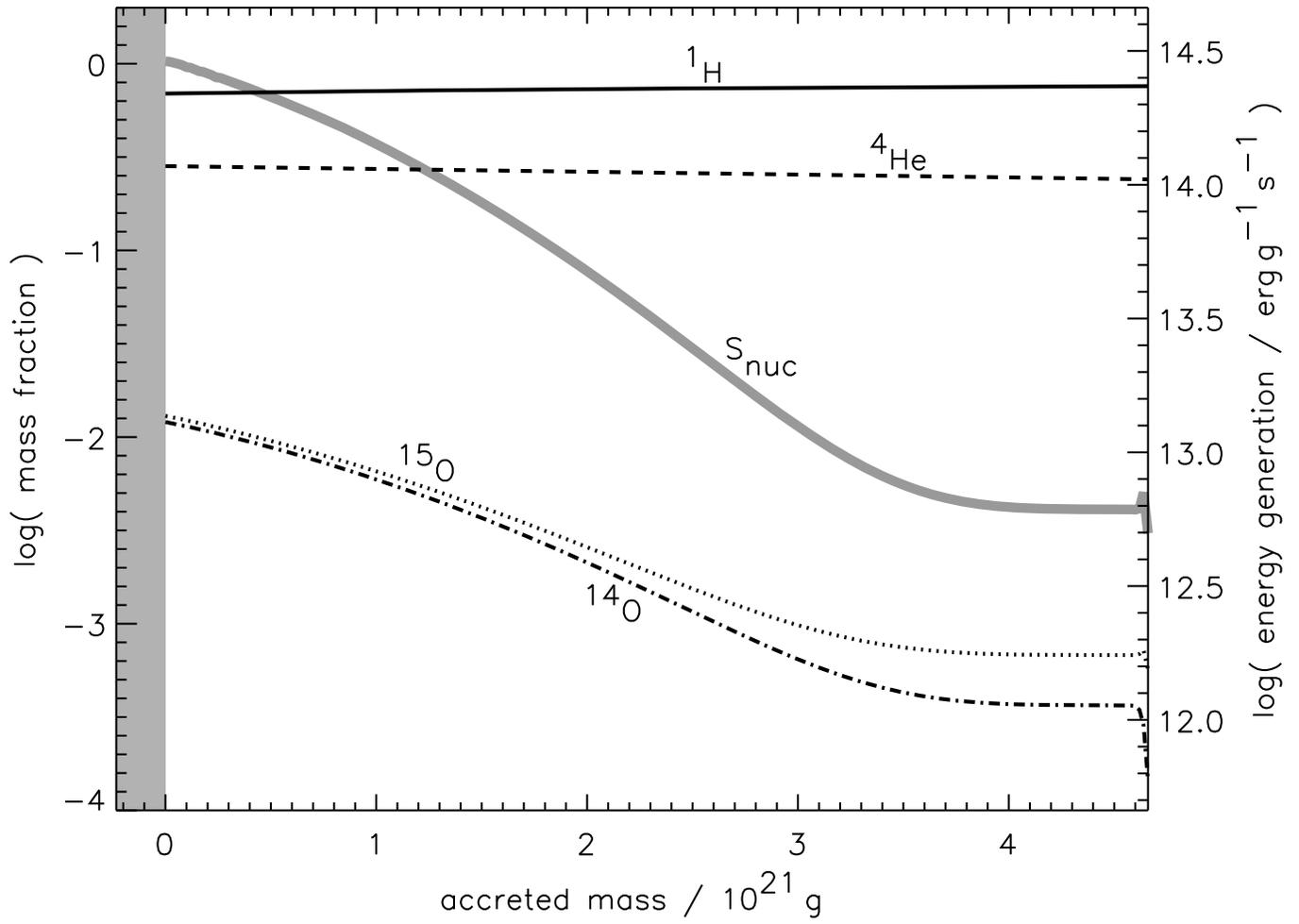} 
\caption{Composition just prior to the first burst of Model~\XRBaiii \
(at $41,640.5\,\Sec$).  Combined hydrogen burning by the beta-limited
CNO-cycle and helium burning are responsible for a rapidly rising
energy production. The total nuclear power here is $3.4\E{35}\,\ergs$
and the time before the burst, about one minute.  \lFig{comp1}}
\end{figure}

\clearpage
%4
\begin{figure}
\centering
\includegraphics[draft=\Draft,angle=0,height=0.8\textheight]{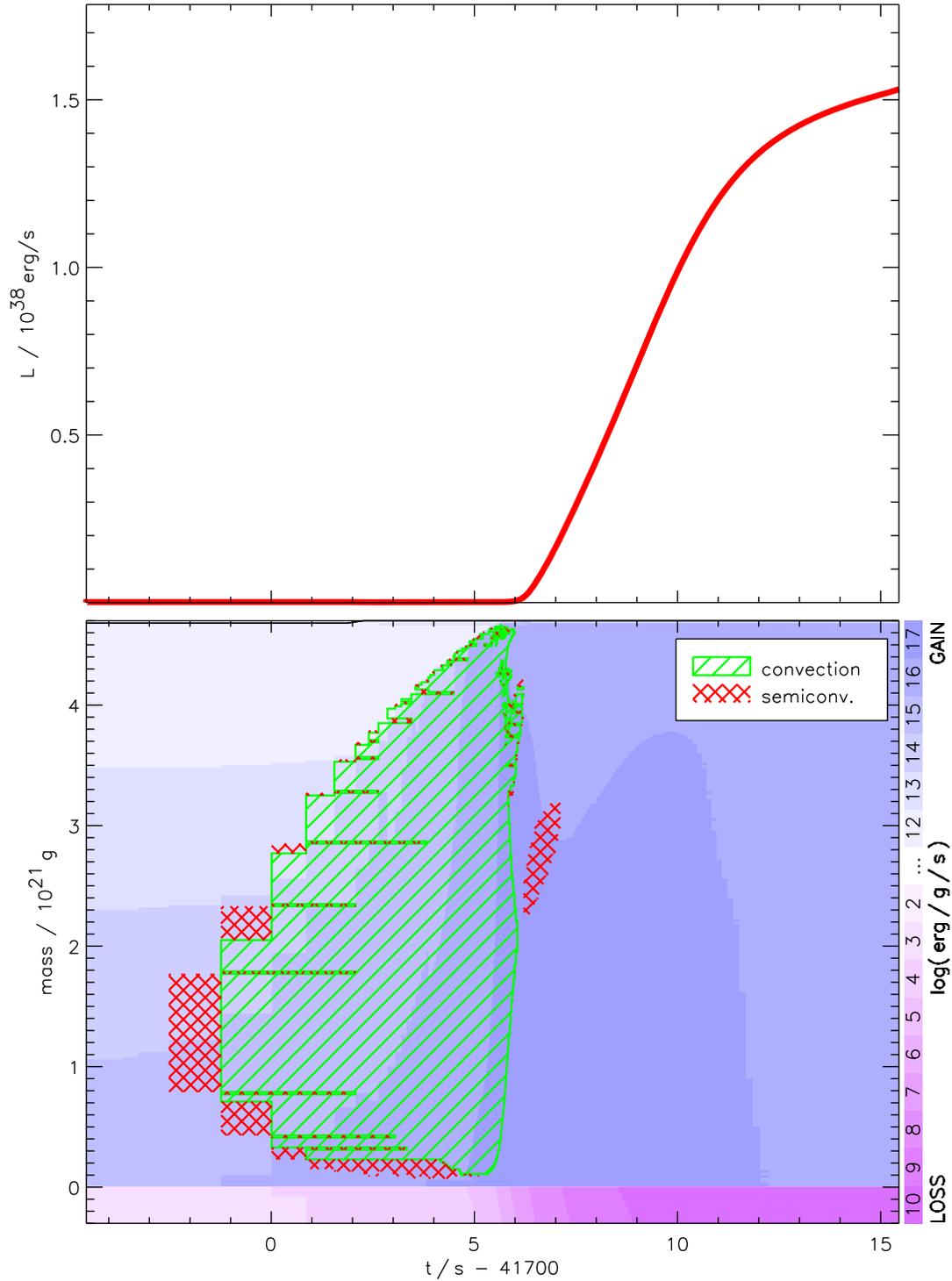}
\caption{The first burst in Model \XRBaiii. The lower panel shows that
ignition initially occurs above the composition interface, but grows
over the next $6\,\Sec$, both outwards and inwards, to encompass the
entire shell. \textsl{Green hatched} regions are convective;
\textsl{red cross hatched regions} are semi-convective.  The observed
burst commences shortly after convection reaches the surface and has
begun to recede so that the luminosity during the rise is chiefly
transported by diffusion.  Times are offset by $41,700\,\Sec$ since
the beginning of accretion.  Any effects due to the spreading of
burning over the surface of the neutron star are ignored. Though given
here for our standard choice, the rise time is sensitive to the
nuclear reaction rates employed in the calculation (\Fig{rist3}). In
this and all subsequent depictions of the light curve, general
relativistic effects have been ignored (\Sect{gr}).  \lFig{Aconv1} }
\end{figure}

\clearpage
%5
\begin{figure} \centering
\includegraphics[draft=\Draft,angle=0,height=0.8\textheight]{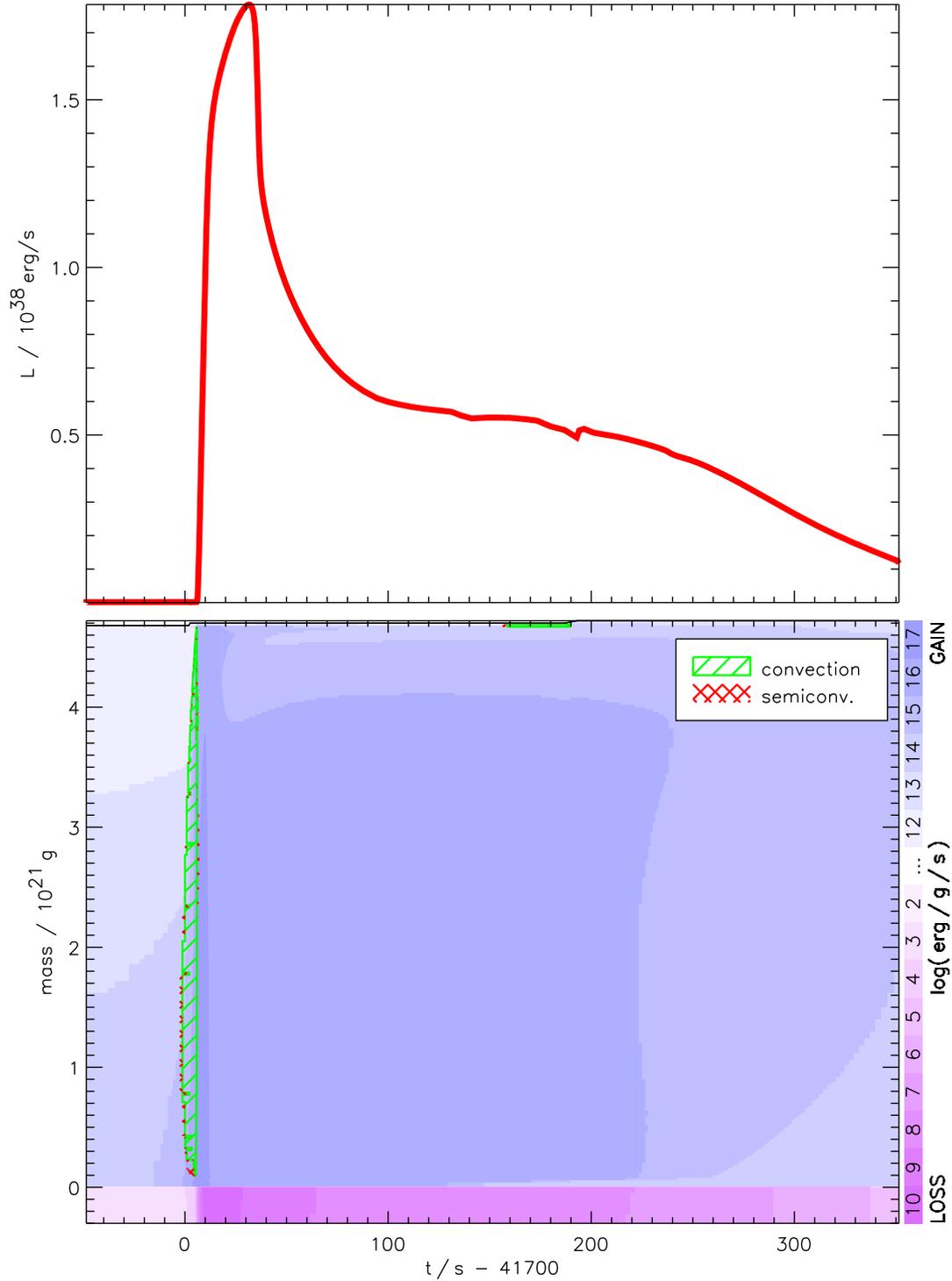}
\caption{ The top figure shows the light curve of burst 1 of
Model~\XRBaiii.  Following the main pulse, lasting about a minute,
there is a long tail, lasting perhaps $5\,\min$, powered by the
continued burning of hydrogen. The shape of the light curve and its
tail are sensitive to nuclear decay rates and proton capture along the
\textsl{rp}-process path (\Fig{weaktest1}).  The bottom panel shows
that appreciable convection only occurs prior to the rise of the pulse
(see also \Fig{Aconv1}).  Shades of blue color indicate the nuclear
energy generation rate while shades of purple indicate energy loss by
neutrinos, both on a logarithmic scale.  The flow of heat into the
neutron star substrate is apparent.  \lFig{lite1}}
\end{figure}

\clearpage
%6
\begin{figure} \centering
\includegraphics[draft=\Draft,angle=0,width=\columnwidth]{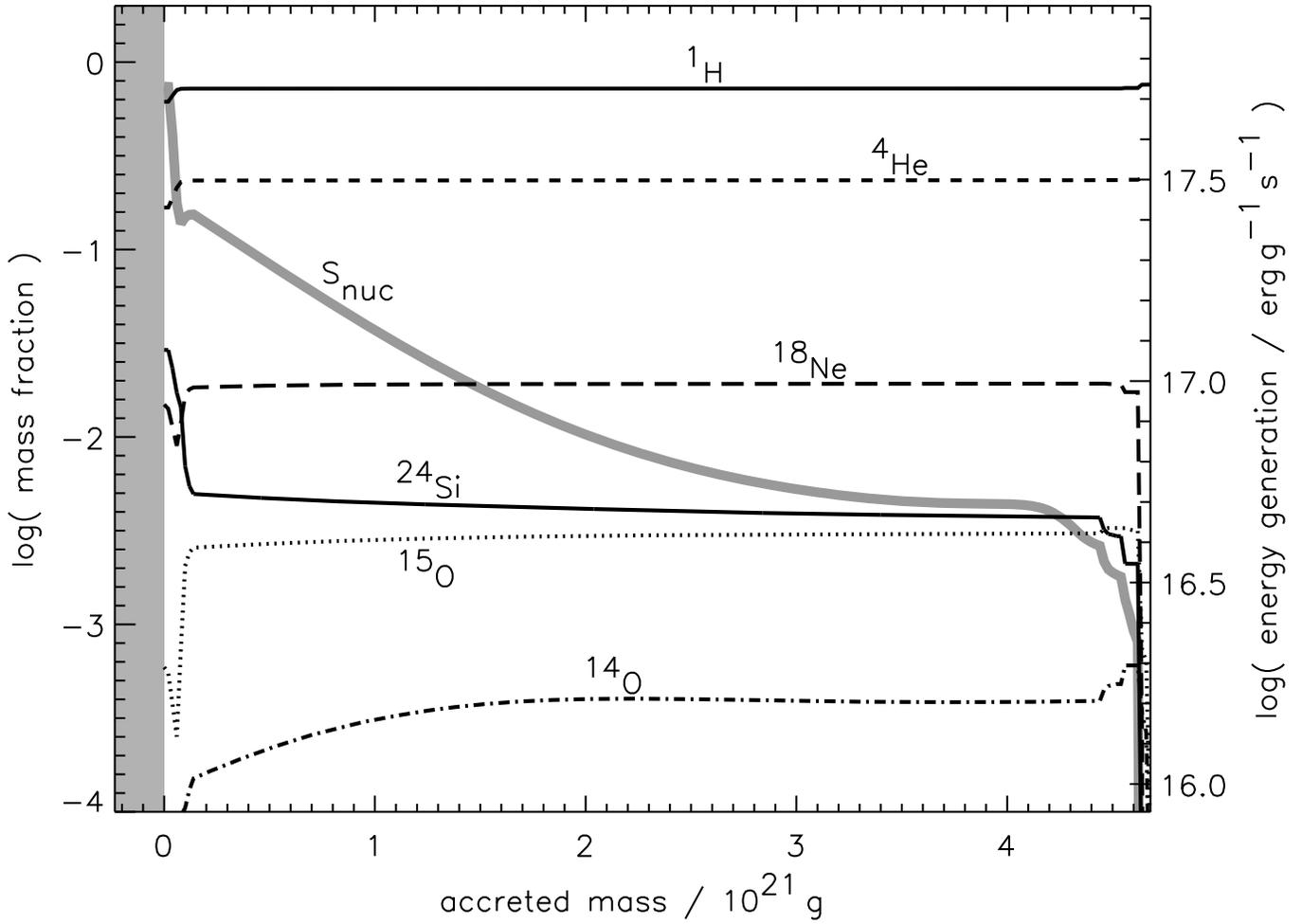} 
\caption{Composition and nuclear energy generation during the
convective phase of burst 1 in Model~\XRBaiii \ (41,705.4 s). The
\textsl{rp}-process is responsible for producing energy and
intermediate mass nuclei, the most abundant two of which are here
$^{18}$Ne and $^{24}$Si.  Nuclei as heavy as \I{64}{Ge} already have
appreciable abundance ($2\E{-4}$ by mass).  The region of near
constant abundances is convective. Nuclear power is $8.3\E{37}\,\erg$
s$^{-1}$, though the luminosity transported to the surface is still
only $1.3\E{35}\,\ergs$, invisible compared with that due to
accretion. The observable burst has not started yet.  \lFig{comp2}}
\end{figure}

\clearpage
%7
\begin{figure} \centering
\includegraphics[draft=\Draft,angle=0,width=\columnwidth]{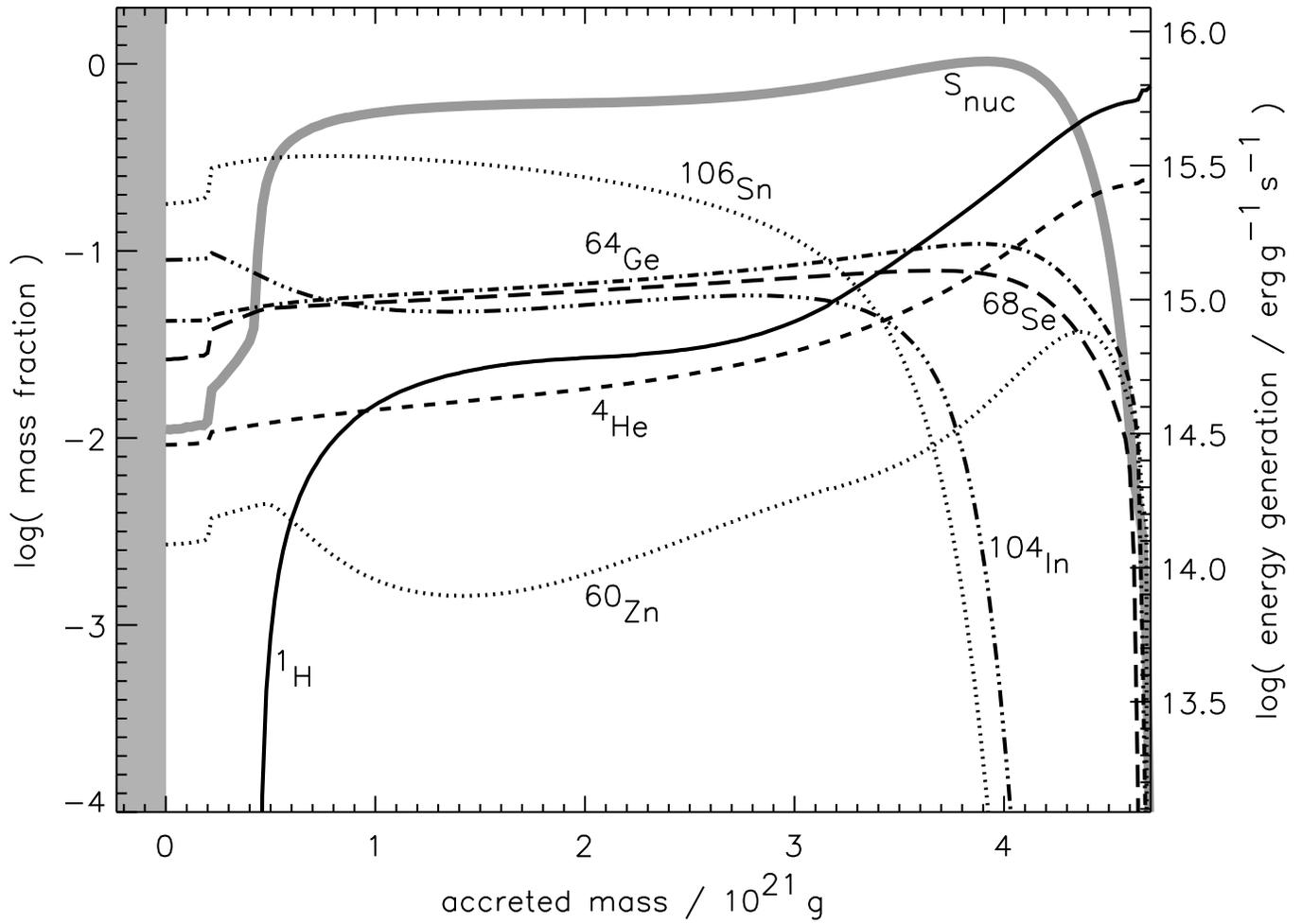} 
\caption{Composition of Model \XRBaiii \ and nuclear energy generation
in the tail of the light curve of burst number 1 ($41,992\,\ergs$,
\Fig{lite1}).  Though depleted at the base, considerable hydrogen
remains at higher altitudes and burning continues at a high rate.  The
ashes at the bottom of the layer are rich in \I{106}{Sn}, the
termination of the \textsl{rp}-process. \lFig{comp3}}
\end{figure}

\clearpage
%8
\begin{figure} \centering
\includegraphics[draft=\Draft,angle=0,width=\columnwidth]{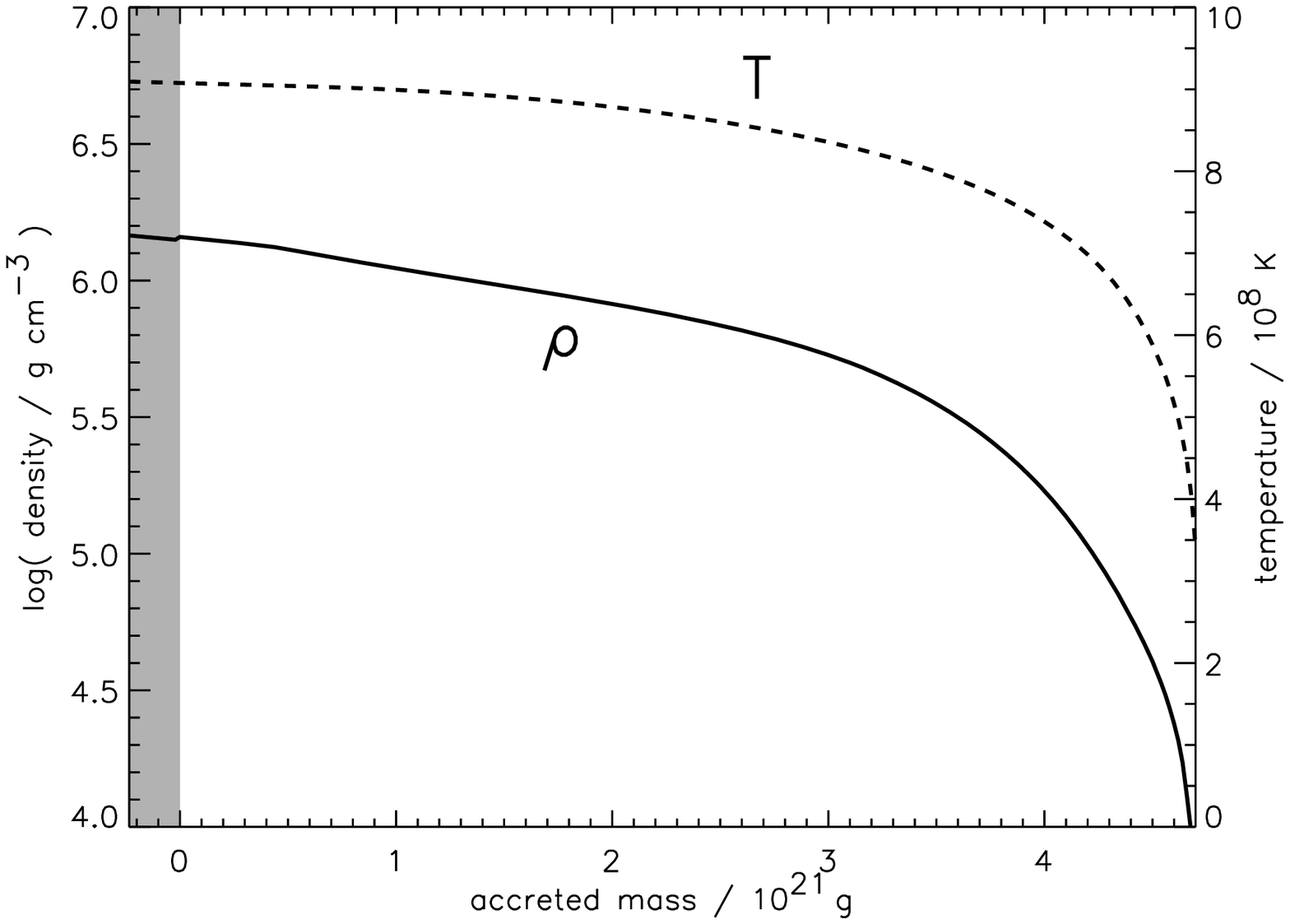} 
\caption{Density and temperature for Model~\XRBaiii \ at the same time
(41,992 s) as \Fig{comp3}. The temperature at the base is
$9.07\E8\,\K$ and the density, $1.44\E6\,\gcc$.
\lFig{dntn1}}
\end{figure}

\clearpage
%9
\begin{figure} \centering
\includegraphics[draft=\Draft,angle=0,width=\columnwidth]{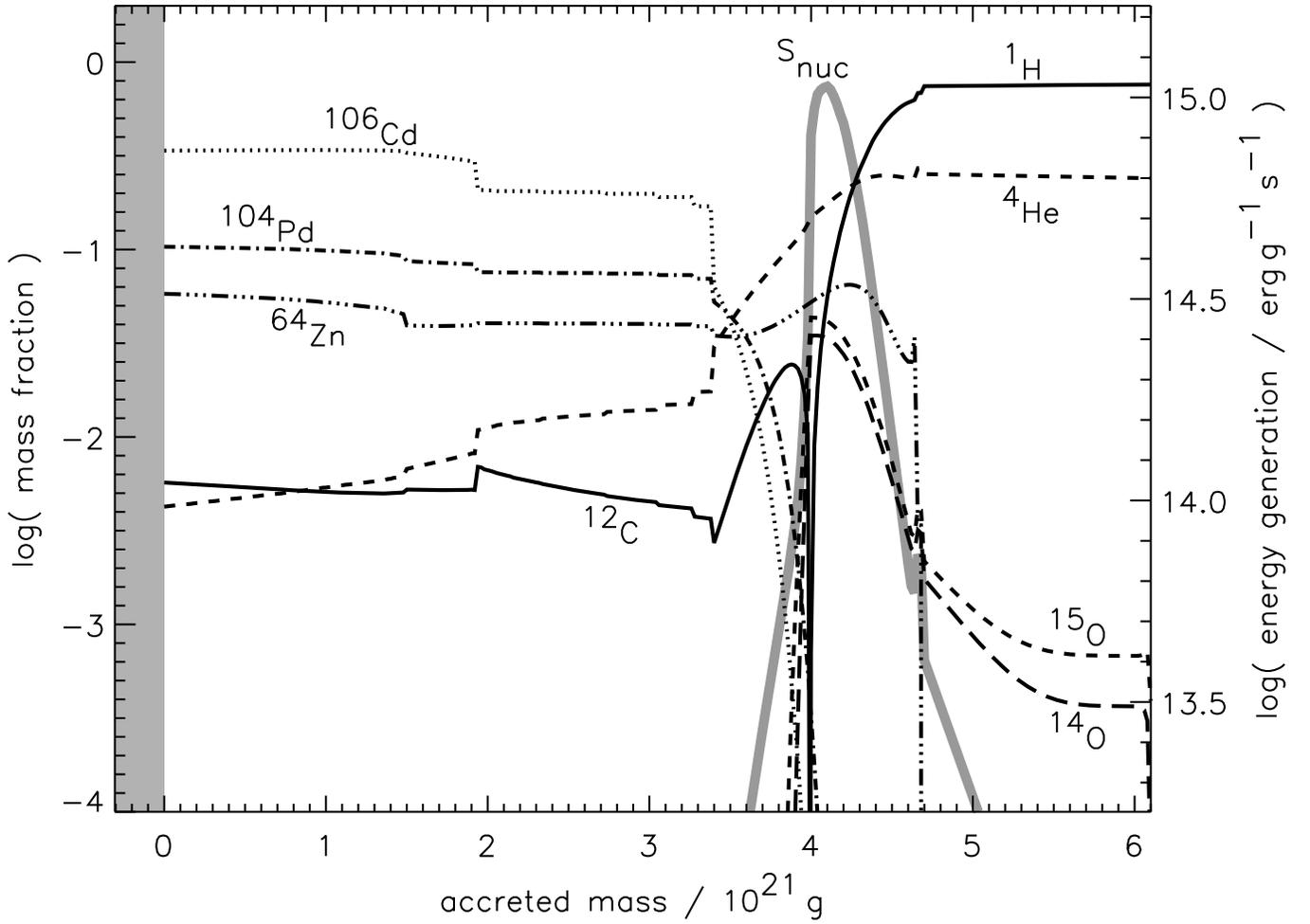} 
\caption{Composition and energy generation at the onset of the second
burst in Model \XRBaiii \ ($5,4592\,\Sec$), $29\,\Sec$ before the
flash.  The ignition of the second burst occurs in the ashes of the
previous one, not at the bottom of the newly accreted layer (which is
located at $4.7\E{21}\,\g$). This occurrence makes the critical mass
for second and subsequent bursts smaller and less sensitive to the
metallicity of the accreted material. Temperature and density at the
location of maximum energy generation are $9.4\E5\,\gcc$ and
$2.9\E8\,\K$ respectively.  A short time later, convection mixes the
heavy ashes, e.g., \I{64}{Zn}, out into the freshly accreted
layer. \lFig{compp2}}
\end{figure}

\clearpage
%10
\begin{figure} \centering
\includegraphics[draft=\Draft,angle=0,height=0.8\textheight]{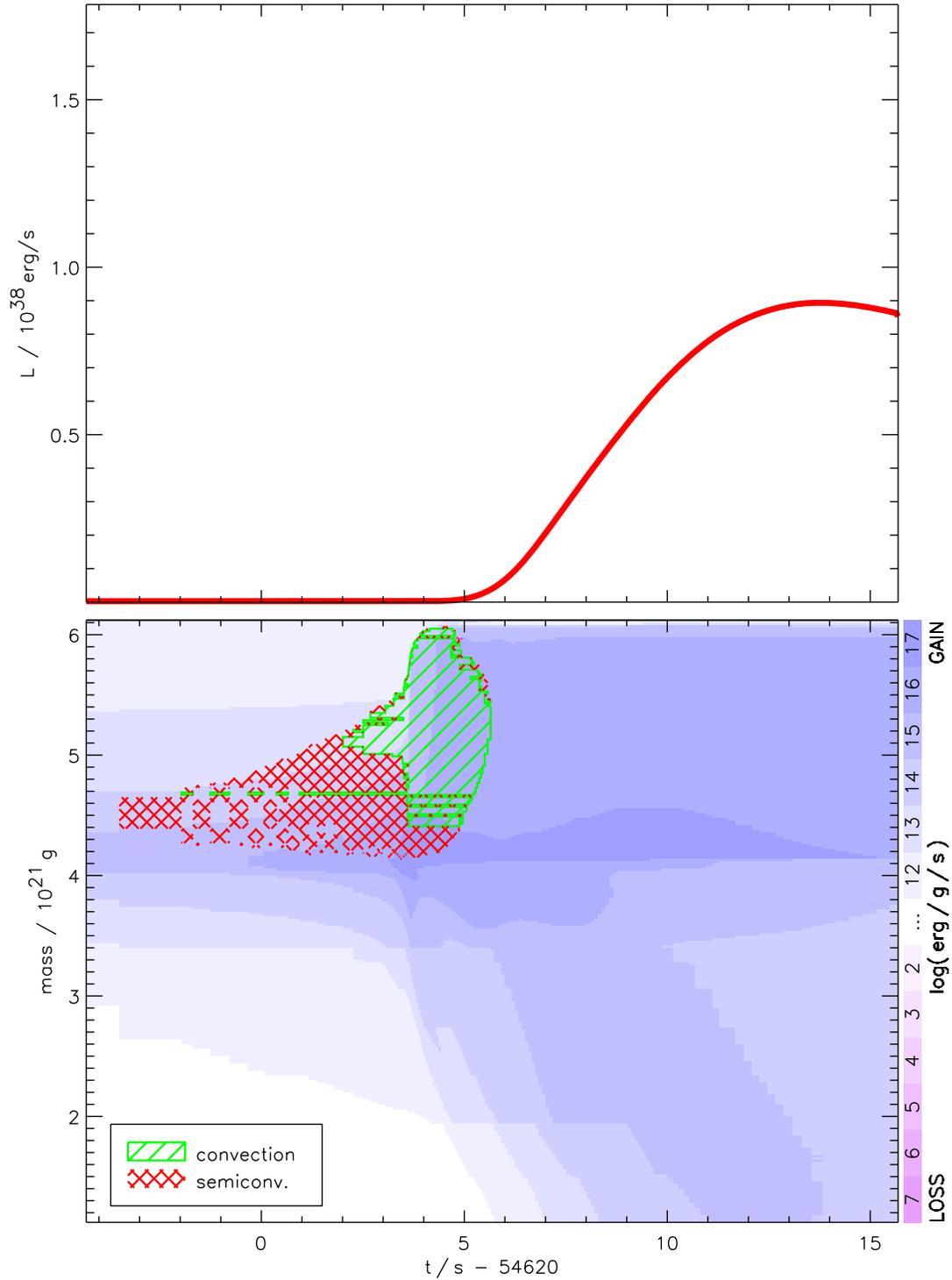}
\caption{Rise time and convection for burst number 2 of Model
\XRBaiii. This is a more typical burst for the model than the one
shown in \Fig{Aconv1}. Once again, convection has ceased by the time
the burst first becomes visible. Following a brief convective stage
lasting about 2 s, well above the bottom of the freshly accreted layer,
the principal burning starts inside the ashes of the previous burst
which end at $4.7\E{21}\,\g$ (\Fig{compp2}).  The intensity of the blue
scale indicates nuclear burning.  Note that appreciable burning occurs
in the ashes of burst number 1 as the heat wave from burst number 2
propagates through. This behavior is in contrast to the neutrino
losses which dominated in \Fig{Aconv1}.  The rise time is sensitive to
the nuclear reaction rates as well as the diffusion time scale
(\Fig{rist3}).\lFig{lite2r}}
\end{figure}

\clearpage
%11
\begin{figure} \centering
\includegraphics[draft=\Draft,angle=0,height=0.8\textheight]{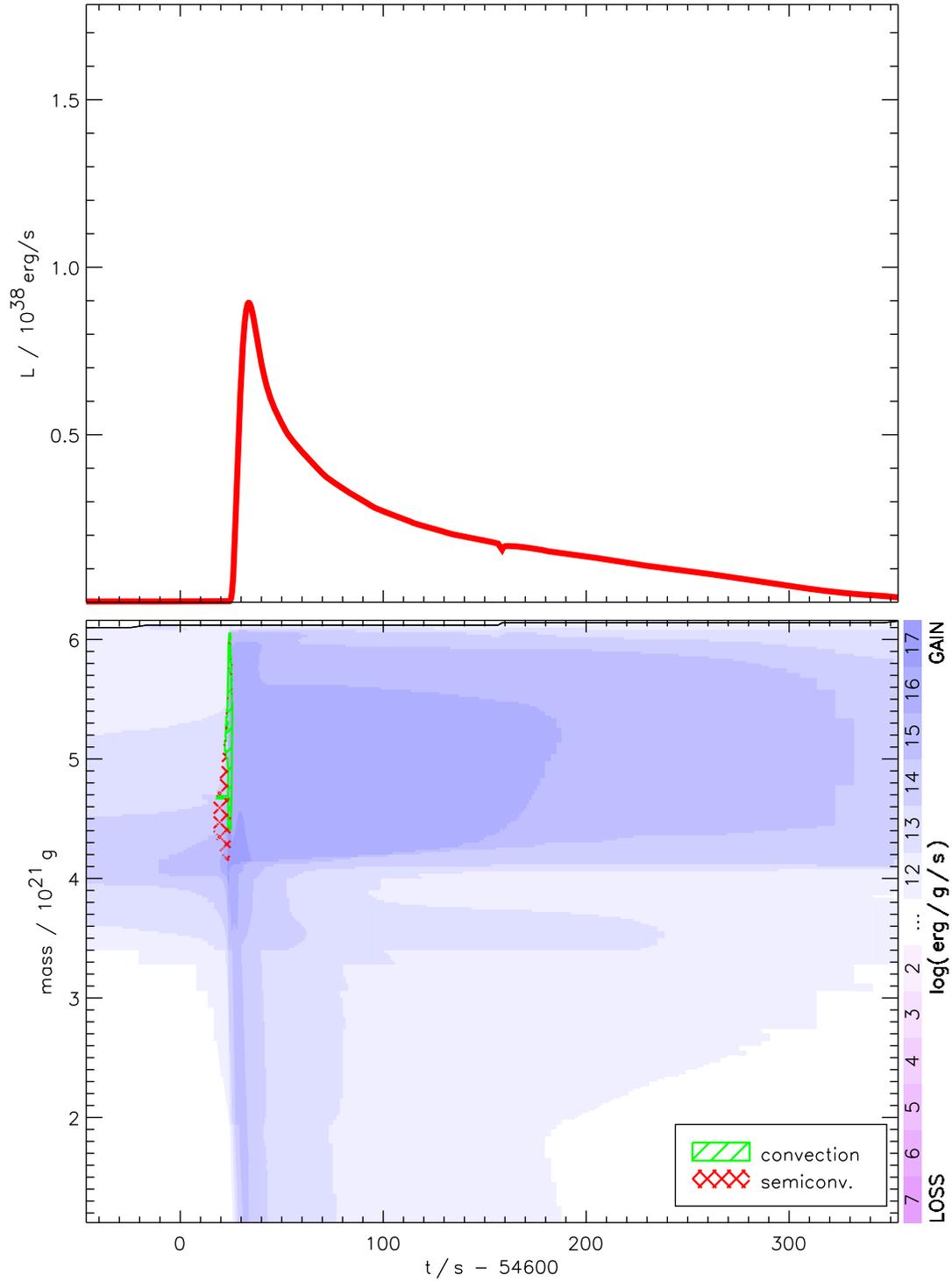}
\caption{Light curve for burst number 2.  This briefer, less
luminous second burst is more typical of all subsequent bursts
in Model~\XRBaiii.  Shades of blue indicate nuclear
burning. \lFig{lite2}}
\end{figure}

\clearpage
%12
\begin{figure} \centering
\includegraphics[draft=\Draft,angle=0,width=\columnwidth]{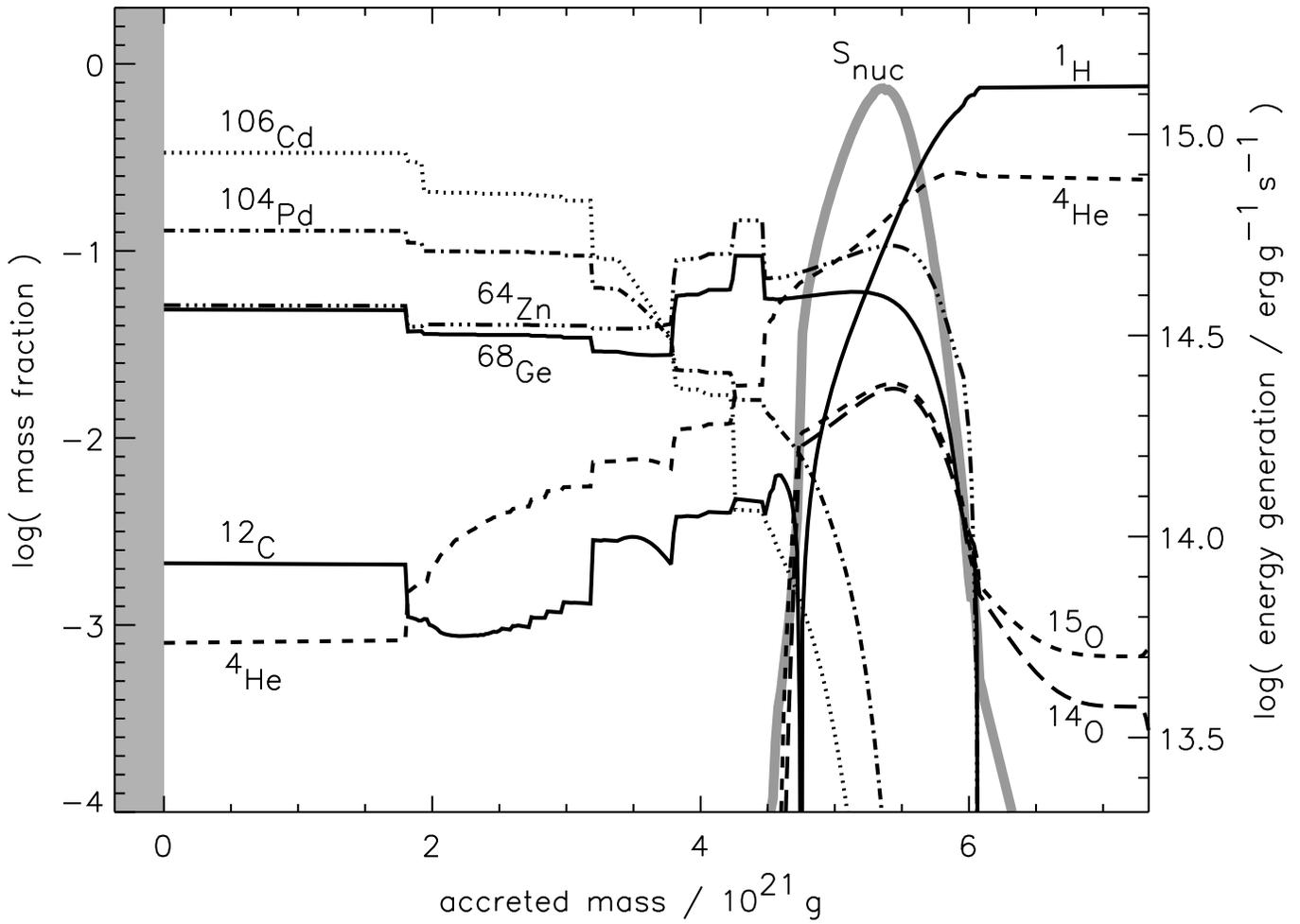} 
\caption{Mass fractions and energy generation at the onset of the
third burst in Model \XRBaiii.  This shows that the conditions at which
the third burst ignites are very similar to those of the
second.  Compare with \Fig{compp2}.  \lFig{compp3}}
\end{figure}

\clearpage
%13
\begin{figure} \centering
\includegraphics[draft=\Draft,angle=90,width=\columnwidth]{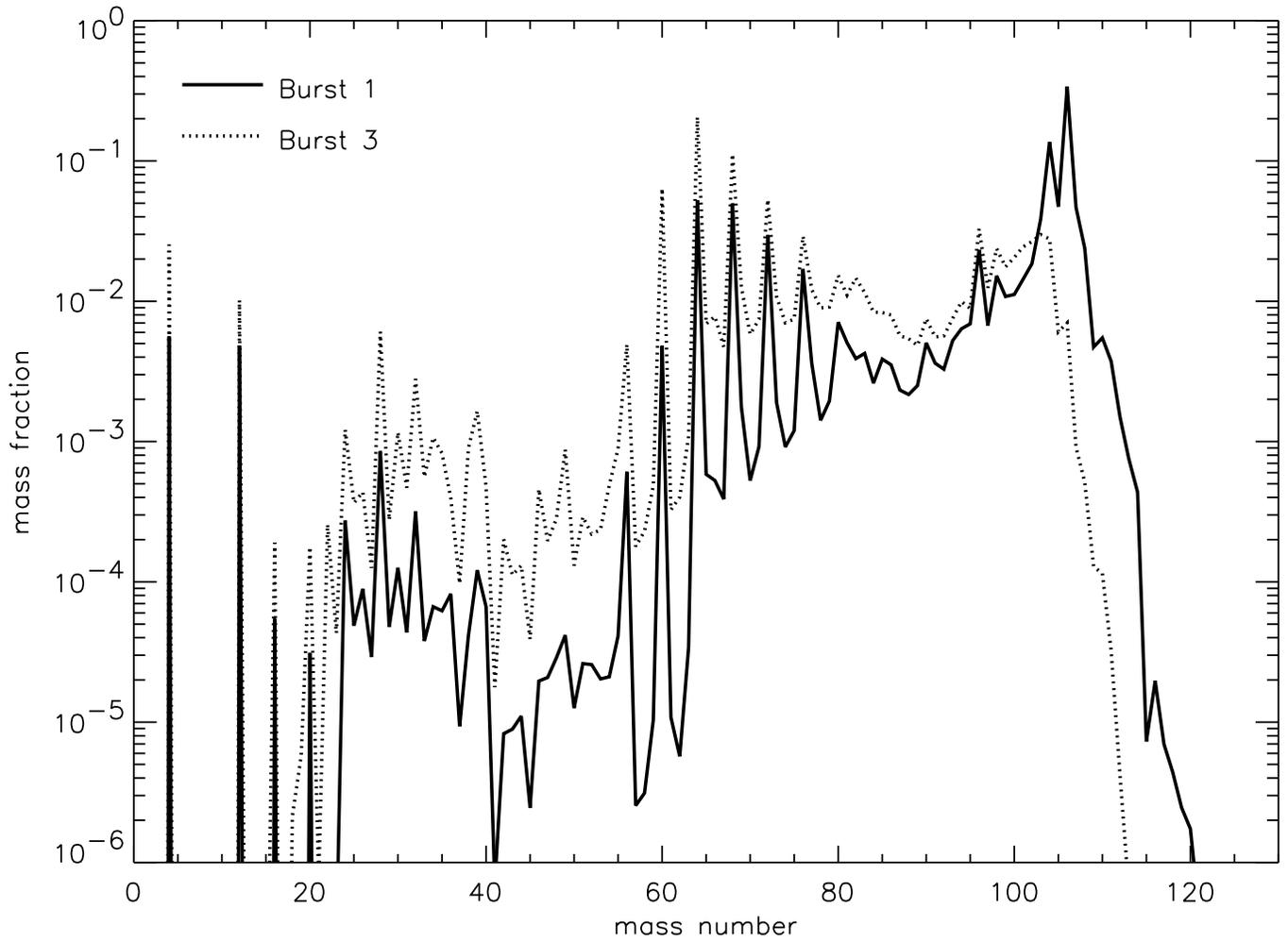} 
\caption{Contrasting abundance distribution at the end of burst 1 and
3 in Model \XRBaiii. In burst 1 the principal products were $^{106}$Cd
and $^{104}$Pd.  In burst 3 they were \I{64}{Zn} and \I{68}{Ge}.
Abundances are evaluated at the base of each accreted layer after
decay.  \lFig{adistp13}}
\end{figure}

\clearpage
%14
\begin{figure} \centering
\includegraphics[draft=\Draft,angle=90,width=\columnwidth]{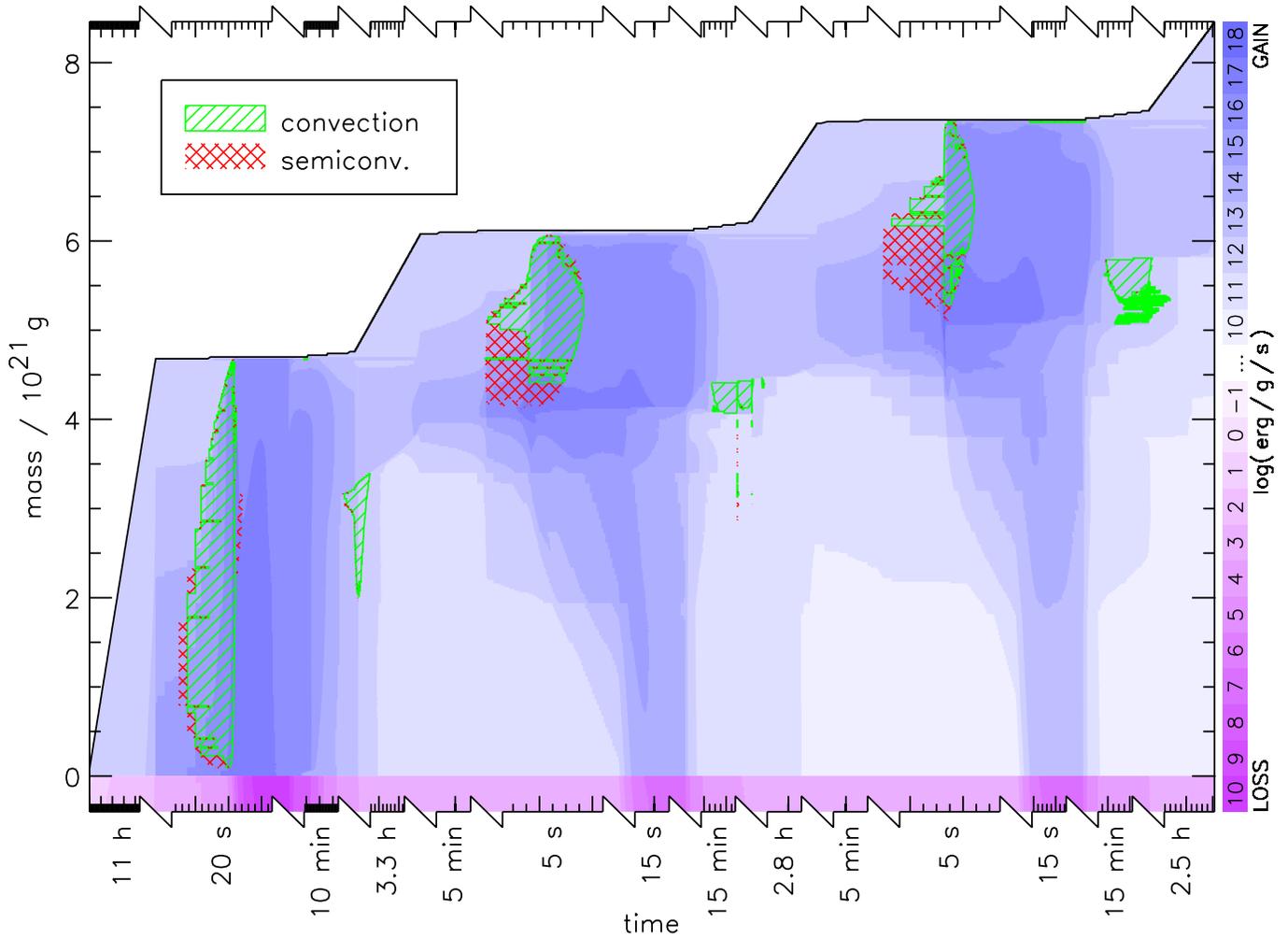}
\caption{Kippenhahn diagram for bursts 1, 2, and 3 in Model \XRBaiii.
\textsl{Green hatching} indicates convection; \textsl{red cross
hatching} semi-convective regions; \textsl{blue shading} shows net
energy generation (nuclear minus neutrino losses); and pink shading
indicates net energy losses (neutrino losses exceeding nuclear energy
generation).  Each level of blue color indicates an increase by one
order of magnitude.  The $y$-axis gives the enclosed mass coordinate
above the assumed neutron star substrate and the thick black line
gives the total mass of accreted material.  The $x$-axis gives time
increasing from left to right, but different parts of the evolution
are plotted on different time scale.  Breaks on the axis indicate a
change of time scale and and below each segment we give the length of
that time interval (not the total time).  The rate of change of the
total mass is inversely proportional to the magnification of time in
each section and is thus also and indicator of the evolutionary time
scale in each section.  Note repeated waves of nuclear burning in the
ashes of previous bursts as heat from the current burst propagates
inwards.  There are also periods of convection in between
bursts. \lFig{conv3}}
\end{figure}

\clearpage
%15
\begin{figure} \centering
\includegraphics[draft=\Draft,angle=0,height=0.8\textheight]{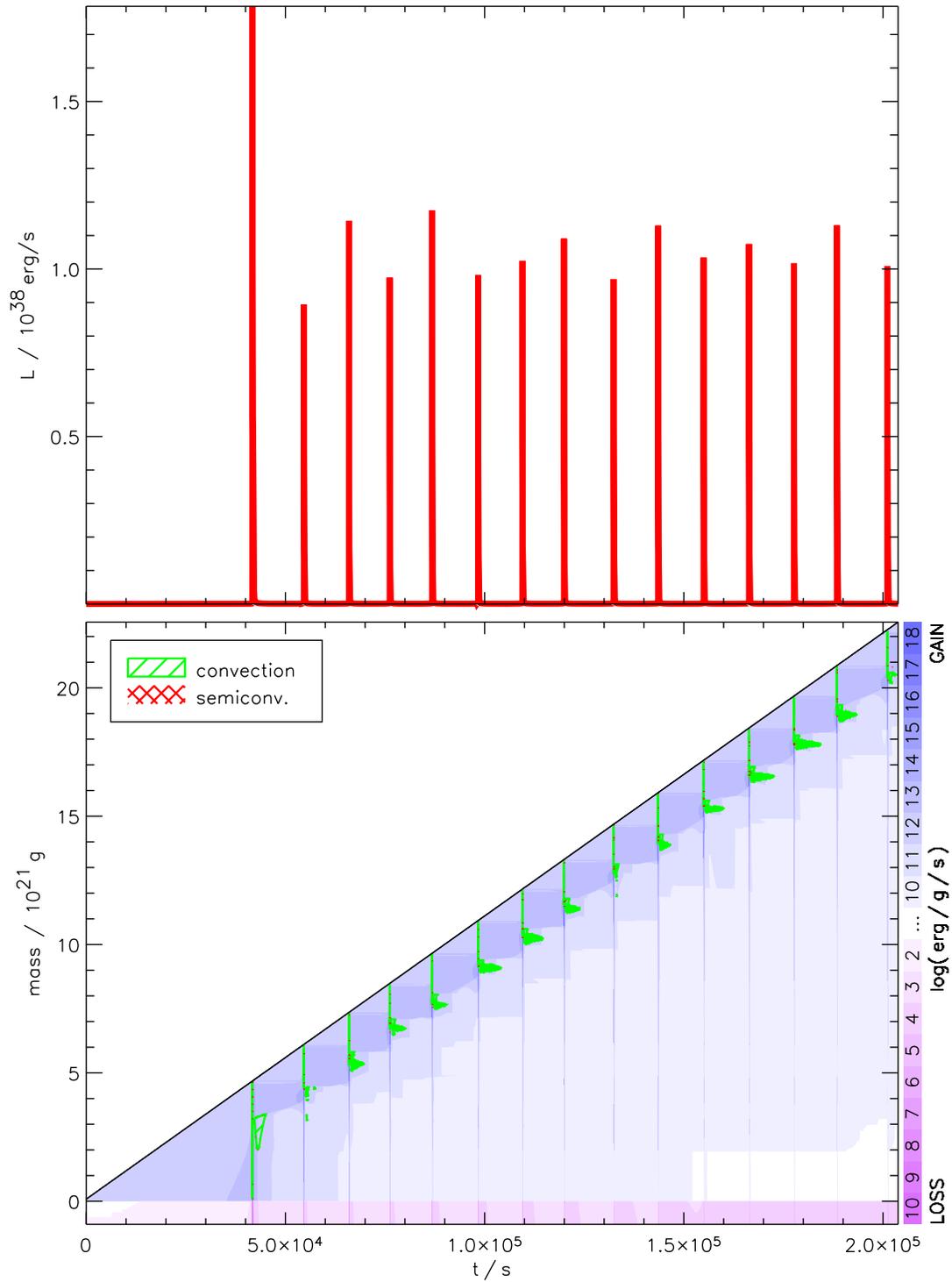}
\caption{Fourteen flashes from Model \XRBaiii.  Note the regularity of the 
last 13.  The first is a start up transient.  Note also the heat flow
and burning in the ashes of earlier bursts. \lFig{Arep2}}
\end{figure}

\clearpage
%16
\begin{figure}
\includegraphics[draft=\Draft,angle=0,width=\columnwidth]{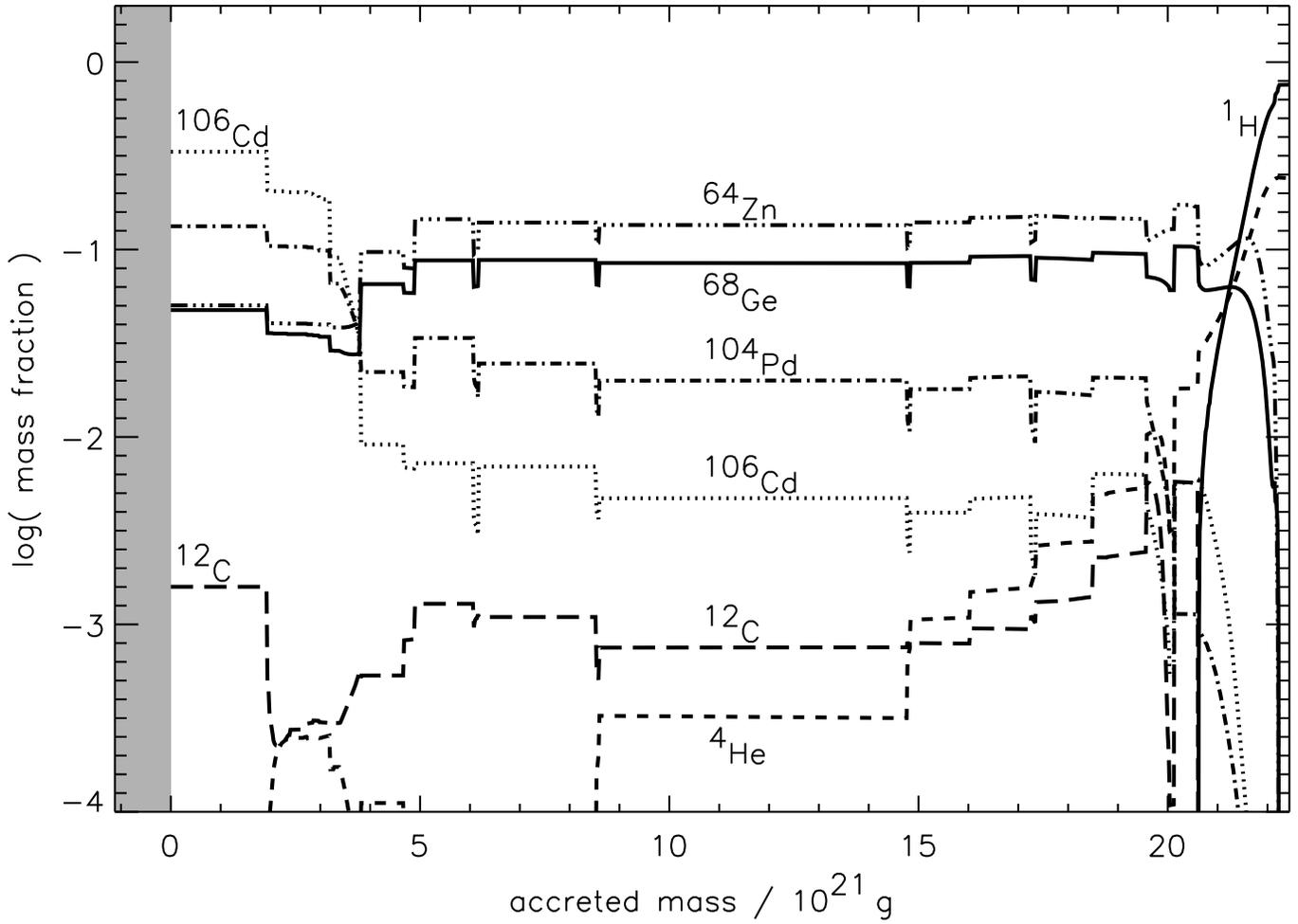}
\caption{Composition of the entire accreted shell after the
14$^{\mathrm{th}}$ burst in Model \XRBaiii.  This is the composition
that will ultimately be merged into the neutron star.  Aside from the
anomalously heavy ashes associated with the first burst, that
composition is mostly \I{64}{Zn} and \I{68}{Ge}.  Very little unburned
carbon remains in these ashes. \lFig{lastcompa3}}
\end{figure}

\clearpage
%17
\begin{figure}
\includegraphics[draft=\Draft,angle=0,width=\columnwidth]{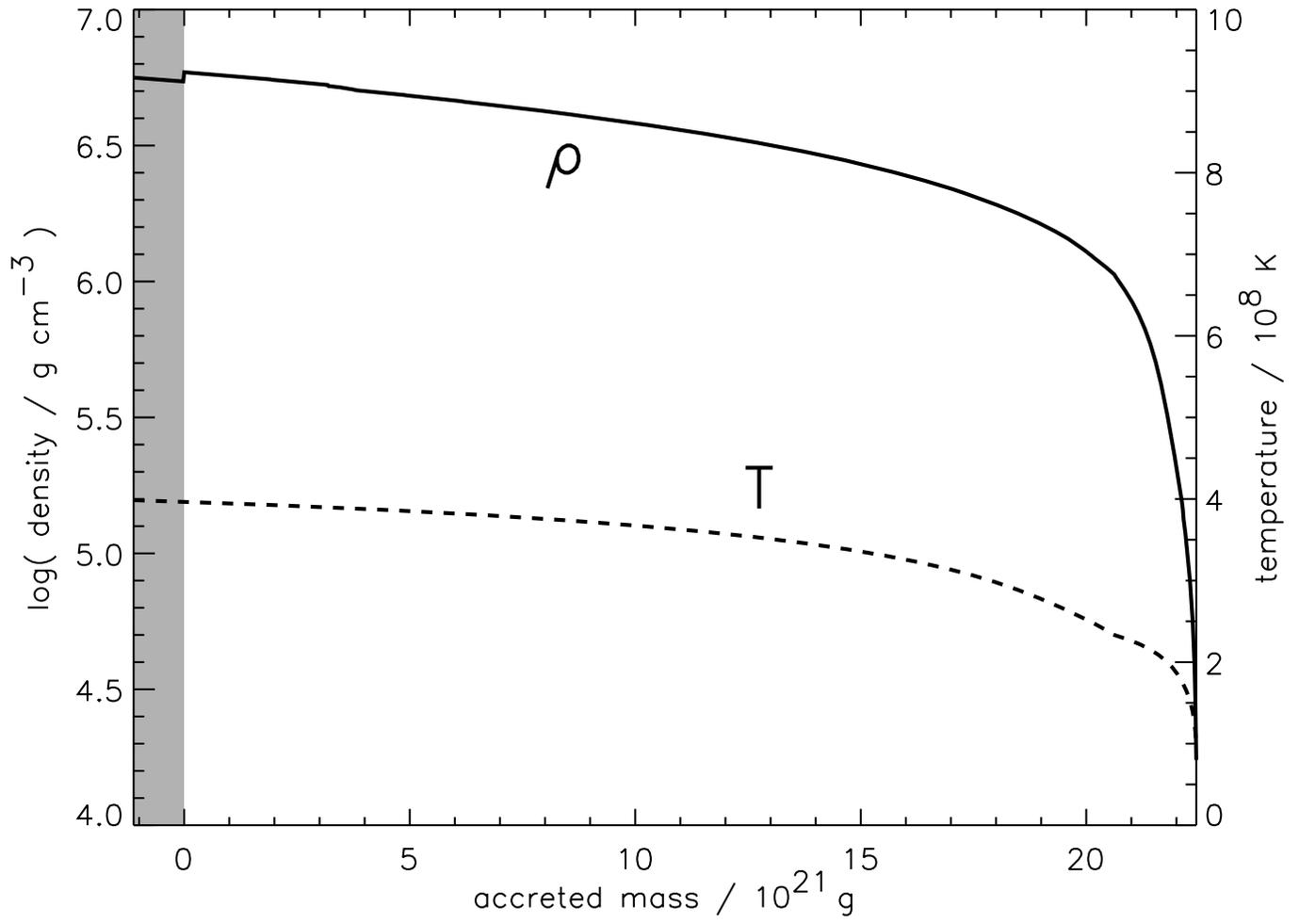}
\caption{Temperature and density at the end of the 14$^{\mathrm{th}}$
burst in Model \XRBaiii.\lFig{lastdntn}}
\end{figure}

\clearpage
%18
\begin{figure}
\includegraphics[draft=\Draft,angle=0,width=\columnwidth]{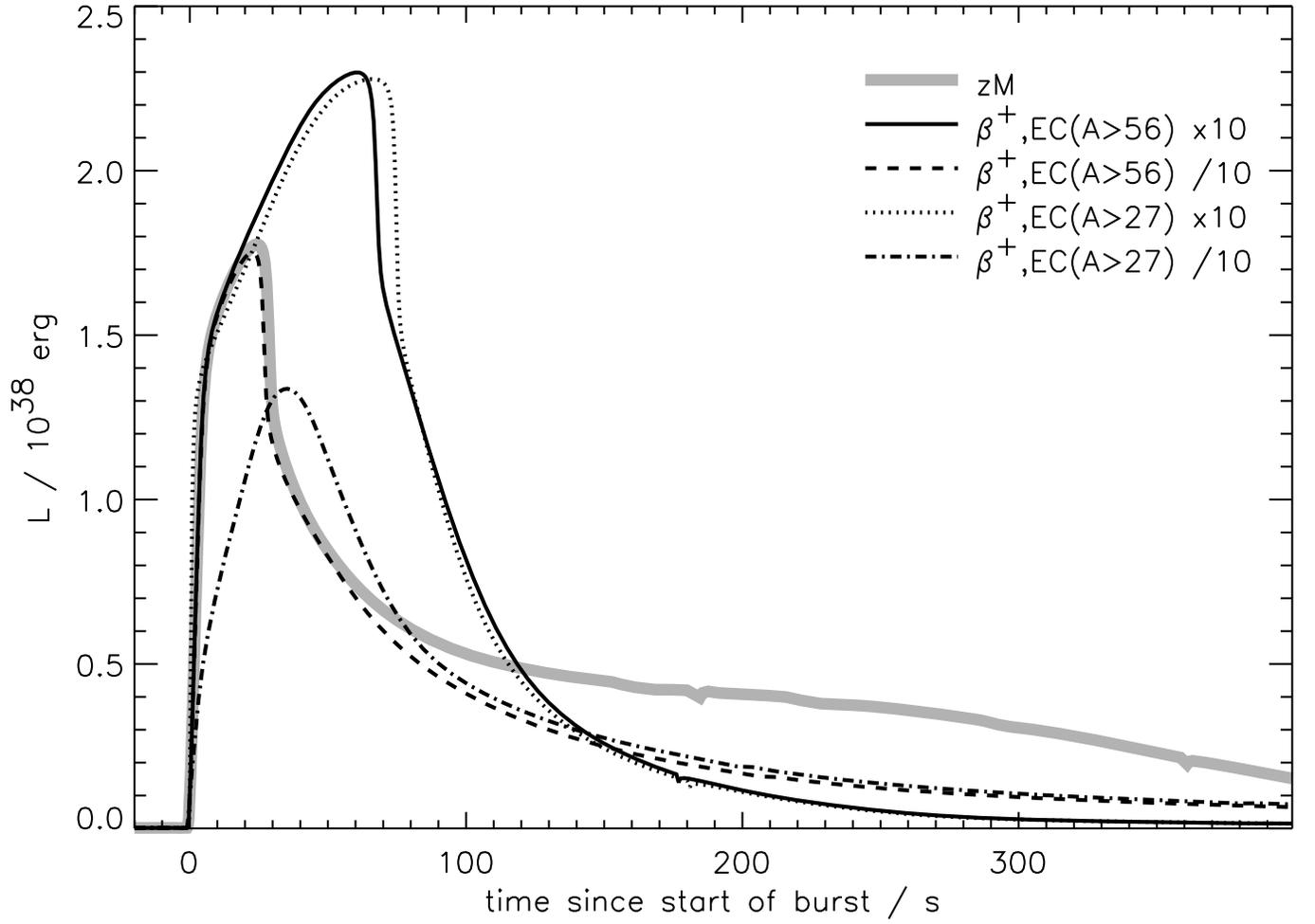}
\caption{Sensitivity of the light curve of the first pulse in Model
\XRBaiii \ to five choices of weak rates - standard, standard times 10
above $A = 56$, standard divided by $10$ above $A = 56$ and similar
modifications above $A = 27$.  Though only the weak rates were
altered, the changes reflect uncertainty in all processes, including
proton capture inhibited by photodisintegration, that affect the
nuclear flow. The shape of the light curve is clearly quite sensitive
to the nuclear data set employed. \lFig{weaktest1}}
\end{figure}

\clearpage
%19
\begin{figure}
\includegraphics[draft=\Draft,angle=0,width=\columnwidth]{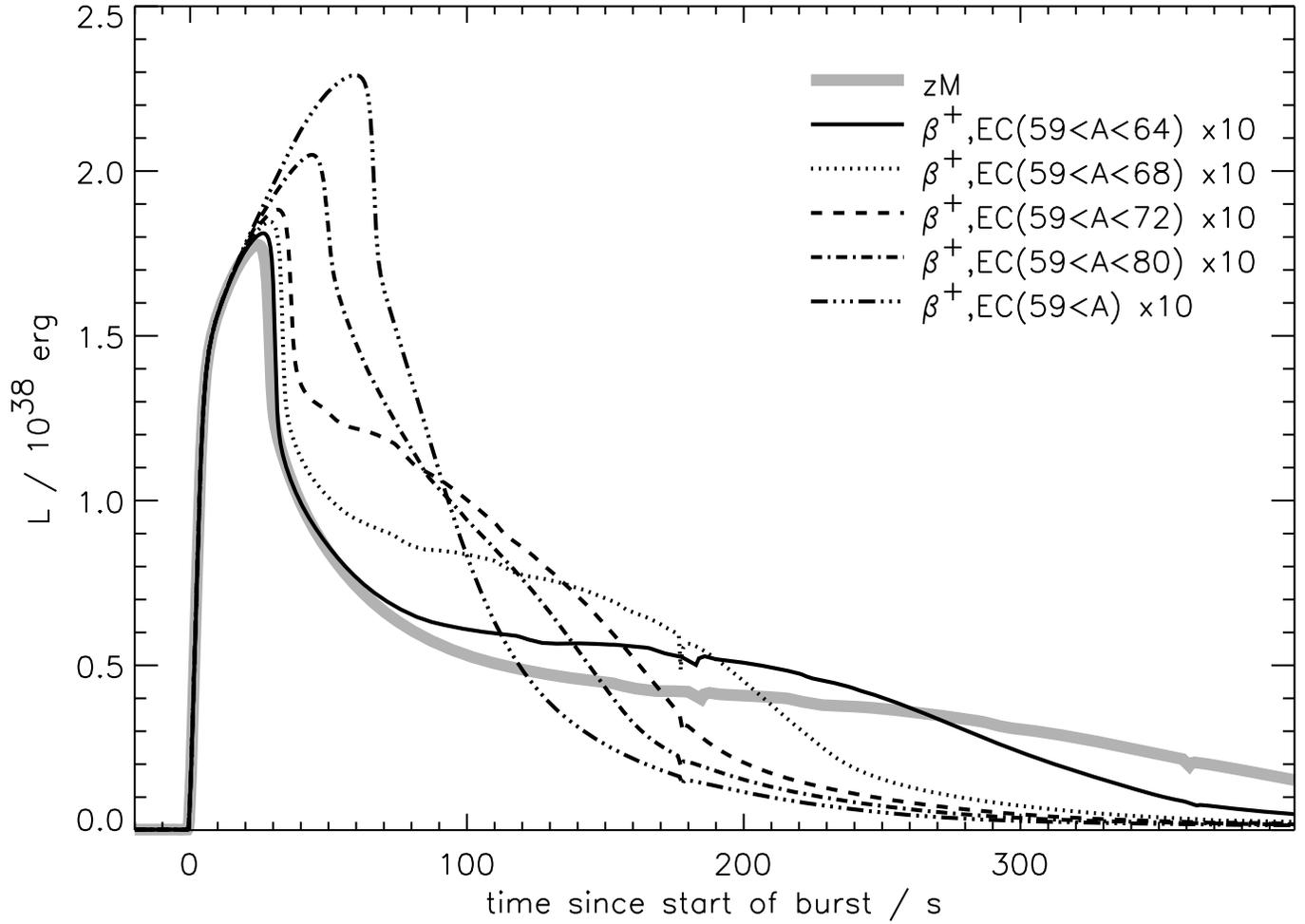}
\caption{Sensitivity of the light curve of the first pulse in Model
\XRBaiii to variations along the waiting points in the vicinity of $A
= 60$, $64$, and $68$. The nominal light curve is shown along with the
result when all weak rates above $A = 59$ are multiplied by $10$ (see
\Fig{weaktest1}; changing $A_{\mathrm{min}}$ from $57$ to $59$ has no
effect). Also shown are the results of progressively adding in
accelerations to flows in the mass ranges $A = 60$--$63$, $64$--$67$,
$68$--$71$, and $72$--$79$.  Details of the flows for these mass ranges
are given in the text.  A separate calculation, not shown, in which
only the decay rate of \I{64}{Ge} was accelerated by 10 is virtually
indistinguishable from the curve \betap, EC($59 < A < 68$) $\times10$.
Factors affecting leakage out of this single nucleus thus dominate the
flow from the iron group to \I{68}{Se}.  The ``blip'' at $180$ s is the
addition of a new surface zone by accretion. \lFig{zn60}}
\end{figure}

\clearpage
%20
\begin{figure}
\includegraphics[draft=\Draft,angle=0,width=\columnwidth]{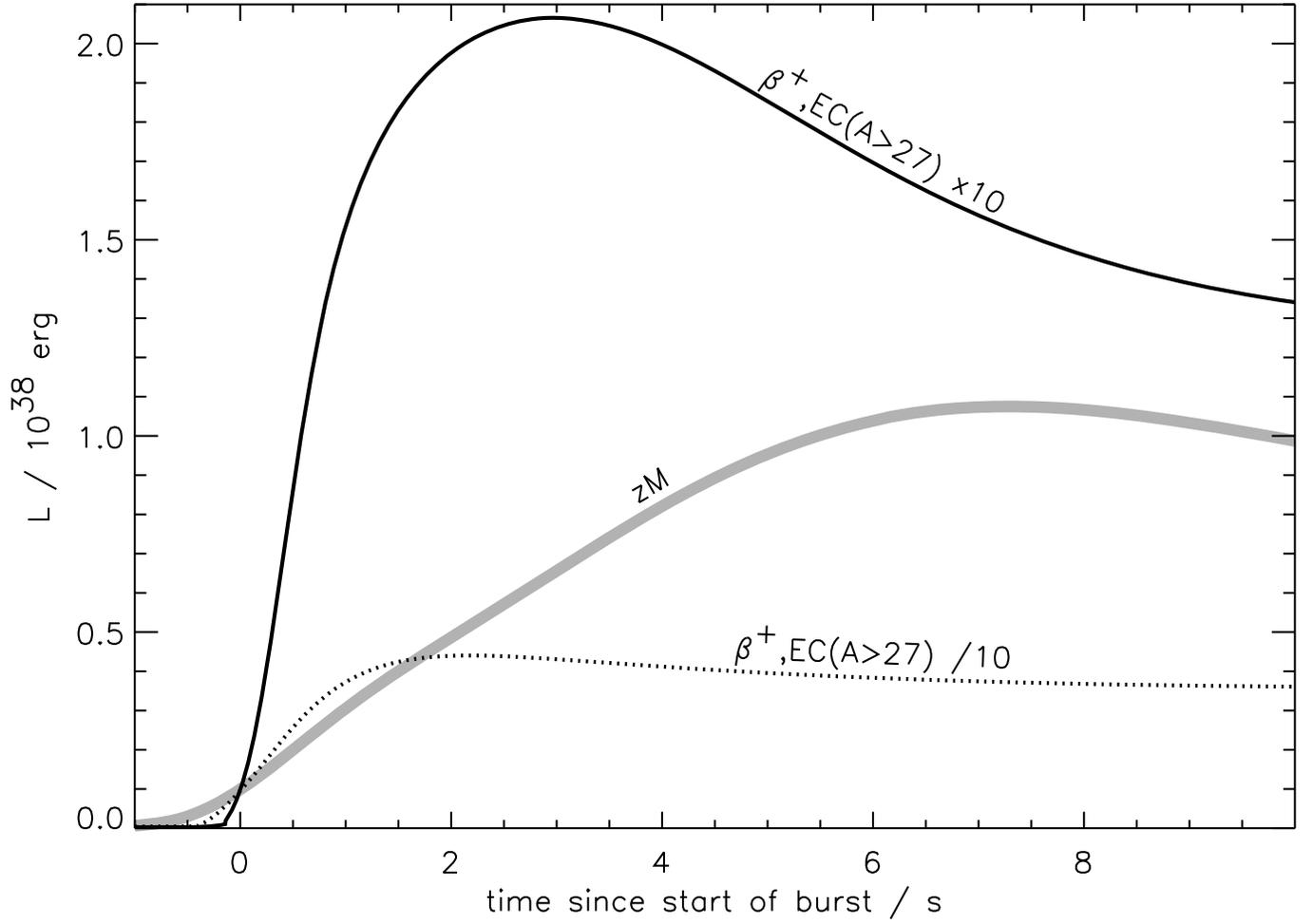}
\caption{Sensitivity of the rise of the light curve of the third pulse
in Model \XRBaiii for three choices of weak rates - standard, standard
times $10$ above $A = 27$, and standard divided by $10$ above $A =
28$. Flows affecting the rise time are discussed in the text.  Time
zero is defined as when each burst reaches $\Ep{37}\,\ergs$.  Because
this is the third burst, there have been cumulative effects from the
altered rates; the critical masses of the burning layers, for example,
are not the same (\Tabs{a3p} - \Tabff{a3pp1}), nor are the total burst
energies.  \lFig{rist3}}
\end{figure}

\clearpage
%21
\begin{figure}
\includegraphics[draft=\Draft,angle=0,width=\columnwidth]{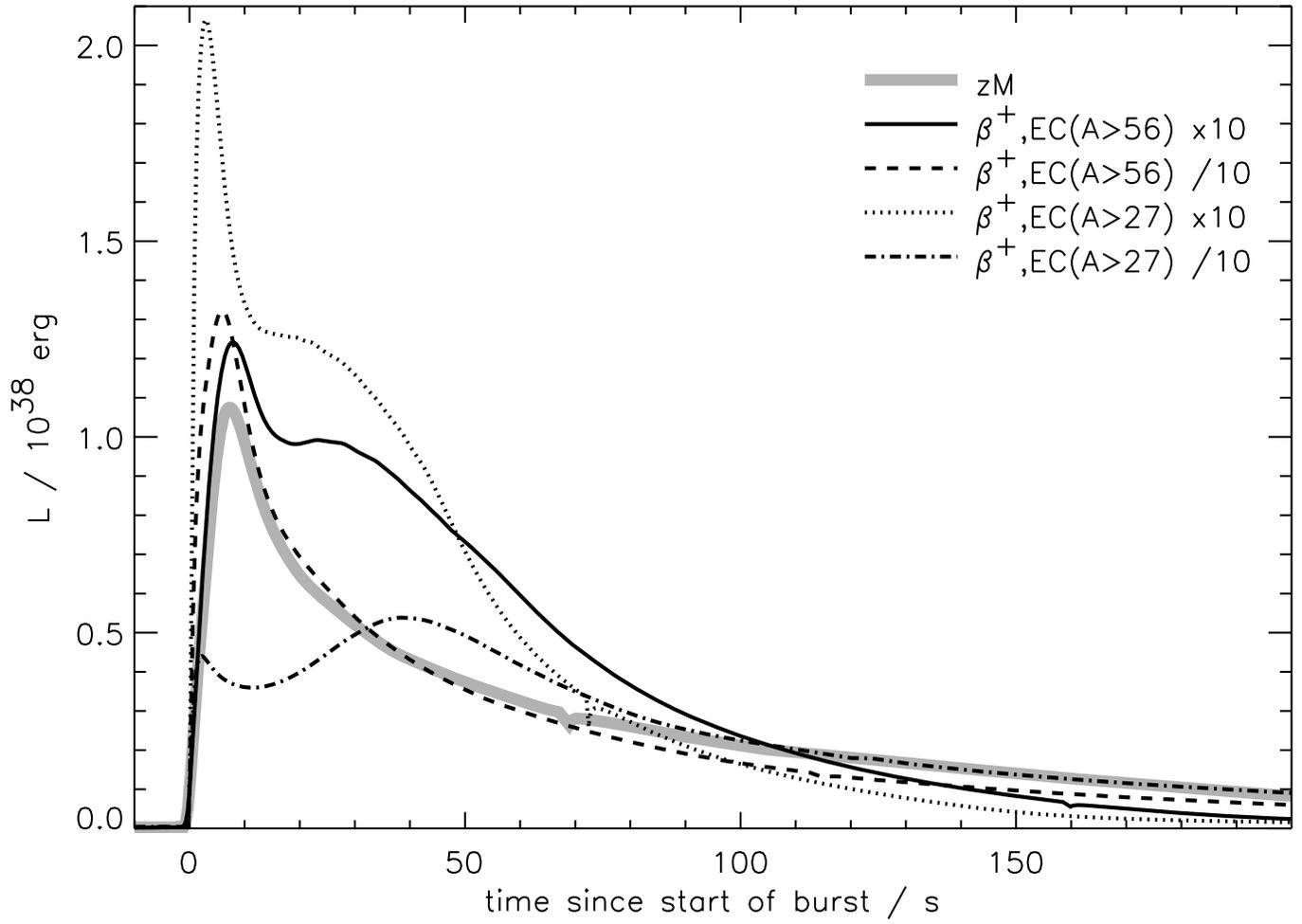}
\caption{Sensitivity of the light curve of the third pulse in Model
\XRBaiii to variation in the nuclear physics employed (see
\Fig{weaktest1}).  This third burst may be more representative than the
first one. \lFig{weaktest3}}
\end{figure}

\clearpage
%22
\begin{figure}
\includegraphics[draft=\Draft,angle=0,width=\columnwidth]{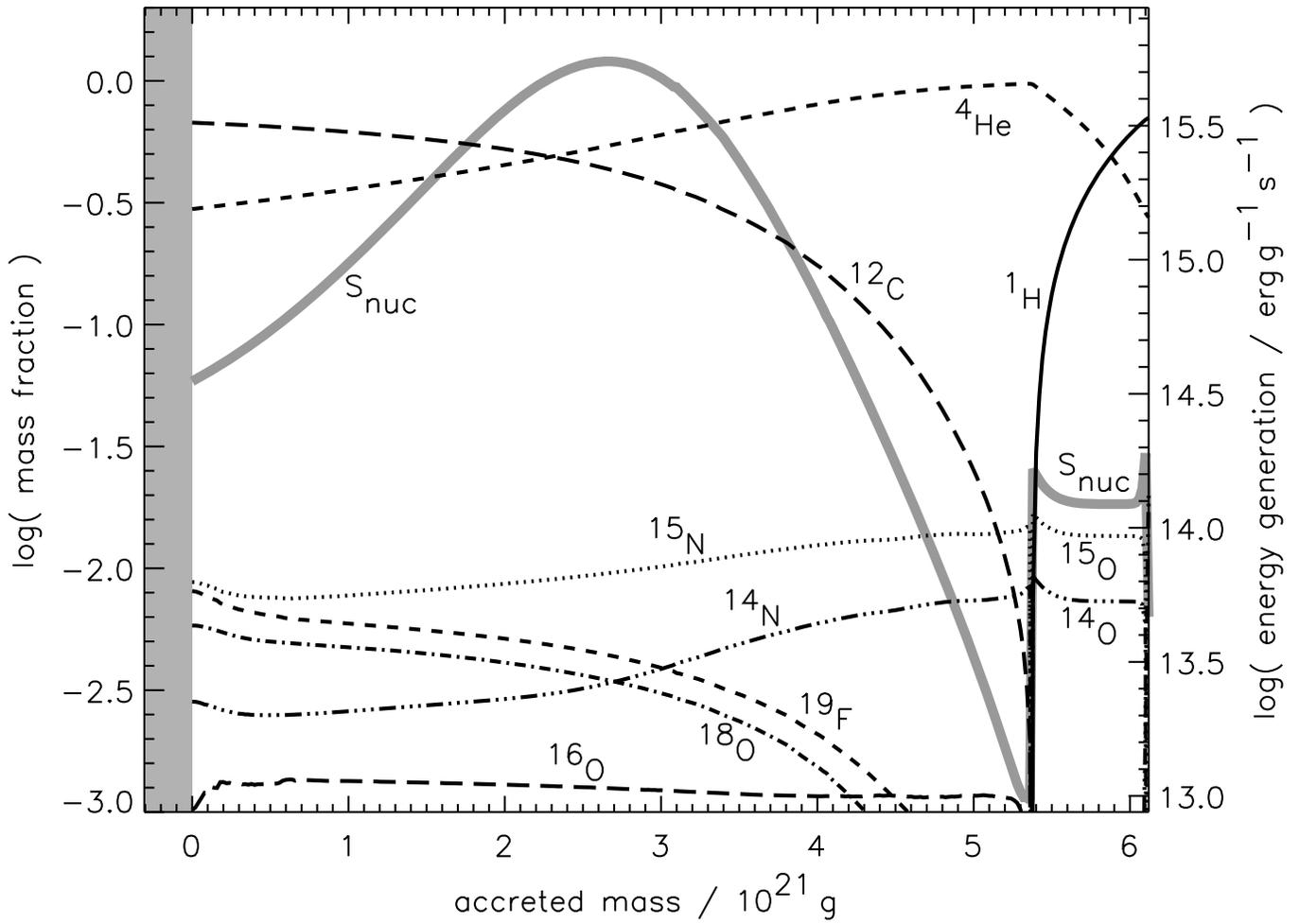}
\caption{Composition at the beginning of the first burst in Model
\XRBavi.  Runaway here occurs in the helium burning shell where a lot
of carbon ($67\,\%$ by mass) has already been synthesized.  Subsequent
convective mixing of the helium and hydrogen layers leads to the
explosive burning of hydrogen in the presence of a large amount of
carbon catalyst.  The final products will not be particularly
heavy. \lFig{compa6a}}
\end{figure}

\clearpage
%23
\begin{figure}
\newcommand{\panelwidth}{0.48\columnwidth}
\includegraphics[draft=\Draft,angle=0,width=\panelwidth]{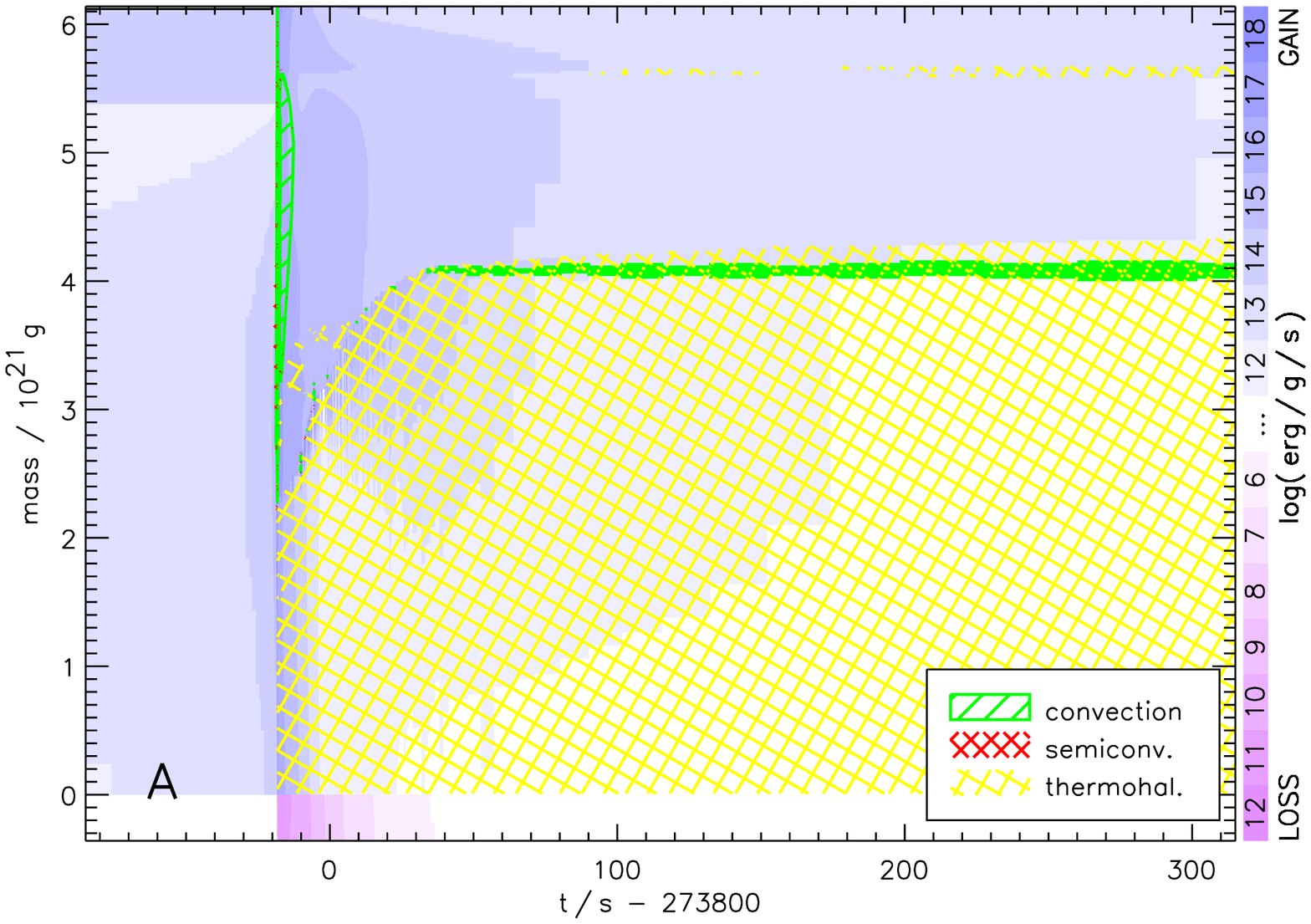}\hfill
\includegraphics[draft=\Draft,angle=0,width=\panelwidth]{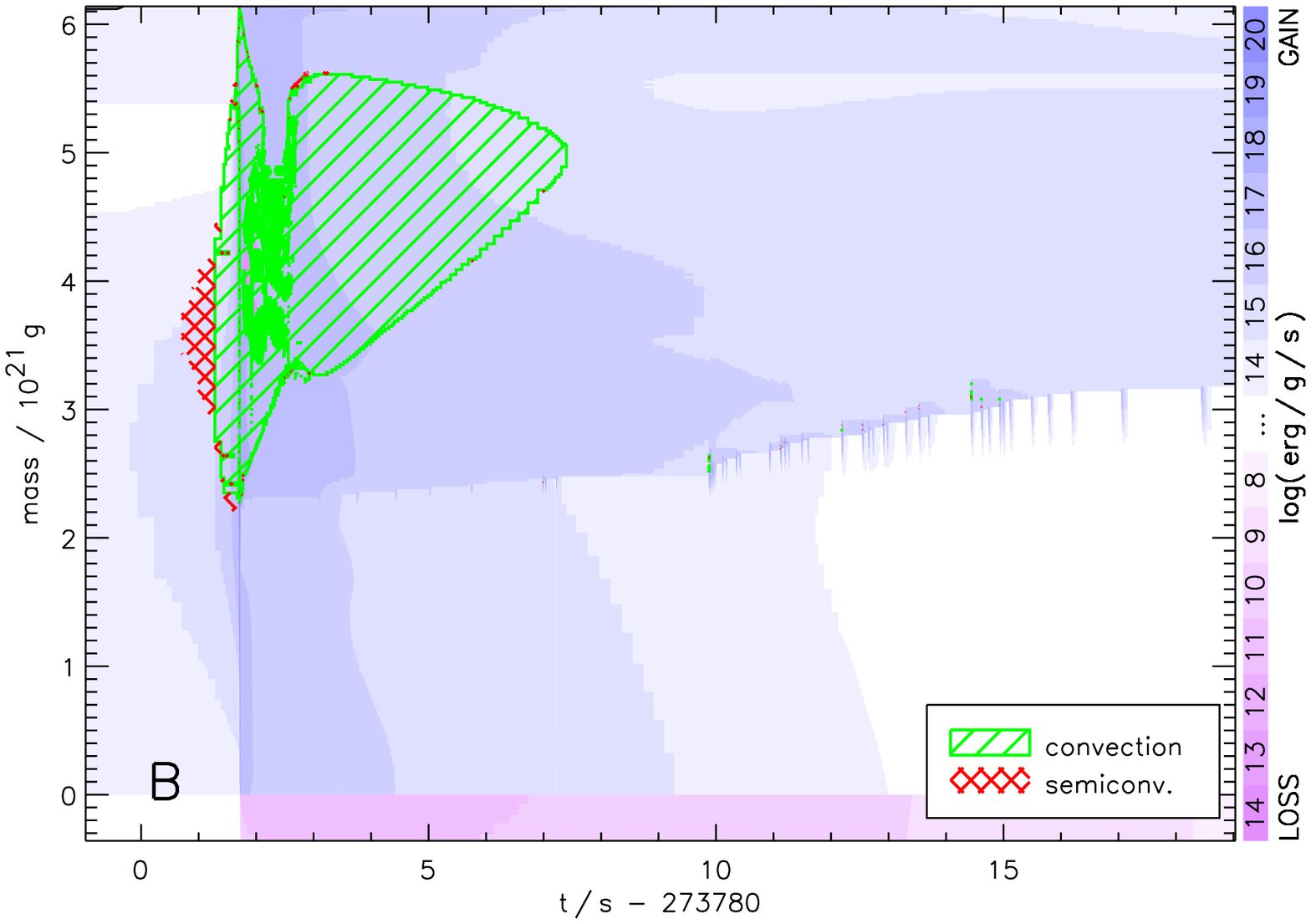}\\[\baselineskip]
\includegraphics[draft=\Draft,angle=0,width=\panelwidth]{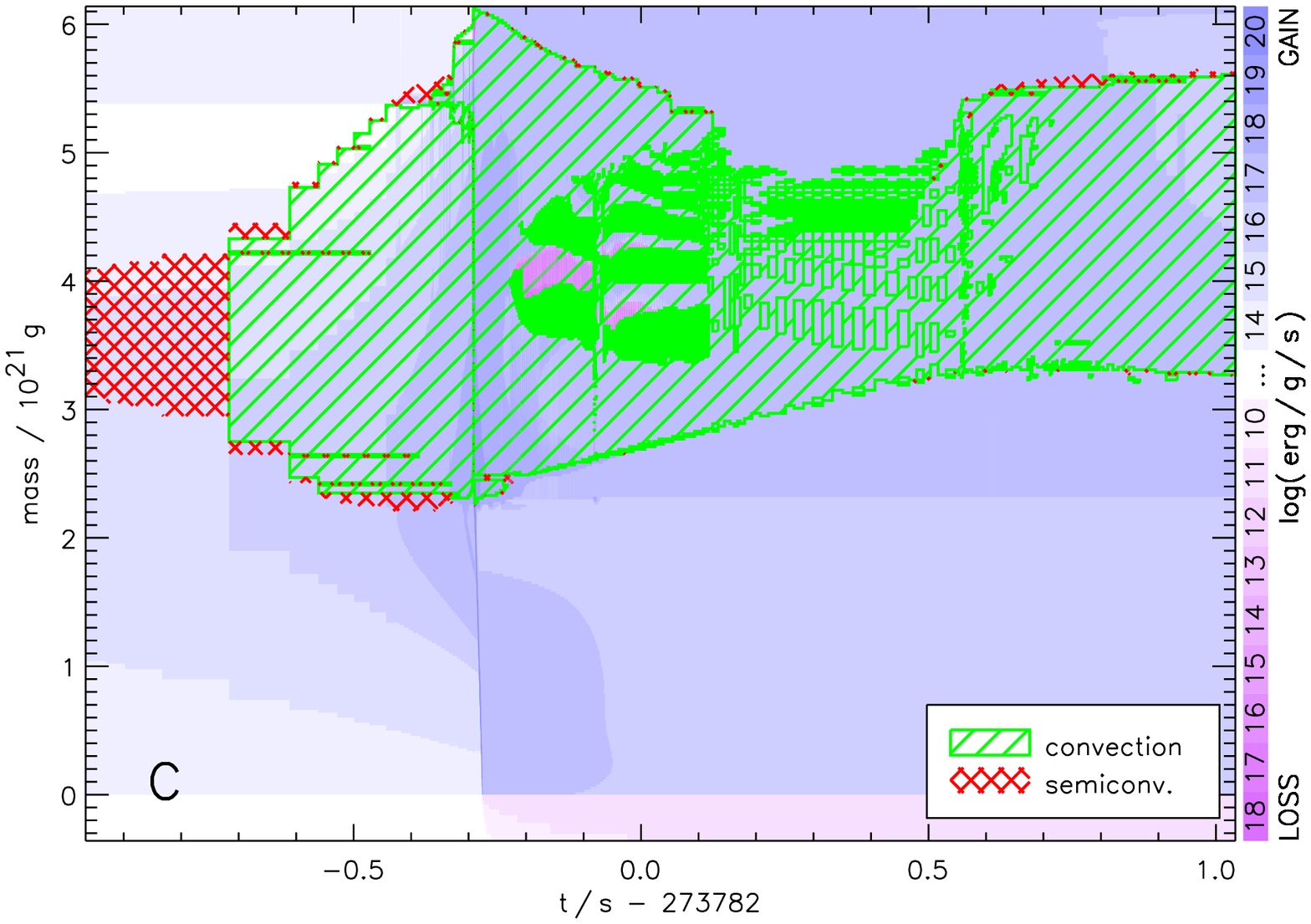}\hfill
\includegraphics[draft=\Draft,angle=0,width=\panelwidth]{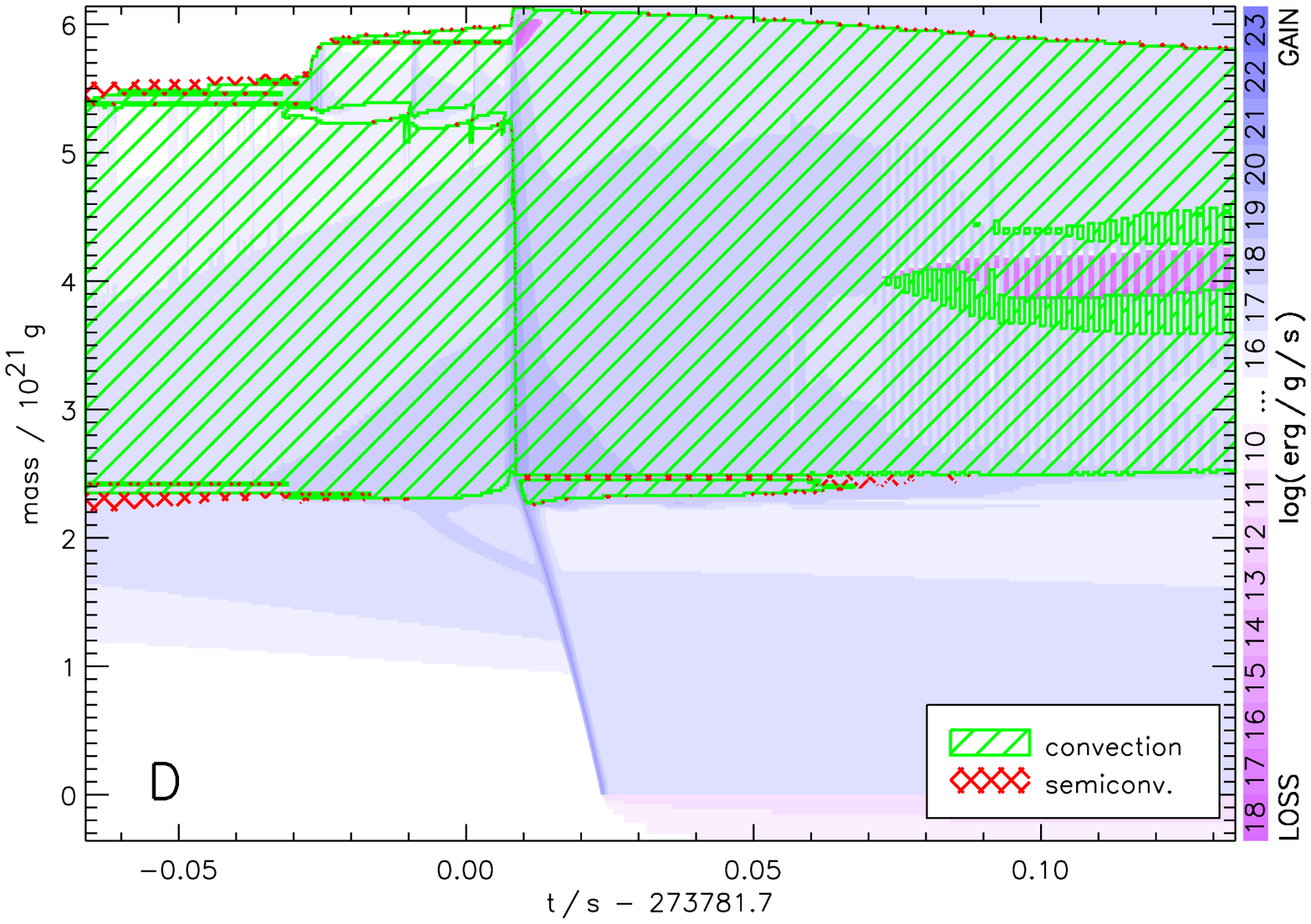}\\[\baselineskip]
\includegraphics[draft=\Draft,angle=0,width=\panelwidth]{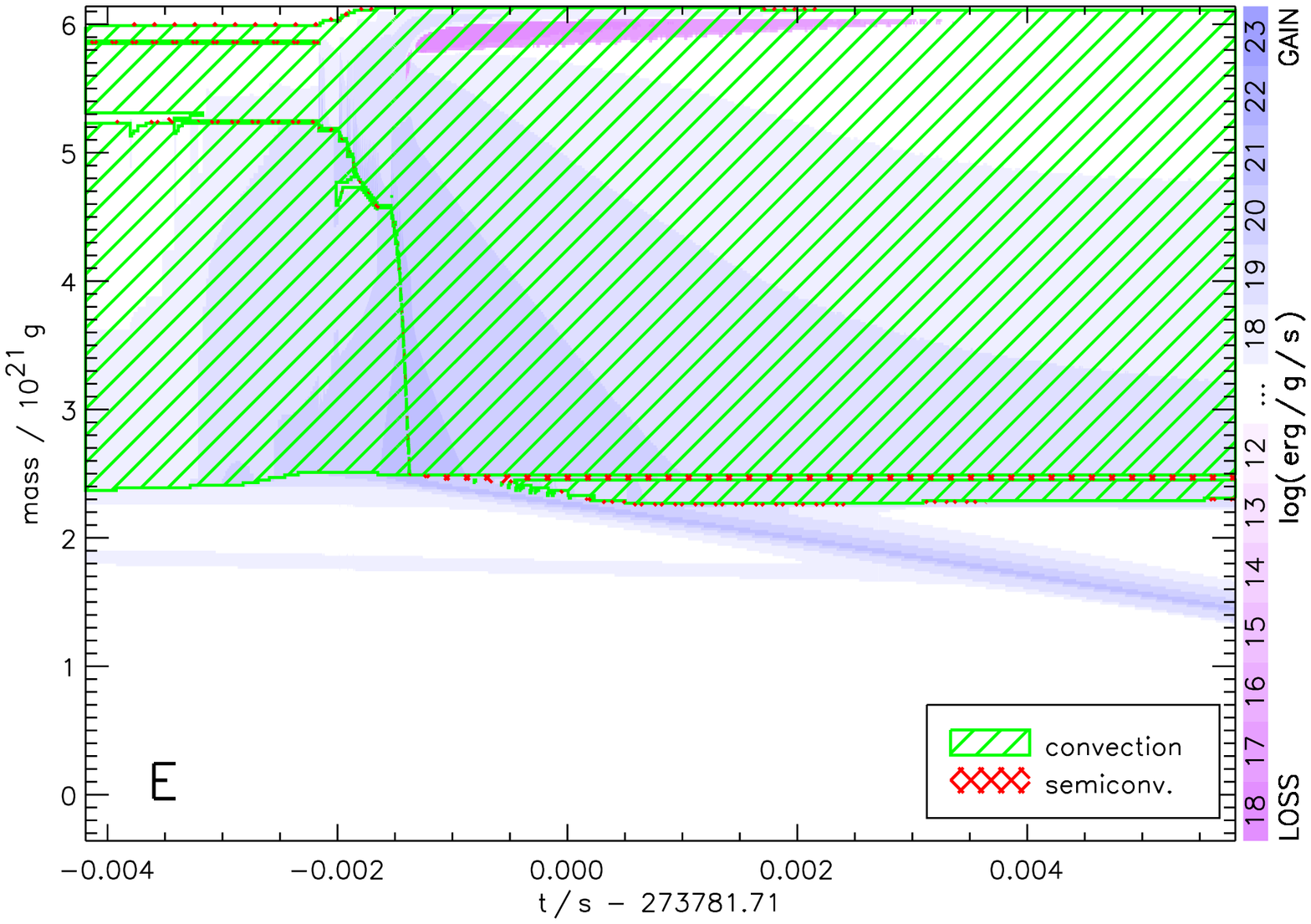}\hfill
\includegraphics[draft=\Draft,angle=0,width=\panelwidth]{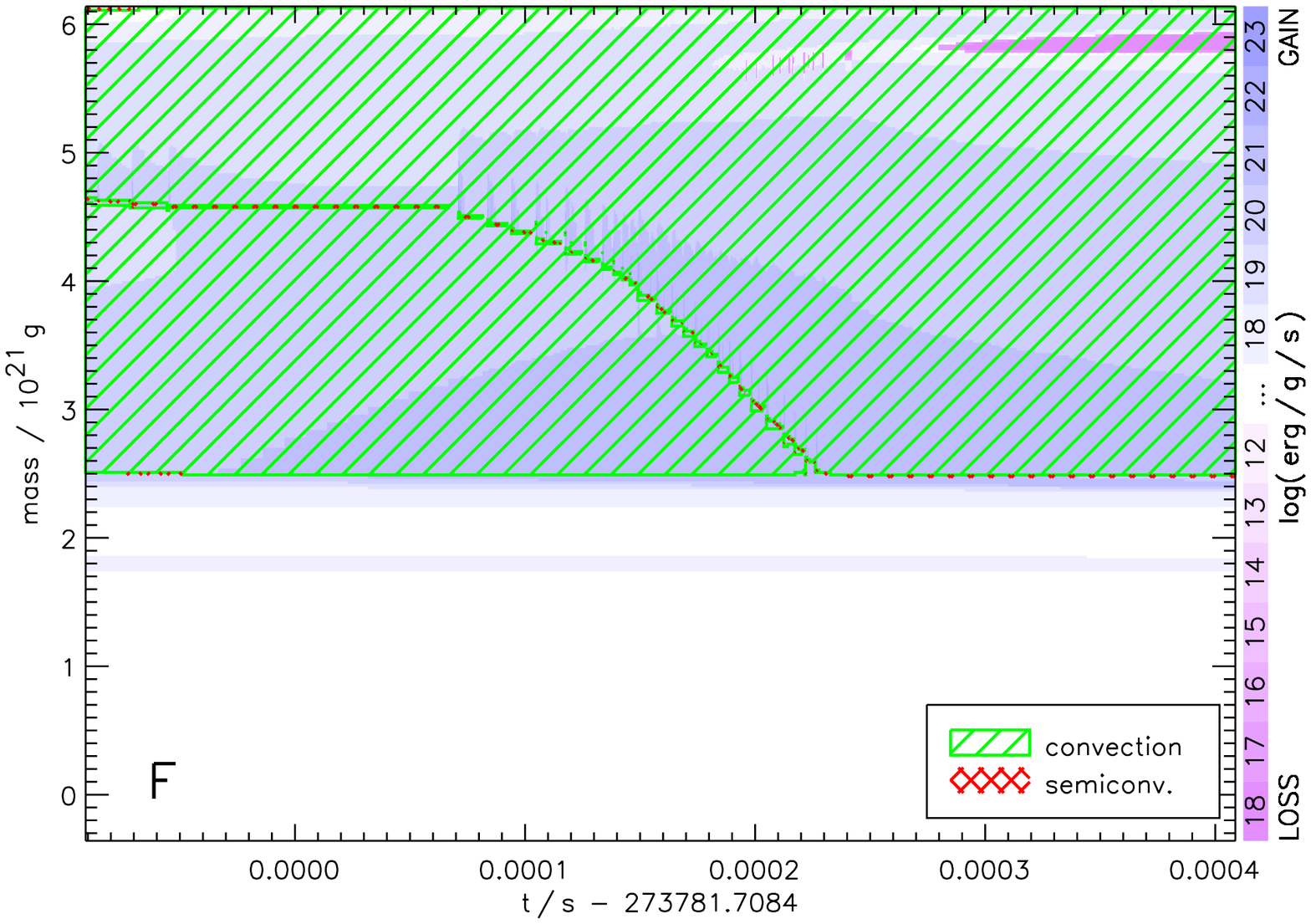}
\caption{Convective structure of the first burst generated in Model
\XRBavi \ viewed on progressively smaller time scales.  Note the
adjustment of the definition of zero time in each frame.  \pan{A:} On
a time scale of minutes the burst is followed by an extended period of
slow burning (indicated by blue color) and mixing by thermohaline
convection (yellow cross-hatching); \pan{B:} The light curve itself
lasts about $13\,\Sec$ (similar to \Fig{lite6a}, which is for the
second burst) and the burning layer is at least partly convective
about half that time.  The runaway and convection both begin well
above the bottom of the accreted layer.  The inner $2.3\E{21}\,\g$ of
carbon-rich material is never convective, but does burn radiatively
(note the dark blue band between $2$ and $4.5\,\Sec$).  \pan{C:} At
still higher resolution, one sees the growth, both inwards and
outwards, of the helium burning shell and its collision with the
hydrogen layer at $-0.32\,\Sec$ at $5.38\E{21}\,\g$.  The collision
causes a mild explosion and rapid growth in the convection zone which
bifurcates into hydrogen-rich and helium convective shells.
\pan{D}--\pan{F:} Subsequent frames, at increasing resolution, show
the inward propagation of the base of the hydrogen convective shell.
Many thousands of stellar models were needed to follow this in this
one burst.  By the end, the hydrogen and helium shells are completely
merged.  In the radiative helium-carbon layer, at about $-0.01\,\Sec$
on \Pan{D}, a weak front of \I{22}{Ne} burning by \I{22}{Ne}\Ran and
\I{22}{Ne}\Rag begins to move inward.  This is followed by a
self-sustaining helium burning flame ignited at $+0.01\,\Sec$.  This
flame leaves behind mostly silicon and carbon, but little helium.
Carbon continues to burn after the flame has passed, mostly between
$-0.28\,\Sec$ and $-0.05\,\Sec$ on \Pan{C}.  The fact that convection
persists (above $m\gtrsim2.5\E{21}\,\g$), carrying high luminosity
almost to the surface for over $0.1\,\Sec$ leads to a very rapid rise
in the light curve.  Unlike Model \XRBaiii, where compositional
``inertia'' plays a big role, the critical masses, convective
structures, compositions, and light curves are very similar in
subsequent bursts from Model \XRBavi.  \lFig{lite6a1}}
\end{figure}

\clearpage
%24
\begin{figure} \centering
\includegraphics[draft=\Draft,angle=0,height=0.5\textheight]{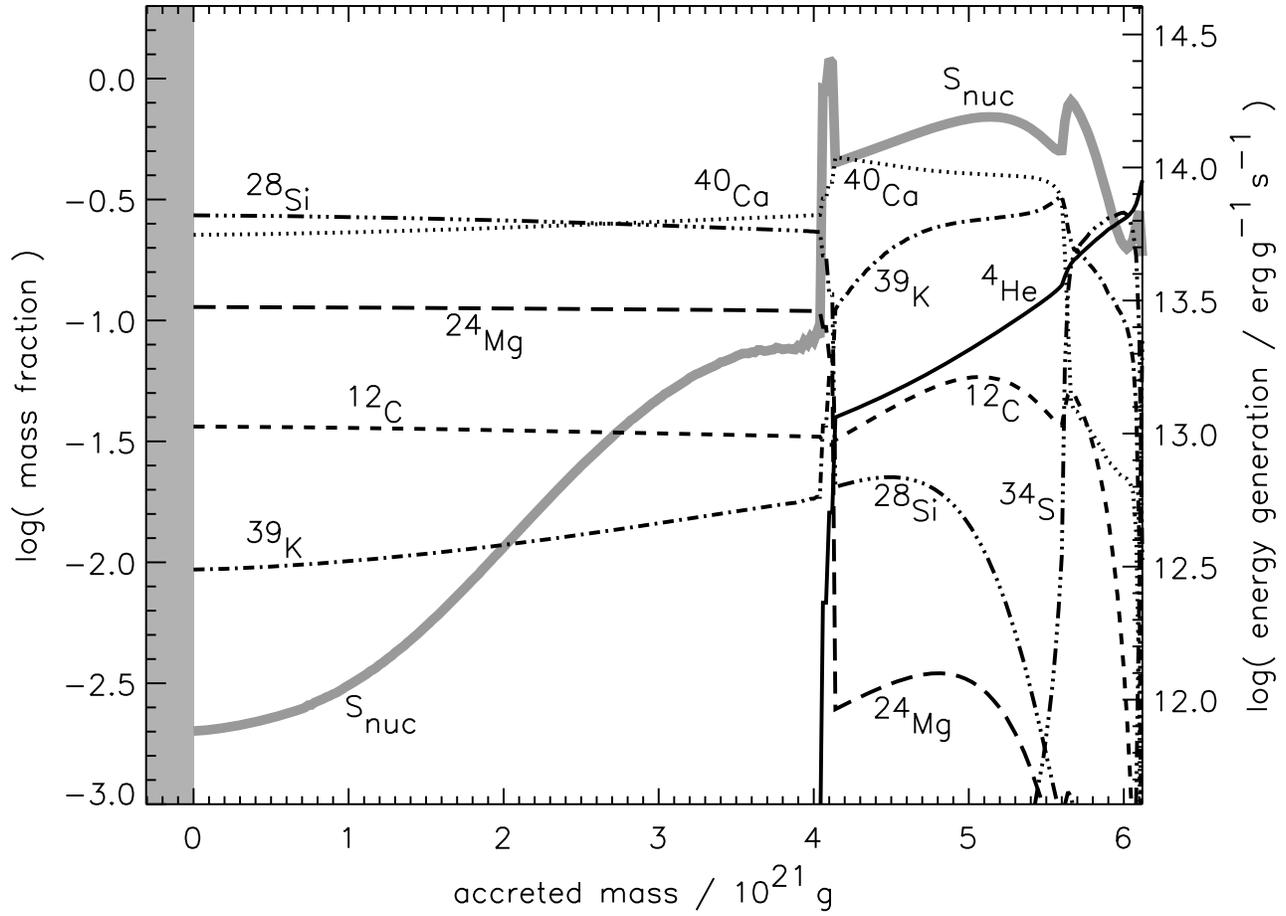}
\caption{Composition following the first burst in Model \XRBavi. The
time is 78 s after the beginning of the burst, well into the tail of
the event whose brilliant display lasted about $30\,\Sec$.  Some slow
burning is still occurring in the helium layer. Interior to
$4\E{21}\,\g$ the layer is well mixed by thermohaline convection due
to a mild inverted gradient in $\bar A$.  Only $4\,\%$ carbon remains
in these inner ashes, but this will survive, since there is no helium
there. Subsequent bursts leave similar ashes. \lFig{compa6f}}
\end{figure}

\clearpage
%25
\begin{figure} \centering
\includegraphics[draft=\Draft,angle=0,height=0.8\textheight]{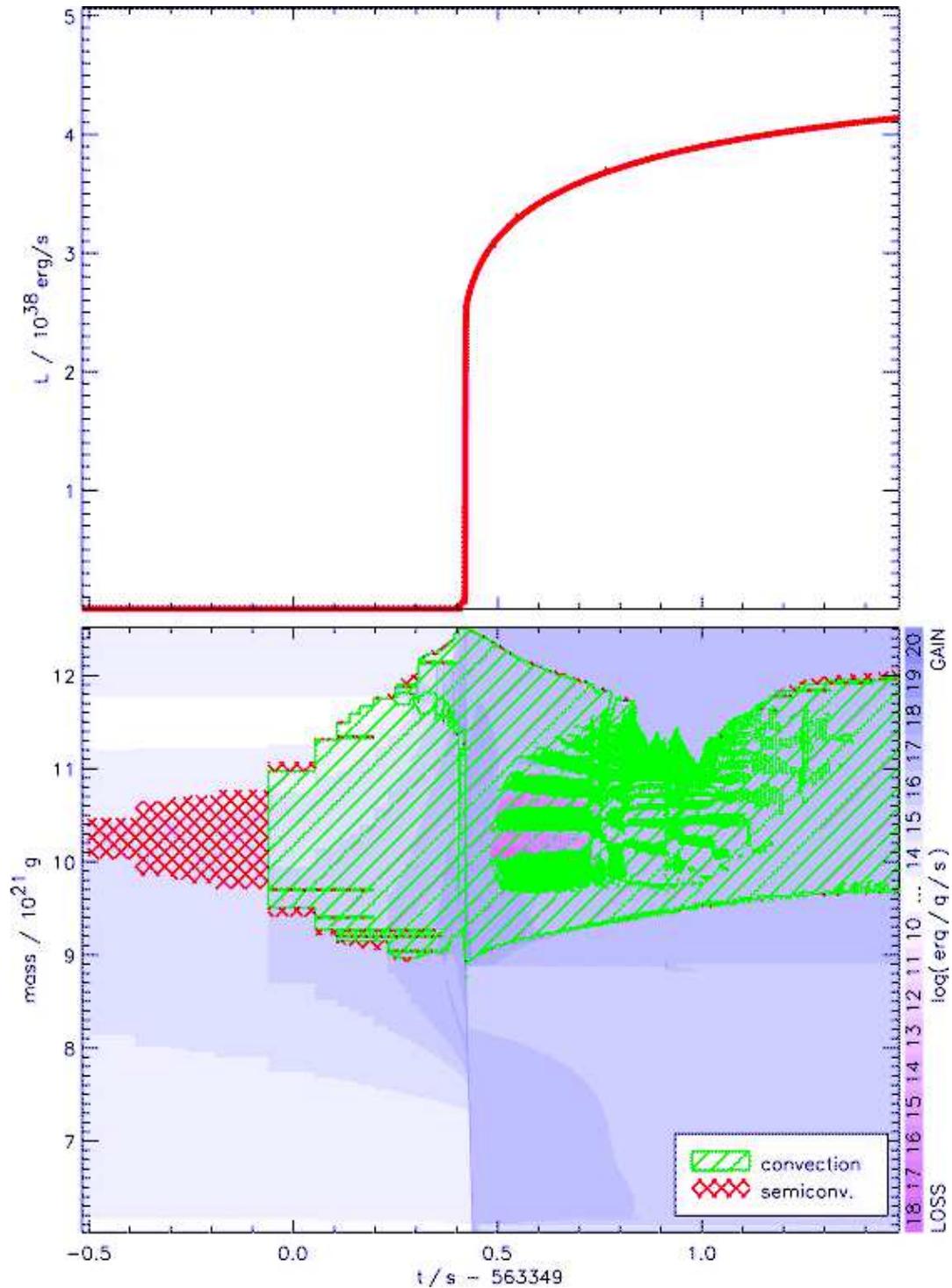}
\caption{Rise of the light curve for the second burst generated in
Model \XRBavi. The light curve of the first burst is similar. The rise
from $\Ep{37}\,\ergs$ to $\E{38}\,ergs$ takes less than 1 ms
(neglecting the propagation time around the neutron star). The rise from
$1\E{38}\,\ergs$ to $2\E{38}\,\ergs$ takes about $100\,\ms$.  Above
the Eddington luminosity, $2\E{38}\,\ergs$, there will be radius
expansion not properly followed in the present study.  The excess
energy will go into driving a wind. \lFig{lite6as}}
\end{figure}

\clearpage
%26
\begin{figure} \centering
\includegraphics[draft=\Draft,angle=0,height=0.8\textheight]{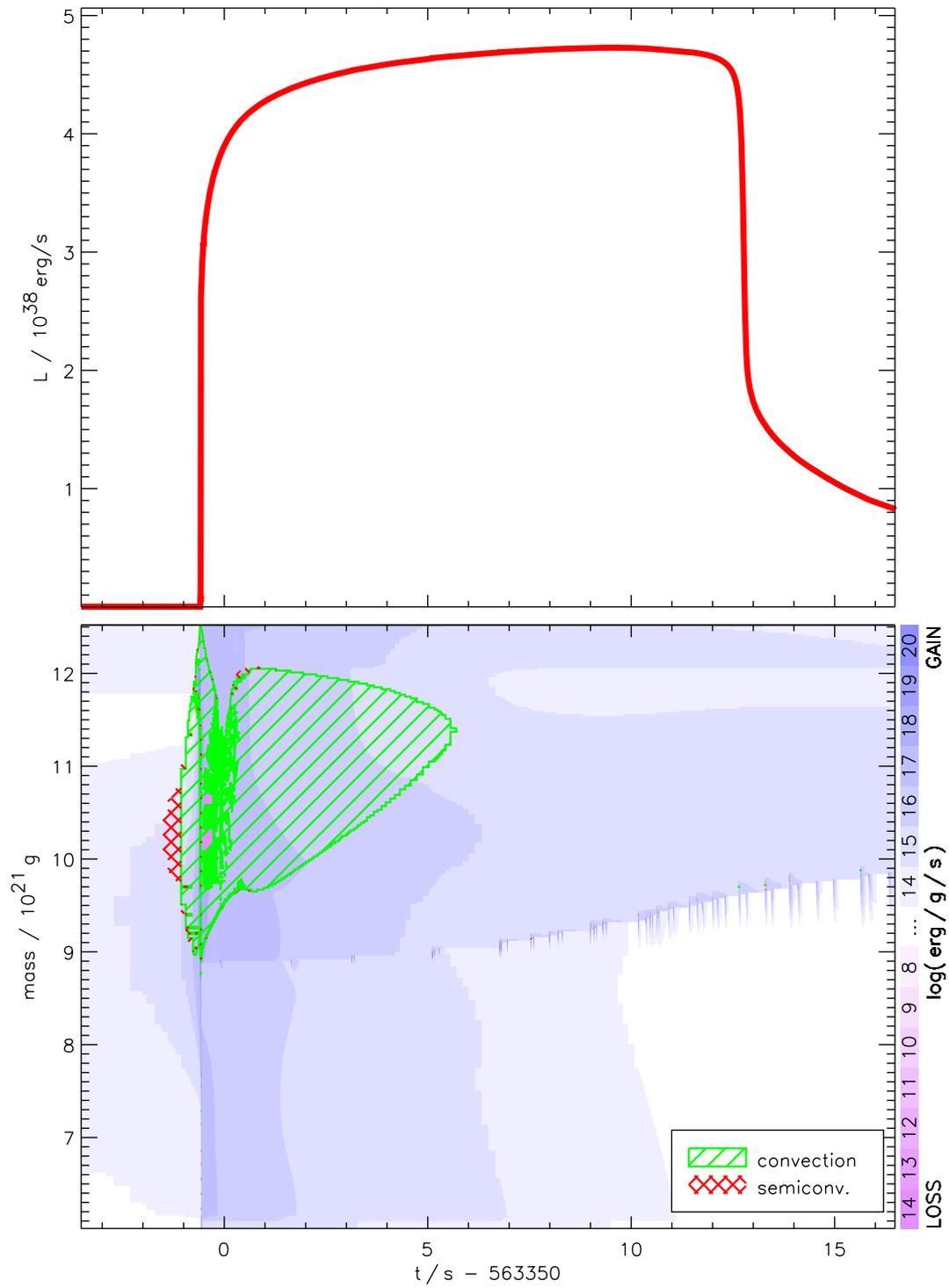}
\caption{Full light curve for the second burst generated in Model
\XRBavi. \lFig{lite6a}}
\end{figure}

\clearpage
%27
\begin{figure} 
\includegraphics[draft=\Draft,angle=90,width=\columnwidth]{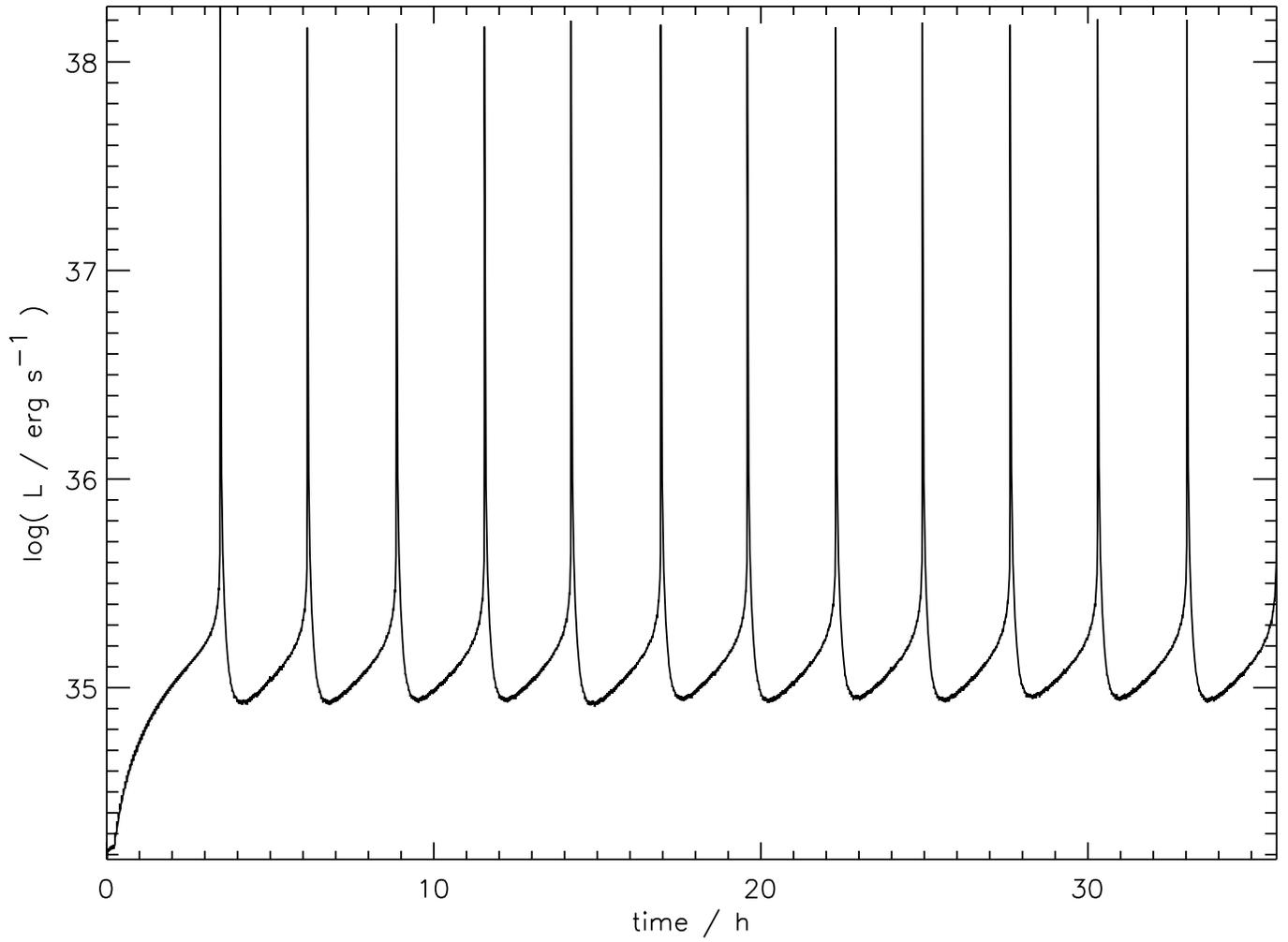}
\caption{Repeated flashes from Model~\XRBav. During the inter-pulse
period the luminosity from accretion ($2.0\E{37}\,\ergs$ ) would
obscure the thermal emission from the cooling ashes that is
shown. \lFig{Brep1}}
\end{figure}

\end{document}